\documentclass[fleqn,usenatbib]{mnras}

\usepackage{newtxtext,newtxmath}

\usepackage[T1]{fontenc}

\DeclareRobustCommand{\VAN}[3]{#2}
\let\VANthebibliography\thebibliography
\def\thebibliography{\DeclareRobustCommand{\VAN}[3]{##3}\VANthebibliography}

\usepackage{graphicx}    
\usepackage{multicol}        
\usepackage{bm}        
\usepackage{pdflscape}    
\usepackage[inline]{enumitem}
\usepackage{geometry}
\usepackage{float}
\usepackage{hyperref}
\usepackage{breqn}
\usepackage{multirow}
\usepackage{cite}
\usepackage{mwe}
\usepackage{booktabs}
\usepackage{mdframed}
\usepackage[dvipsnames]{xcolor}
\usepackage[export]{adjustbox}
\usepackage{tabularx}

\definecolor{oceanboatblue}{rgb}{0.0, 0.47, 0.75}

\newcommand{\kms}{\,km\,s$^{-1}$} 

\newcommand{\velociraptor}{\textsc{VELOCIraptor}}
\newcommand{\hi}{\textsc{Hi}}

\newcommand{\mhi}{$M_{\rm HI}$}
\newcommand{\shark}{\textsc{Shark}}
\newcommand{\lcdm}{$\Lambda \rm CDM$}
\newcommand{\surfs}{\textsc{surfs}}
\newcommand{\M}{${\rm M}_{\odot}$}
\newcommand{\solarValue}[1]{$10^{\rm #1}$\M}
\newcommand{\subsuperscript}[3]{$#1^{\rm #3}_{\rm #2}$}

\title[HI-Stacking: Systematics at play]{Unveiling the atomic hydrogen--halo mass relation via spectral stacking}

\author[G. Chauhan et al.]{Garima Chauhan,$^{1,2}$\thanks{Contact e-mail: \href{mailto:garima.chauhan@icrar.org}{garima.chauhan@icrar.org}}
Claudia del P. Lagos,$^{1,2}$
Adam R. H. Stevens,$^{1,2}$
Mat\'ias Bravo,$^{1}$
Jonghwan Rhee,$^{1,2}$\newauthor{
Chris Power,$^{1,2}$
Danail Obreschkow,$^{1,2}$
Martin Meyer$^{1,2}$}
\\
$^{1}$International Centre for Radio Astronomy Research, The University of Western Australia, 35 Stirling Highway, Crawley, WA 6009, Australia \\
$^{2}$ARC Centre of Excellence for All Sky Astrophysics in 3 Dimensions (ASTRO 3D)}

\date{Accepted XXX. Received YYY; in original form ZZZ}

\pubyear{2020}

\begin{document}
\label{firstpage}
\pagerange{\pageref{firstpage}--\pageref{lastpage}}
\maketitle

\graphicspath{{Figures/}}

\begin{abstract}
Measuring the \hi--halo mass scaling relation (HIHM) is fundamental to understanding the role of \hi\ in galaxy formation and its connection to structure formation. While direct measurements of the \hi\ mass in haloes are possible using \hi-spectral stacking, the reported shape of the relation depends on the techniques used to measure it (e.g. monotonically increasing with mass versus flat, mass-independent). Using a simulated \hi\ and optical survey produced with the \shark\ semi-analytic galaxy formation model, we investigate how well different observational techniques can recover the intrinsic, theoretically predicted, HIHM relation. We run a galaxy group finder and mimic the \hi\ stacking procedure adopted by different surveys and find we can reproduce their observationally derived HIHM relation. However, none of the adopted techniques recover the underlying HIHM relation predicted by the simulation. We find that systematic effects in halo mass estimates of galaxy groups modify the inferred shape of the HIHM relation from the intrinsic one in the simulation, while contamination by interloping galaxies, not associated with the groups, contribute to the inferred \hi\ mass of a halo mass bin, when using large velocity windows for stacking. The effect of contamination is maximal at \subsuperscript{M}{vir}{} $\sim 10^{12-12.5}\rm M_{\odot}$. Stacking methods based on summing the \hi\ emission spectra to infer the mean \hi\ mass of galaxies of different properties belonging to a group suffer minimal contamination but are strongly limited by the use of optical counterparts, which miss the contribution of dwarf galaxies. Deep spectroscopic surveys will provide significant improvements by going deeper while maintaining high spectroscopic completeness; for example, the WAVES survey will recover $\sim 52$\% of the total \hi\ mass of the groups with \subsuperscript{M}{vir}{} $\sim$ \solarValue{14} compared to $\sim 21$\% in GAMA.

\end{abstract}



\section{Introduction}
\label{sec:Introduction}

Since the establishment of the $\Lambda$ cold dark matter model (hereafter \lcdm) as the standard cosmological model, our understanding of galaxy formation and evolution is shaped by the idea that galaxies form, evolve, and merge within host dark matter (DM) haloes.
It is well known that gravitational instability led to the growth of overdensities in the primordial matter density field, further leading gas to decouple from DM and dissipate, falling to the gravitational potential centre, and then cooling and forming stars, which eventually grow into galaxies \citep[see][]{Baugh2006, Benson2010}. Thus, the growth, internal properties, and spatial distribution of galaxies and DM haloes are closely linked \citep[see review by][]{Wechsler2018-galaxyDMhaloes}. This link between galaxy properties and DM haloes has led to the establishment of some fundamental scaling relations, which characterise the dependence of the abundance on baryons to the host halo mass. The stellar--halo mass relation (SMHM) is the most well-studied of these relations \citep[e.g.][]{Behroozi2010-Stellar-halo-mass, Moster2010-StellarHalo-mass}. Our focus in this paper is instead on the less-explored neutral atomic hydrogen--halo mass scaling relation (HIHM). Neutral atomic hydrogen (\hi) plays a fundamental role in our understanding of galaxy formation and evolution. Not only is it the raw fuel for galaxies, it also is an excellent tracer of galactic interactions and the underlying DM distribution \citep[e.g.][]{Gunn:1972, Chung:2009, Denes:2016}. 

Unlike the SMHM relation, the HIHM relation is much more complex, which we can see from the correlation between \hi\ mass and stellar mass in galaxies, which is characterised by a large scatter \citep[e.g.][]{Catinella2010, Brown2015TheGalaxies, Brown2017ColdClusters}, and thus, understanding the HIHM complexity is of particular interest. The form of the HIHM relation has been explored in various empirical studies, using for example, \hi\ clustering measurements \citep{Padmanabhan2017ConstrainingHydrogen, Obuljen2019TheFromALFALFA}, which infer the HIHM relation by determining the \hi\ associated with a halo by matching it with haloes that have the same clustering strength. This requires a fair amount of modelling, as the shape of the clustering of DM haloes of a given mass can significantly deviate from that of the \hi\ clustering. To solve this, additional halo properties have been invoked, e.g. assembly age \citep[see][]{Guo2017ConstrainingClustering}.

A more direct way of measuring the \textit{mean} \hi\ content of groups is by employing \hi\ spectral stacking, whereby one co-adds the \hi\ detections and non-detections associated with an optically identified group in an attempt to boost the signal-to-noise ratio, thereby recovering a more statistically significant detection. This does not need prior assumptions about relationships between halo properties and \hi\ as required when using \hi\ clustering. Recently, \citet[][]{Guo2020} calculated the \hi\ content of groups identified in an optical redshift survey by using an estimate of the halo radius to choose the sky-projected area within which \hi\ emission was stacked. \citet[][]{Guo2020} used SDSS \citep[The Sloan Digital Sky Survey;][]{York2000-SDSS} as their spectroscopic survey and ALFALFA \citep[Arecibo Legacy Fast ALFA; ][]{Giovanelli2005, Haynes2018TheCatalog} for their \hi\ stacking measurements. They first divided their galaxy groups (as were presented in the \citealp{Lim2017-SDSS-group}) according to the number of member galaxies, and then derived the HIHM relation for each subset. They found a monotonically increasing HIHM relation, which starts to plateau at \subsuperscript{M}{vir}{} $\geq$ \solarValue{13.2}. At those high halo masses, \citet[][]{Guo2020} found that all selections in the number of member galaxies produce the same relation (due to haloes of these masses always having a large number of member galaxies). \citet[][]{Guo2020} also found that as they select haloes to have a high occupation of galaxies (i.e. $N_{\rm g} \geq 5$, $N_{\rm g} \geq 6$), the inferred HIHM relation flattens across the whole halo mass range ($\gtrsim$ \solarValue{12}).   

Another way of estimating the \hi\ content of a group, employed by Rhee et al. (in prep), is through stacking the \hi\ spectra of individual galaxies identified as belonging to that group. Rhee et al. (in prep) used the GAMA \citep[Galaxy And Mass Assembly;][]{Driver2011-GAMA,Liske2015-GAMA} survey as their spectroscopic survey and, the ASKAP DINGO \citep[Deep Investigation of Neutral Gas Origin;][]{Meyer-2009-DINGO} early science observations for their \hi\ stacking. In this method, the spectra of galaxies are divided in host halo mass bins (based on the group catalogue presented in \citealp{Robotham2011GalaxyG3Cv1}), and then stacked to measure the mean \hi\ mass of central/satellite galaxies within that halo mass bin. To then measure a total \hi\ mass for the haloes in a bin, the mean \hi\ mass is simply multiplied by the number of central/satellite galaxies belonging to groups in that halo mass bin, followed by summing up the total \hi\ mass of central/satellite galaxies in a halo mass bin. Unlike \citet[][]{Guo2020}, Rhee et al. (in prep) find a non-monotonic HIHM relation, which dips at \subsuperscript{M}{vir}{} $\sim$ \solarValue{11.8}, followed by an increasing \hi\ mass from \subsuperscript{M}{vir}{} $\gtrsim$ \solarValue{12.8}.

These are innovative new ways to further our understanding of the HIHM relation and provide valuable constraints for it. Although, novel in approach, these two techniques yield very different measurements of the HIHM relation. For example, there is a difference of $\sim 0.5$ dex in the measured mean \hi\ mass at \subsuperscript{M}{vir}{} $\sim$ \solarValue{12.8} between \citet[][]{Guo2020} and Rhee et al. (in prep). It has so far been unclear whether this is simply due to the difference in their methodologies. We explicitly address this in this paper.

The HIHM relation is of particular interest to theorists as well, and has been extensively investigated using various theoretical models, ranging from semi-analytic models of galaxy formation \citep[SAMs; e.g.][]{Baugh2018-PMillennium, Spinelli2019_Marta, Chauhan2020} to hydrodynamical simulations \citep{Villaescusa-Navarro2018}. Simulations have the power to investigate not only the average HIHM relation, but also the scatter associated with it. Although the shape and scatter of the HIHM relation is model-dependent \citep[][]{Chauhan2020}, it has been reported independently that the shape and scatter of the HIHM relation is heavily influenced by feedback from Active Galactic Nuclei (AGN) \citep[see][]{Baugh2018-PMillennium, Spinelli2019_Marta, Chauhan2020}. \citet{Spinelli2019_Marta} find that the HIHM relation depends on the detailed assembly history of haloes, which agrees with inferences based on  \hi\ clustering studies in  \citet{Guo2017ConstrainingClustering}. \citet{Chauhan2020} find that the scatter around the HIHM relation can be constrained instead using halo spin and the ratio of subhalo mass to host halo mass.

It is evident that progress has been made in our understanding of the HIHM relation on both observational and theoretical fronts, independently. On both of these fronts, some challenges need to be overcome. For example, \hi\ stacking is a very powerful tool for making \hi\ measurements of groups, but it heavily relies on group finders and halo mass estimates based on optical redshift surveys, which may sometimes provide inaccurate results due to survey limitations \citep{Campbell2015-groupfinder-comparison, Bravo2020}. Interestingly, simulations and models, which do not suffer from observational limitations, show a range of different HIHM relations \citep[see figure 2 in ][]{Chauhan2020}, which are significantly different from each other and depend on the models employed in the simulations. Here, we try to bridge the gap between the observed and predicted HIHM relations, by producing a mock survey made using the state-of-the-art semi-analytic model of galaxy formation, \shark\ \citep{Lagos2018-Shark, Chauhan2019}, and deriving mock-observed HIHM relations that mimic the techniques of \citet[][]{Guo2020} and Rhee et al. (in prep). Our main aim is to see if we can reproduce the observed HIHM relation using our mock survey and try to understand the limitations and advantages of the stacking techniques used in the observational HIHM literature. As we will show, the HIHM relation that we derive from the mock survey is notably different from \shark's intrinsic HIHM relation. We explore the origin of this difference, and test whether there is an ``optimal-'' observing method that can minimise it.

The structure of this paper is as follows. Section~\ref{sec:simulated-galaxy-catalogue_chap5} details the construction of the mock survey and mock group catalogue used for our analysis. Section~\ref{sec:virial-mass-allocation_chap5} compares the halo mass estimates made for groups by different methods against the intrinsic halo mass of groups. In Section~\ref{sec:HI-SIMULATION_chap5}, we delve into the systematic effects involved in \hi~stacking measurements and compare different \hi~stacking techniques with the intrinsic prediction of \shark. In Section~\ref{sec:Why-parameters-cause-change}, we discuss the causes of the discrepancy seen between the intrinsic and the mock-observed HIHM relation. We draw our conclusions in Section~\ref{sec:conclusion_chap5}.

\section{The simulated galaxy catalogue}
\label{sec:simulated-galaxy-catalogue_chap5}

In order to make a fair comparison with the \hi~stacking done observationally, it is imperative to replicate stacking procedures on our simulated ``mock surveys". For this, it is necessary to build a simulated lightcone that replicates the limitations of the optical surveys that serve as a base for the \hi~stacking experiments. In the following subsections, we describe how these are built.

\subsection{The \shark\ semi-analytical model of galaxy formation}
\label{sec:shark-sam_chap5}

We use the semi-analytical model of galaxy formation `\shark'\ \citep{Lagos2018-Shark} to provide us with the simulated galaxies for our \hi~stacking experiment. SAMs, such as \shark, use halo merger trees, which are produced from a cosmological DM-only $N$-body simulation, to follow the formation and evolution of galaxies by solving a set of equations that describe exchange of mass, metals and angular momentum produced by a series of physical processes. In the following sections, we describe the DM-only $N$-body simulation (\surfs) that \shark\ is run on, followed by a description of the baryon physics  included in \shark. We end the section with a brief description of how the spectral energy distribution (SED) of galaxies is computed in \shark.


\subsubsection{The \surfs\ N-body suite}
\label{subsubsec:surfs_chap5}

We use the \surfs\ suite of DM-only $N$-body simulations \citep{Elahi_SURFS}, which consists of $N$-body simulations of differing volumes, from $40$ to $210$ \subsuperscript{h}{}{-1}\ cMpc (comoving megaparsec) on a side, and particle numbers ranging from $\sim 130$ million up to $\sim 8.5$ billion. The simulations adopt the \lcdm\ \citet{PlanckXIII} cosmology, which assumes total matter, baryon and dark energy densities of $\Omega_{ m} = 0.3121$, $\Omega_{ b} = 0.0491$ and $\Omega_{ \Lambda } = 0.6879$, respectively, and a dimensionless Hubble parameter of $h = 0.6751$. In this paper, we use the L210N1536 (hereafter referred to as `medi-\surfs') simulation, which has a box size of $210$  \subsuperscript{h}{}{-1} cMpc, \subsuperscript{1536}{}{3}\ DM particles with a mass of $2.21 \times 10^{8}$ \subsuperscript{h}{}{-1}\M\ and a softening length of $4.5$ \subsuperscript{h}{}{-1} ckpc (comoving kiloparsec). \surfs\ contains $200$ snapshots for each simulation, with a typical time span between snapshots in the range $6-80$ Myr. 

Merger trees and halo catalogues were constructed using the phase-space finder \velociraptor\footnote{{\url{https://github.com/icrar/VELOCIraptor-STF/}}} \citep{Elahi2019-Velociraptor, Canas2019-VELOCIRAPTOR} and the halo merger tree code \textsc{treefrog}\footnote{\href{https://github.com/pelahi/TreeFrog}{\url{https://github.com/pelahi/TreeFrog}}} \citep{Poulton_Treefrog2018,Elahi2019-TreeFrog}, developed to work with \velociraptor. It has been shown in \citet{Poulton_Treefrog2018} that \textsc{treefrog} + \velociraptor\ lead to well behaved merger trees, with orbits that are well reconstructed. We refer the readers to \citet{Lagos2018-Shark} for more details on how the merger trees and halo catalogues are constructed for \shark, and to \citet{Elahi_SURFS,Elahi2019-Velociraptor, Canas2019-VELOCIRAPTOR, Poulton_Treefrog2018} for more details on the \velociraptor\ and \textsc{treefrog} software.

\subsubsection{Baryon physics in \shark}
\label{subsubsec:baryon-physics-shark_chap5}

\shark\footnote{\href{https://github.com/ICRAR/shark}{\url{https://github.com/ICRAR/shark}}} is an open-source, flexible and highly modular SAM that models the key physical processes of galaxy formation and evolution. These include \begin{enumerate*}[label=(\roman*)]
    \item the collapse and merging of DM haloes;
    \item the accretion of gas onto haloes, which is governed by the DM accretion rate;
    \item the shock heating and radiative cooling of gas inside DM haloes, leading to the formation of galactic discs via conservation of specific angular momentum of the cooling gas;
    \item the formation of a multi-phase interstellar medium and subsequent star formation (SF) in galaxy discs;
    \item the suppression of gas cooling due to photo-ionisation;
    \item chemical enrichment of stars and gas;
    \item stellar feedback from evolving stellar populations;
    \item the growth of supermassive black holes (SMBH) via gas accretion and merging with other SMBHs;
    \item heating by AGN;
    \item galaxy mergers driven by dynamical friction within common DM haloes, which can trigger bursts of SF and the formation and/or growth of spheroids; and
    \item the collapse of globally unstable discs leading to bursts of SF and the creation and/or growth of bulges. 
\end{enumerate*} 

\shark\ also includes several different prescriptions for gas cooling, AGN feedback, stellar and photo-ionisation feedback, and SF. \shark\ adopts a universal \citet[][]{Chabrier2003_IMF} IMF (Initial Mass Function). \shark\ uses these models to compute the exchange of mass, metals, and angular momentum between the key baryonic reservoirs in haloes and galaxies, which include hot and cold halo gas, the galactic stellar and gas discs and bulges, central black holes, as well as the ejected gas component that tracks the baryons that have been expelled from haloes. 

The models and parameters used in this study are the \shark\ defaults, as described in \citet{Lagos2018-Shark} and used in \citet{Chauhan2019,Chauhan2020} to study the \hi\ content of galaxies and haloes. These have been calibrated to reproduce the $z = 0,\ 1,\ \text{and}\ 2$ stellar mass functions; the $z = 0$ black hole--bulge mass relation; and the disc and bulge mass--size relations. This model also successfully reproduces a range of observational results that are independent of those used in the calibration process. These include the total atomic and molecular hydrogen--stellar mass scaling relations at $z=0$; the cosmic star formation rate (SFR) density evolution up to $z \approx 4$; the cosmic density evolution of the atomic and molecular hydrogen at $z \leq 2$ or higher in the case of the latter; the mass--metallicity relations for gas and stellar content; the contribution to the stellar mass by bulges; and the SFR--stellar mass relation in the local Universe. \citet{Davies2019} show that \shark\ reproduces the scatter around the main sequence of star formation in the SFR--stellar mass plane; \citet{Chauhan2019} show that \shark\ can reproduce the \hi~mass and velocity widths of galaxies observed in the ALFALFA survey; and \citet{Amarantidis2019} show that the predicted AGN luminosity functions (LFs) agree well with observations in X-rays and radio wavelengths.

In all SAMs, including \shark, the galaxies are assumed to be described by two components, namely a disc and a bulge. The point of difference between these two components is the mechanism involved in their formation, with the discs building their stellar mass by consuming the gas accreted from the halo onto the galaxy, and the bulges forming stars from the gas that is dumped into them during global disc instabilities and galaxy mergers, also acquiring the stellar mass of the satellite in the latter case. Both discs and bulges form stars based on the surface density of molecular hydrogen, with the only difference being that the bulges have $10$ times higher efficiency of converting molecular gas into stars than discs. This follows observational findings that show starburst galaxies to have a higher efficiency than normal star-forming galaxies \citep[][]{Sargent2014-MolecularGasRedshiftIndependence}. In the default \shark\ model, the pressure relation of \citet{Blitz2006} is used to estimate the radial breakdown between atomic and molecular gas in the galaxies. It should be noted that in \shark\ the stripping of halo gas from galaxies when they become a satellite is instantaneous. Along with the halo gas stripping, the satellite galaxy is also cut-off from cosmological accretion, which will lead to an eventual exhaustion of gas in the galaxy via continuing star formation.   

\subsubsection{Spectral Energy Distribution of galaxies in SHARK}
\label{sec:SED-generation_chap5}

The electromagnetic spectrum produced by the integrated contribution of gas, dust and stars in galaxies, provides information about the formation and evolution of the observed galaxies. This integrated electromagnetic spectrum, also referred to as spectral energy distribution (SED), encompasses information of a  galaxy's stellar population, its interstellar medium (ISM) and the dust distribution \citep[see][for a review]{Conroy2013-review_SED}. As previously mentioned, for \hi~stacking, the groups identified by optical surveys are matched with their \hi\ counterparts, to get an estimate of the \hi\ content of the group contained in a halo. In order to have a fair comparison between simulations and observations, we need to compute luminosities of all \shark\ galaxies, which can then be used to apply optical survey limits and group finding for these mock surveys.  

The process of computing the SEDs for galaxies in \shark\ is detailed in \citet{Lagos2019-SED}, but here we briefly describe how is it done. In \shark, we have access to the star formation histories (SFH) and metallicity histories (ZH) for all the galaxies produced at every simulation snapshot. To produce the SEDs, we use two packages: \textsc{ProSpect}\footnote{\href{https://github.com/asgr/ProSpect}{\url{https://github.com/asgr/ProSpect}}} and \textsc{Viperfish}\footnote{\href{https://github.com/asgr/Viperfish}{\url{https://github.com/asgr/Viperfish}}}. \textsc{ProSpect} \citep{Robotham2020-prospect} combines the GALEXev stellar synthesis libraries \citep{Bruzual2003} and/or EMILES \citep{Vazdekis2016} with the two-component dust attenuation model of \citet{Charlot2000} and dust re-emission using the templates of \citet{Dale2014}, which covers a rest-frame wavelength of up to $1,000 \mu m$. \textsc{Viperfish} is a wrapper for \textsc{ProSpect}, which enables the use of \shark\ SFHs and ZHs to generate a series of desired SEDs through target filters. 

An important novelty of the SED modelling in \shark\ is the way the \citet{Charlot2000} parameters are informed from galaxy properties. Using a 3D radiative transfer analysis of galaxies in the EAGLE hydrodynamical simulations \citep{Schaye2015-EAGLE-reference}, \citet[][]{Trayford2020-EAGLE} found that the resulting attenuation curves could be well fitted with a parametrization \'{a}la \citet{Charlot2000}, in which the parameters scale with the dust surface density and are independent of redshift. \citet{Lagos2020} adopted this parametrisation, and hence every galaxy has its own set of Charlot \& Fall attenuation parameters based on dust surface density. When combined with \textsc{ProSpect} and \textsc{Viperfish}, \citet[][]{Lagos2019-SED} showed that \shark\ can reproduce the panchromatic emission of galaxies throughout cosmic time; most notably, \shark\ reproduces the number counts from the GALEX UV to the JCMT $850$-microns bands, the redshift distribution of sub-millimetre galaxies, and the ALMA bands number counts. \citet{Bravo2020} showed that \shark\ reproduces the optical colour distribution of galaxies across a wide range of stellar masses and redshift reasonably well, along with the fraction of passive galaxies as a function of stellar mass.

\subsection{Mock catalogues of simulated galaxies}
\label{subsec:mock-catalogues_chap5}

To produce mock catalogues that mimic the surveys used for the \hi~stacking experiments in the observations, we embed the galaxy population generated by \shark\ in a survey volume by applying the ALFALFA survey's angular and radial selection function. 

\begin{table}
  \begin{center}
    \caption{Mock-survey parameters}
    \label{tab:table_lightcone}
    \begin{tabular}{p{3.5cm}|p{3.5cm}}
      \toprule 
      \textbf{Lightcone parameter} & \textbf{Value} \\
       \midrule 
      Area coverage & $6900$ \subsuperscript{\rm deg}{}{2}\\
      Redshift range & $0-0.1$ \\
      Apparent magnitude limit  & $r$-band magnitude $\leq$ 19.5 \\
                                & (applied in post processing)\\
      Stellar-mass limit & \subsuperscript{M}{\star}{} $\geq$ \solarValue{5} \\
      \bottomrule 
    \end{tabular}
    \label{tab:mock-survey-parameters}
  \end{center}
\end{table}
To construct the lightcone, we use the code \textsc{stingray} (Obreschkow et al, in prep; \citealp{Chauhan2019}), which is an extended version of the lightcone builder code used in \citet{Obreschkow_2009_ligthcone}. \textsc{stingray} tiles simulation boxes together to build a 3D field along the line-of-sight of an observer. The galaxies are drawn from simulation snapshots corresponding to the closest lookback time, which for this analysis ranges over $z=0-0.1$. To ensure that we have a large statistical sample, we set the area of our lightcone to be $\sim 6900\  \text{\rm deg}^2$ containing all galaxies with \subsuperscript{M}{\star}{} $\geq$ \solarValue{5}. Once this lightcone is built, we compute the SEDs of galaxies as described in Section \ref{sec:SED-generation_chap5}, and apply in post-processing an apparent $r$-band magnitude cut-off of $19.8$ mag. See Table \ref{tab:mock-survey-parameters} for these specifications.

The choice of the magnitude limit is based on the GAMA survey, as it is the deeper of the two optical surveys used in the comparison with observations in this study (the other being SDSS). Accompanied by a large multi-wavelength data set, the GAMA \citep[][]{Driver2011-GAMA, Liske2015-GAMA} survey is a spectroscopic campaign aimed at measuring the redshifts of galaxies with $r$-band AB magnitudes $< 19.5$ at $>98$ per cent completeness up to $z \sim 0.25$. The survey consists of five fields, amounting to a total sky area of $230$ \subsuperscript{\rm deg}{}{2} with $\sim 300,000$ galaxies targeted using the Anglo-Australian Telescope. 


\begin{figure*}
  \centering
  \begin{minipage}{0.49\textwidth}
    \includegraphics[width=\linewidth, trim=0.3cm 0.3cm 0.3cm 0.2cm, clip]{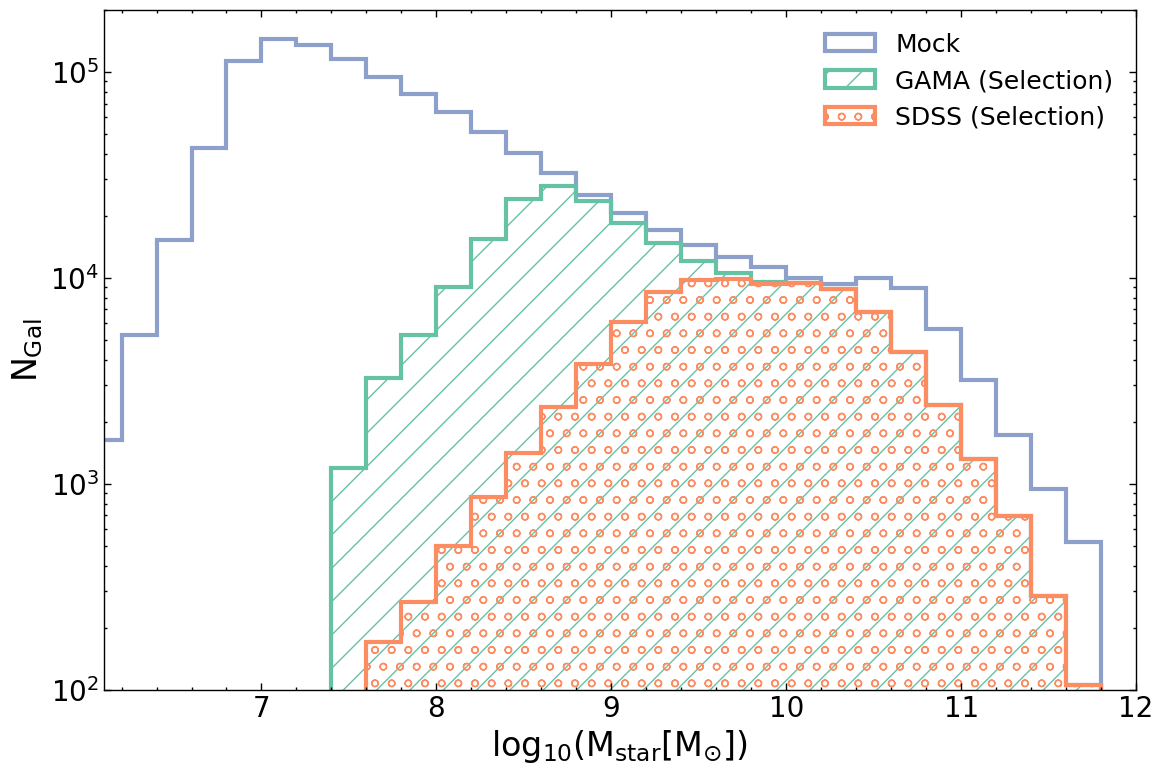}
  
  \end{minipage}
  \hfill
  \begin{minipage}{0.49\textwidth}
    \includegraphics[width=\linewidth, trim=0.3cm 0.3cm 0.2cm 0.2cm, clip]{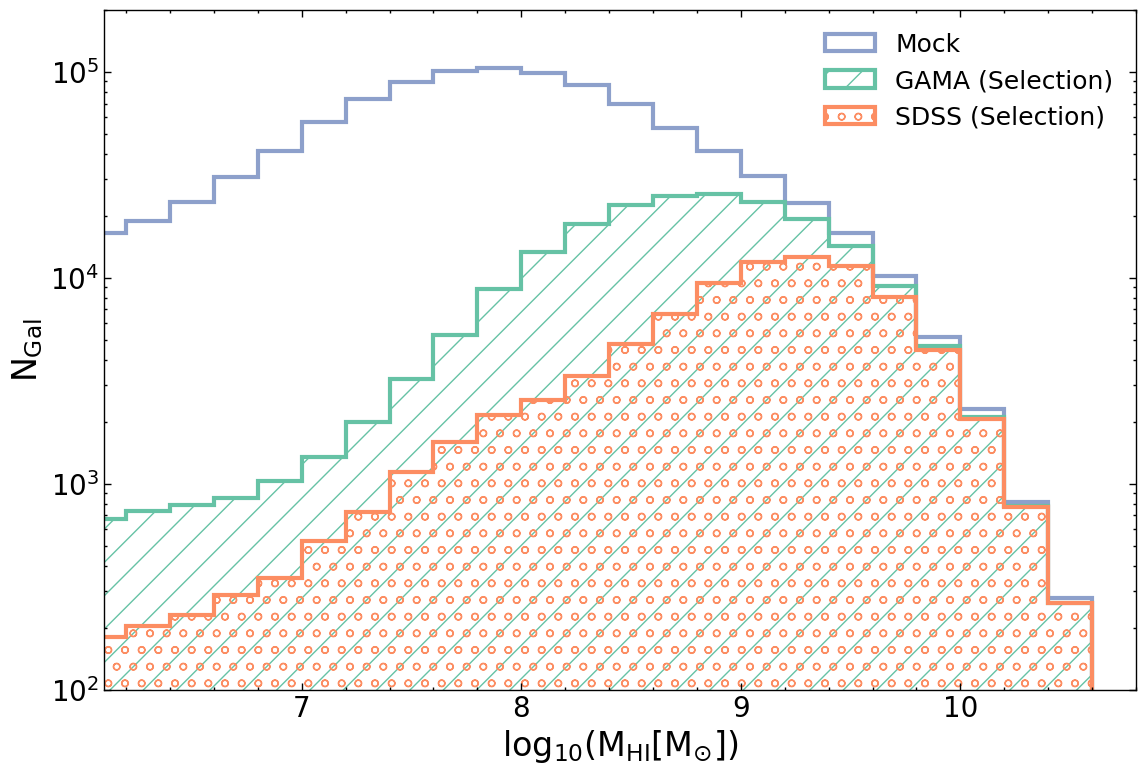}
  \end{minipage}
\caption{Stellar mass (\textit{left panel}) and \hi\ mass (\textit{right panel}) distributions of the central galaxies in the full lightcone constructed using \textsc{stingray}, with the galaxies generated with \shark. `Mock' includes all the centrals present in the lightcone, with `GAMA (selection)' and `SDSS (selection)' corresponding to the GAMA-like, and the SDSS-like mocks (which include the corresponding flux magnitude cut and group finder definition of centrals). See Sections \ref{subsec:mock-catalogues_chap5} and \ref{subsec:group finding-algorithm_chap5} for full description of the mocks.}
\label{fig:stellar_hi_comparison}
\end{figure*}

\subsection{Group-Finding Algorithm}
\label{subsec:group finding-algorithm_chap5}

The concordance \lcdm\ model predicts that all galaxies form and evolve in DM haloes. Galaxy groups are the observable tracer of DM haloes offering direct insight into the physics that occur inside them. Thus, assigning galaxies to groups becomes imperative to further our understanding of galaxy formation and evolution. The galaxies are assigned to groups by employing `group-finding algorithms' which aim to associate galaxies with common DM haloes.


Many large spectroscopic surveys use group finders based on the friends-of-friends (FoF) method, which identifies galaxy systems as member galaxies that are linked by some adopted linkage criteria. There are differences in the way the halo masses of these groups are later estimated, which we will elaborate on in Section \ref{sec:virial-mass-allocation_chap5}. The group-finding algorithm used for our mock survey is the same group finder that was used to construct the GAMA galaxy group catalogue \citep[G$^3$C;][]{Robotham2011GalaxyG3Cv1}. Here we give a brief description of the process of assigning galaxies to groups. 

\begin{enumerate}[label=(\roman*)]
    \item \textit{Step 1:} A luminosity correction is applied to the galaxies that are fed to the group finder. As our mock survey is similar to GAMA in magnitude completeness, a luminosity correction is needed for some \shark\ galaxies, whose luminosities are above the brightest galaxies observed by GAMA. This luminosity correction is required as the group finder produces a luminosity function by integrating the luminosities of galaxies ranging from minimum to the highest detection limit \citep[see][for details]{Robotham2011GalaxyG3Cv1,Bravo2020}, and having brighter than GAMA leads to an integration error.
    
    \item \textit{Step 2:} A FoF algorithm is used to allocate the luminosity-corrected galaxies into groups. Since we work in redshift space, separate linking lengths need to be defined along the line of sight and in projection. These linking lengths are denoted by $l_{z}$ and $l_{p}$, respectively. The linking lengths used by the group finder to define a group scale as a function of the observed density contrast. This leads to linking parameters that depend on both the position of the group and the faint magnitude limit of the survey. These parameters therefore need to be tuned to the survey completeness \citep[see Equation 1-7 in ][]{Robotham2011GalaxyG3Cv1}.
    
    \item \textit{Step 3:} Once galaxies are assigned to a group, an initial estimate of the centre of the group is made by calculating the centre of luminosity. Then the process is iterated after removing the most distant galaxy, and recalculating the centre of luminosity, until only two galaxies are left, of which the brightest is defined as the group central, and the rest of the galaxies are flagged as satellites ranked in the order of their distance from the centre.
\end{enumerate}

For the rest of the paper, we use the `mock group catalogue' generated using the procedure described above for our mock-stacking results. Our mock group catalogue recovers galaxy groups up to $z=0.075$. 
In this paper, we compare the \hi-stacking results presented in \citet{Guo2020} and Rhee et al (in prep) with our mock group catalogue, which is tailored to the completeness of the optical surveys used by these authors. For comparison with Rhee et al. (in prep), we use our current GAMA-like group catalogue, as their \hi~stacking uses GAMA groups as their external optical catalogue. 

\citet{Guo2020}, on the other hand, used the SDSS group catalogue \citep{Lim2017-SDSS-group} as their optical catalogue (see Section~\ref{subsec:obs-stacking-ALFALFA}). SDSS \citep[][]{York2000-SDSS} is a major multi-spectral and spectroscopic survey that covers over $35$ per cent of the sky, and is complete to an apparent $r$-band magnitude of $17.77$ mag. The latest group catalogue, as used by \citet[][]{Guo2020}, contains galaxies from the SDSS DR7 Main Galaxy Sample \citep{Albareti2017-SDSS_DR7}. To make our mock group catalogue more comparable to the SDSS group catalogue, we remove all galaxies that have an $r$-band AB magnitude $> 17.77$, and adapt the number of galaxies associated with each group accordingly. A caveat of using this approach is that the groups in the SDSS-like selection will actually be better recovered than the recovery of groups in the original SDSS catalogue, i.e. the groups recovered will be better matched to the underlying mock. This is due to us using a deeper survey to define our initial groups, and then selecting galaxies that are above the SDSS magnitude limit within those groups. We leave the creation of a consistent group catalogue for the flux limit of SDSS to future work, as this implies re-calibrating to the observed $r$-band luminosity distribution as described in \citet[][]{Bravo2020}.

To confirm that the group finder recovers the underlying baryon distribution, in Figure \ref{fig:stellar_hi_comparison} we plot the distribution of the stellar and \hi\ masses of the central galaxies present in the lightcone for all of the galaxies in \shark\ catalogue (using all the central galaxies in the lightcone), comparing this against the same distributions after applying the FoF group finder and GAMA/SDSS flux limits. We would expect the high-mass end in all three criteria to match, as the group finder is expected to recover the most luminous centrals.
In Figure \ref{fig:stellar_hi_comparison}, we define \textit{mock} as the underlying mock survey described in Section \ref{subsec:mock-catalogues_chap5}, containing all the centrals (as defined by \shark). \textit{GAMA-selection} refers to the centrals that are defined by the group finder that has been run on the mock survey. \textit{SDSS-selection} includes all the centrals (as defined by the group finder) that are above the SDSS magnitude limit. A shift in the high-mass end of the stellar-mass distribution (left panel) is seen, with the GAMA and SDSS selection not matching with the lightcone distribution. This shift is a result of introducing luminosity corrections to the galaxies that were passed to the group finder, as described earlier. A consequence of this correction (lowering \shark\ galaxies luminosities) is that the stellar masses of those galaxies are also decreased assuming a 1:1 light-to-mass ratio
\citep[][]{Bravo2020}, resulting in the mismatch seen at the high stellar masses. Taking this into account, we can see that the SDSS selection is complete for \subsuperscript{M}{star}{} $\gtrsim$ \solarValue{9.5}, while the GAMA selection matches the underlying lightcone distribution for \subsuperscript{M}{star}{} $\gtrsim$ \solarValue{8.5}. This is to be expected, as the sensitivity of both surveys differs by approximately $2$ mag. The high-mass end of the \hi~mass distribution (right panel) of the centrals matches up in all the three cases, confirming that the centrals with \subsuperscript{M}{HI}{} $\geq$ \solarValue{9.5} are recovered irrespective of the survey magnitude limitations. Note that the correction to luminosities and stellar masses above is not applied to the \hi\ masses ( as the \hi\ masses of galaxies are not determined by the optical luminosity of the galaxy), hence the agreement here is very good. We see the numbers declining at \subsuperscript{M}{HI}{} $\leq$ \solarValue{9.5}, with the SDSS-like selection having a lower number of galaxies than the GAMA-like selection in the same \hi\ mass bin. This happens because at fixed \subsuperscript{M}{HI}{}, the GAMA limit has the capability of going lower in stellar mass than the SDSS limit, capturing better those low-stellar-mass galaxies that are relatively gas-rich.

\section{Comparing halo mass allocation by group finders with \textsc{shark}}
\label{sec:virial-mass-allocation_chap5}

One of the major possible systematic effects in measuring the HIHM relation is the estimation of the halo mass of a galaxy group. This is easier in simulations than it is for observations. For example, in \shark, halo mass is determined from the \velociraptor\ halo catalogues, using well-defined quantities such as \subsuperscript{M}{200}{crit} (hereafter \subsuperscript{M}{vir}{}). In this section, we briefly describe how the halo masses are assigned by group finders and then compare them with the halo masses intrinsic to \shark.


\subsection{Estimating the halo mass of a galaxy group}
\label{subsec:halo-mass-obs-allocation_chap5}

As stated earlier, it is possible to use certain properties of galaxies in groups to obtain a mapping to properties of the underlying DM halo.
Therefore, once the galaxy groups have been defined, the next step is to assign properties to the underlying DM halo, namely its mass. The way that halo masses are assigned to galaxy groups varies based on the group finder and the preferences of the surveyor. Here we describe the two ways that halo mass estimates were made for SDSS \citep[abundance matching; ][]{Yang2005-SDSS-groupfinder, Lim2017-SDSS-group} and GAMA \citep[dynamical mass estimation; ][]{Robotham2011GalaxyG3Cv1}.

\subsubsection{Abundance Matching estimate}
\label{subsubsec:abundance-matching-explaination_chap5}

It is expected that the total luminosity of the galaxies in a halo scales with the virial mass of the halo. Using this relation, we can match up the number densities of galaxies and haloes, a process termed  `abundance matching'. Note that this process assumes no scatter in the luminosity--halo relation. We use the luminosity--halo relation obtained from \citet{Lim2017-SDSS-group} to estimate the halo masses of the groups belonging to our SDSS-like mock to make a more one-to-one comparison to the \hi~stacking based on SDSS groups. \citet{Lim2017-SDSS-group} developed their luminosity--halo relation by comparing their fit from the mean luminosity--halo relation, derived from SDSS and 2dfGRS \citep[The 2dF Galaxy Redshift Survey; ][]{Colless2001-2dfgrs} with the results obtained from the EAGLE hydrodynamical simulation \citep{Schaye2015-EAGLE-reference,Crain2015-EAGLE-reference}. 

In order to assign halo masses to the galaxy groups in our catalogue, we use the following relation for isolated centrals:
\begin{dmath}
 {\rm log} \frac{M_{\rm h}}{M_{\odot}h^{-1}} = 10.595 + 4.370 \times 10^{-4} {\rm exp} \left({\rm log} \frac{L_{\rm c}}{L_{\odot} h^{-2}} \frac{1}{1.214}\right),
\label{eq:abundance-matching}
\end{dmath}

\noindent where \subsuperscript{M}{h}{} and \subsuperscript{L}{c}{} refer to the halo mass and the luminosity of the central galaxy, respectively. In order to estimate the halo masses of galaxy groups, \citet{Lim2017-SDSS-group} use the `GAP correction' based on \citet{Lu2016-GAP_correction} but modified using EAGLE. The GAP correction uses a combination of the luminosity of the central galaxy and the difference in the luminosity to the $n^{\rm th}$ brightest galaxy in the group to estimate a correction to the halo mass of the galaxy group. We have used the same procedure as \citet{Lim2017-SDSS-group} (see their Equations 9-10) to assign halo masses to our galaxy group using the SDSS best-fitting parameters for the GAP correction \citep[see Table 2 in][]{Lim2017-SDSS-group}. This means, our estimates are close to the SDSS group catalogue used for the \hi~stacking in \citet{Guo2020}.   

\subsubsection{Dynamical Mass estimate}
\label{subsubsec:dynamical-mass-explaination_chap5}

With the information gathered from the galaxy--galaxy links made by the FoF algorithm, we can recover the group velocity dispersion ($\sigma$) and radius ($R$), which are key for estimating the dynamical mass of the group. Galaxy groups are assumed to be virialized systems, in which case we expect the dynamical mass to scale as $M \propto \sigma^2 R$. \citet{Robotham2011GalaxyG3Cv1} use the dynamical mass estimate to assign a halo mass to the galaxy group, using the following relation: 
\begin{equation}
    M_{\rm dyn} = \frac{A}{G} (\sigma_{\rm dyn})^2 R_{\rm dyn},
    \label{eq:dynamical-mass}
\end{equation}

\noindent where $M_{\rm dyn}$ is the dynamical mass of the system, $\sigma_{\rm dyn}$ and $R_{\rm dyn}$ are the velocity dispersion among the galaxies in the group and radius of the FoF group, respectively. $G$ is the gravitational constant and $A$ is a scaling factor dependent on the number of galaxies in a group and its redshift \citep[see Equation 19 and Tables 2--4 in][]{Robotham2011GalaxyG3Cv1}.  

Unlike abundance matching, where the DM halo properties are reliant on the halo distribution obtained from \lcdm\ predictions, the dynamical mass method depends more on the physically measurable properties of galaxy groups to estimate the properties of their halo. This makes the dynamical mass method more physical in nature and gives us more information about the dynamics inside a halo. Note that a velocity dispersion can only be strictly computed with more than 2 galaxies. Hence, for isolated galaxies (i.e., those that are not assigned to any groups), abundance matching is the only way of assigning them halo masses.

  
\begin{figure*}
  \centering
\begin{minipage}{0.48\textwidth}
    \includegraphics[height=5.5cm, width=7.5cm, trim=0.3cm 0.3cm 0.3cm 0.2cm, clip]{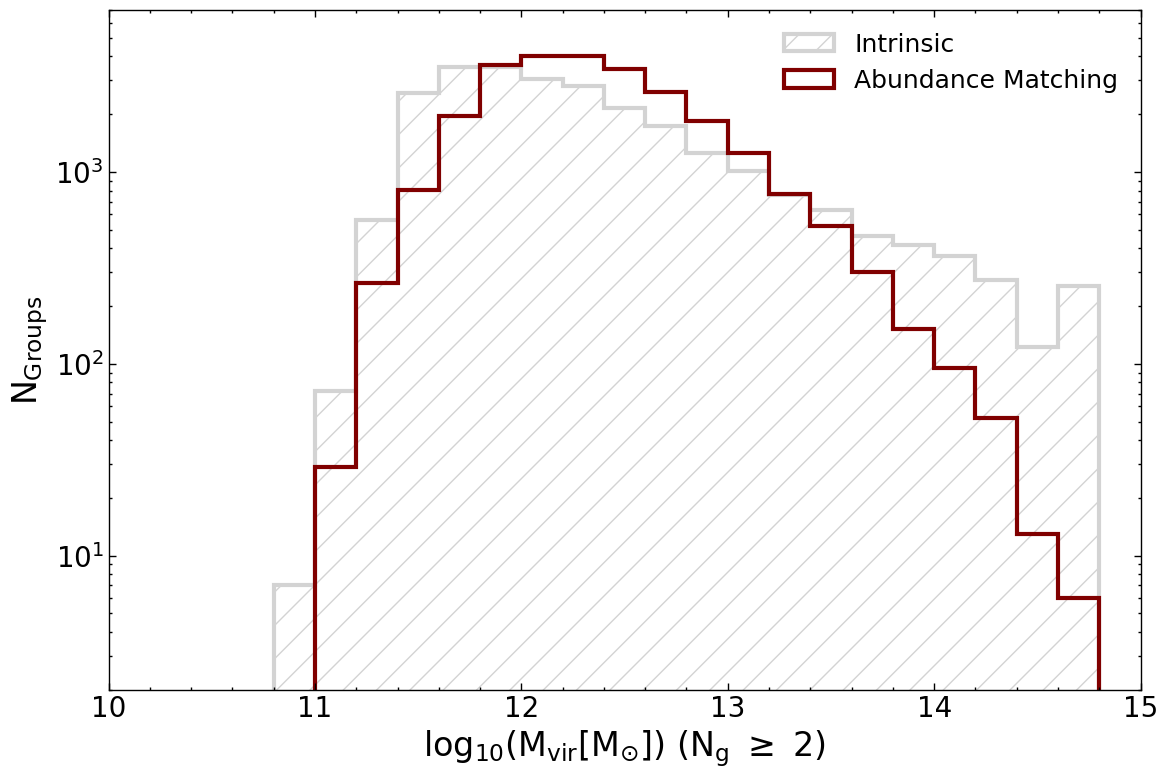}
  
  \end{minipage}
  \hfill
  \begin{minipage}{0.48\textwidth}
    \includegraphics[height=5.5cm, width=7.5cm, trim=0.3cm 0.3cm 0.3cm 0.2cm, clip]{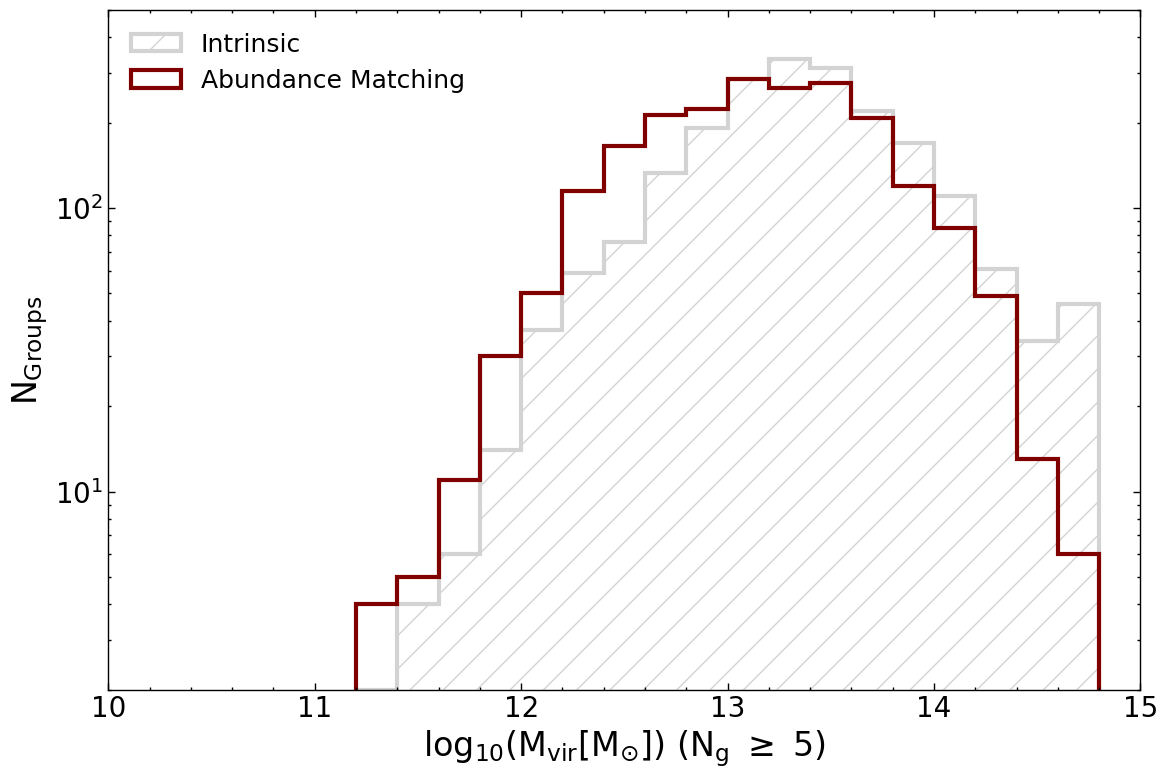}
  \end{minipage}
  
  \begin{minipage}{0.48\textwidth}
    \includegraphics[height=5.5cm, width=7.5cm, trim=0.3cm 0.3cm 0.3cm 0.2cm, clip]{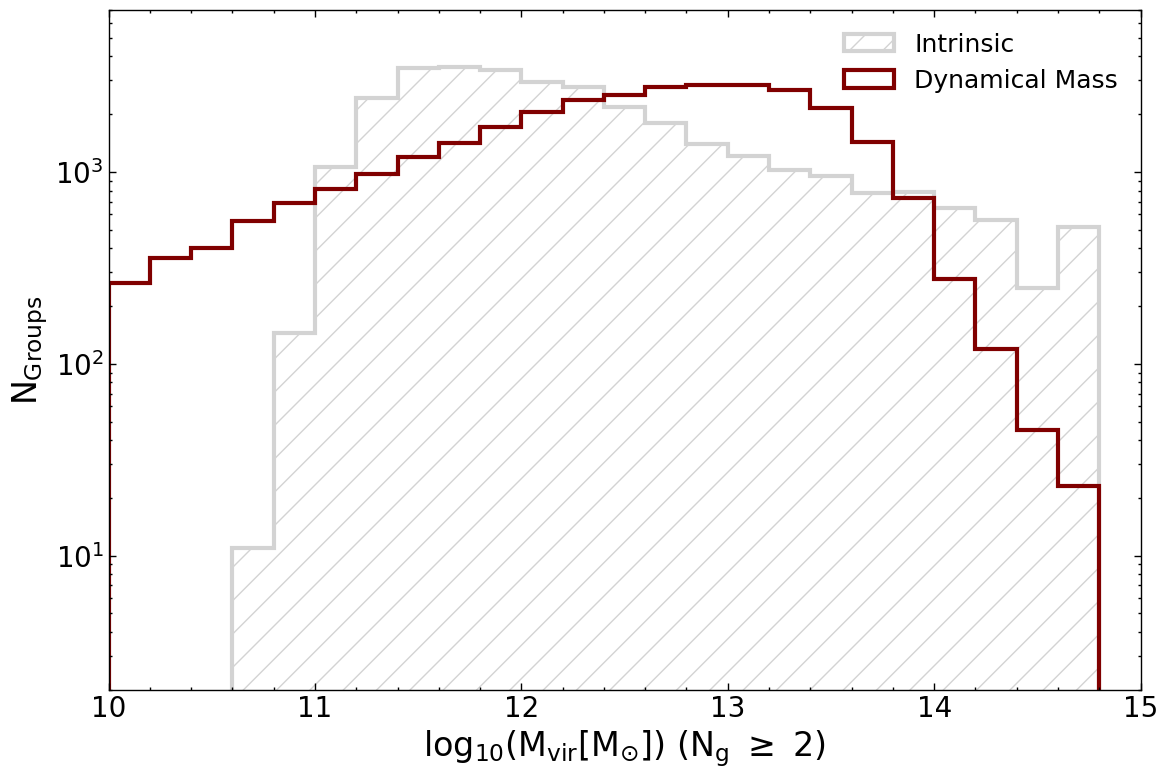}
    \end{minipage}
  \hfill
  \begin{minipage}{0.48\textwidth}
    \includegraphics[height=5.5cm, width=7.5cm, trim=0.3cm 0.3cm 0.3cm 0.2cm, clip]{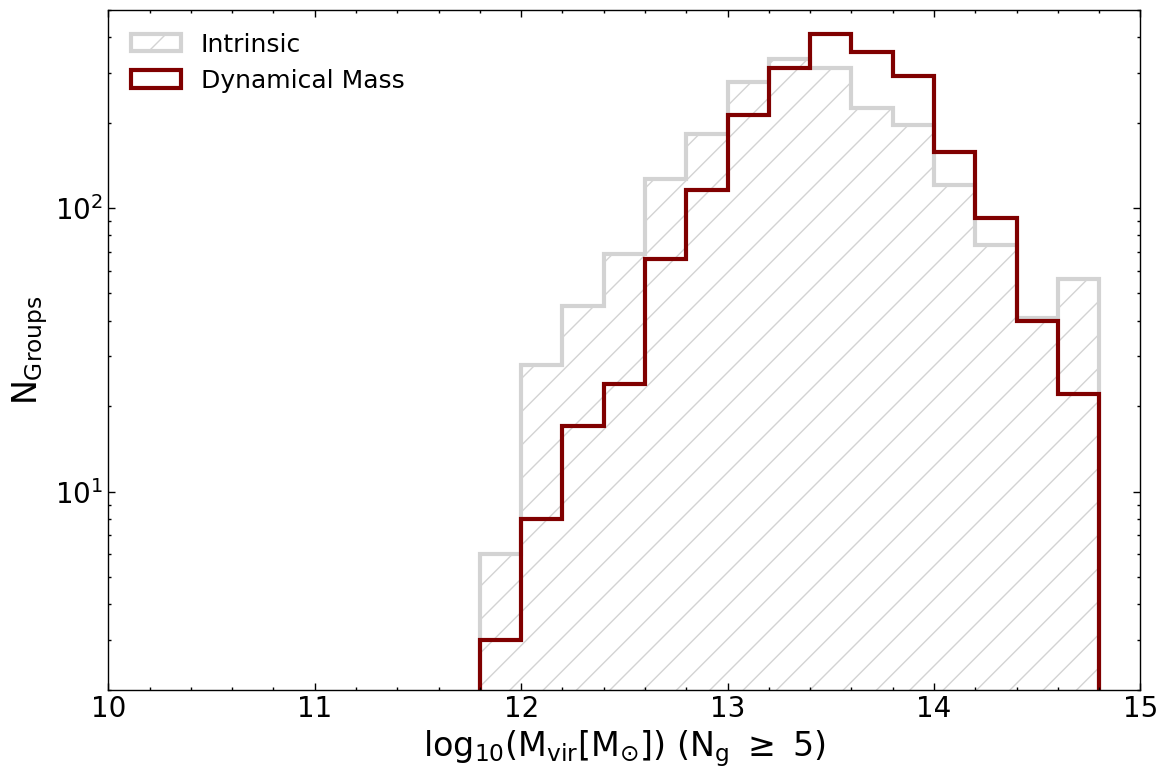}
  \end{minipage}
\caption[Halo mass distribution comparison between \shark\ output and the estimates made by abundance matching and dynamical mass.]{Halo mass distribution comparison between the \shark\ output (light grey) and the estimates made by abundance matching (\textit{upper panel}) and dynamical mass (\textit{lower panel}). The \textit{left panel} comprises all groups with $N_{\rm g} \geq 2$, whereas the \textit{right panel} shows the distribution of groups with at least five galaxies, $N_{\rm g} \geq 5$. The distribution obtained for abundance matching is similar in shape to the intrinsic distribution for both group membership criteria at \subsuperscript{M}{vir}{} $\leq$ \solarValue{13.5} and \subsuperscript{M}{vir}{} $\leq$ \solarValue{14}, for $N_{\rm g} \geq 2$ and $N_{\rm g} \geq 5$, respectively. The number of groups at the high-mass end is significantly smaller for abundance matching than \shark. The distribution derived from dynamical mass is very different from the intrinsic distribution for $N_{\rm g} \geq 2$ because the dynamical mass estimate is not well-constrained for galaxy groups with $N_{\rm g} \leq 5$ \citep{Robotham2011GalaxyG3Cv1}. When $N_{\rm g} \geq 5$ groups are selected, the dynamical mass estimates produce a distribution that closely resembles that of \shark.
}
\label{fig: histograms-surveys-intrinsic-all}

\end{figure*}

\begin{figure*}
  \centering
\begin{minipage}{0.49\textwidth}
    \includegraphics[width=\linewidth, trim=0.3cm 0.3cm 0.3cm 0.2cm, clip]{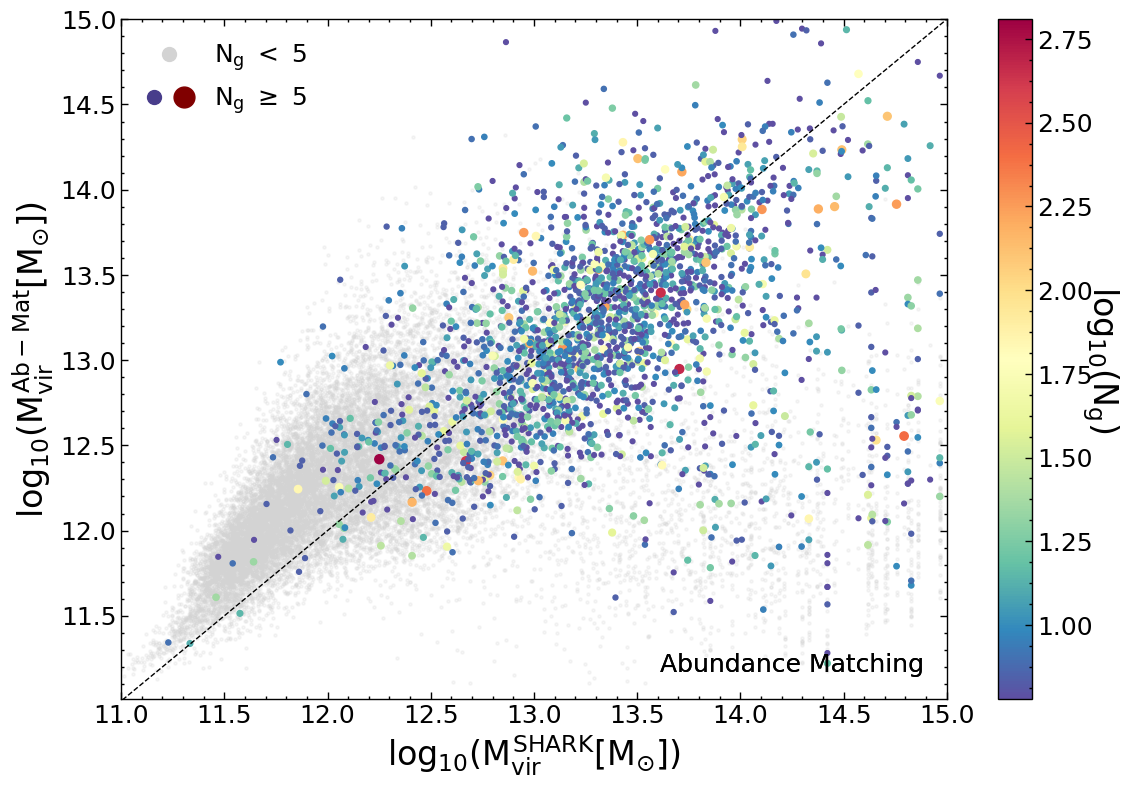}
  
  \end{minipage}
  \hfill
  \begin{minipage}{0.49\textwidth}
    \includegraphics[width=\linewidth, trim=0.3cm 0.3cm 0.3cm 0.2cm, clip]{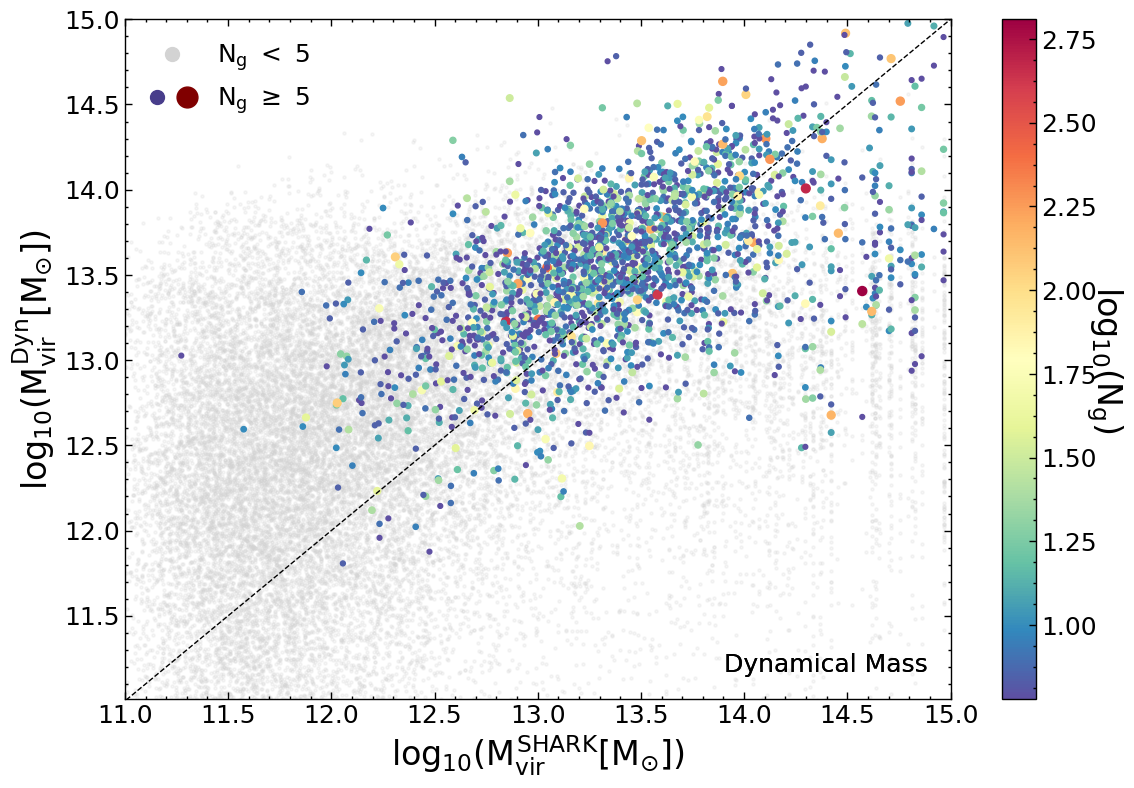}
    \end{minipage}
    
  \begin{minipage}{0.49\textwidth}
    \includegraphics[width=\linewidth, trim=0.3cm 0.3cm 0.3cm 0.2cm, clip]{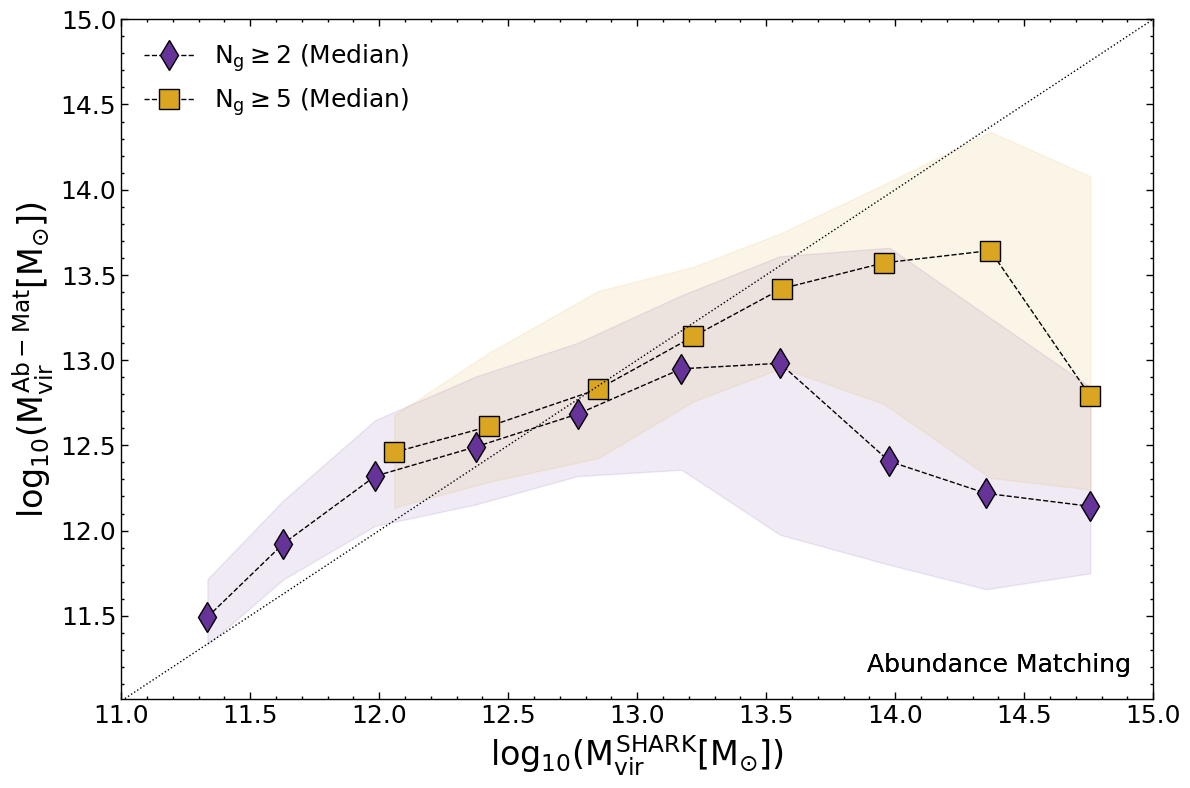}
  \end{minipage}
  \hfill
  \begin{minipage}{0.49\textwidth}
    \includegraphics[width=\linewidth, trim=0.3cm 0.3cm 0.3cm 0.2cm, clip]{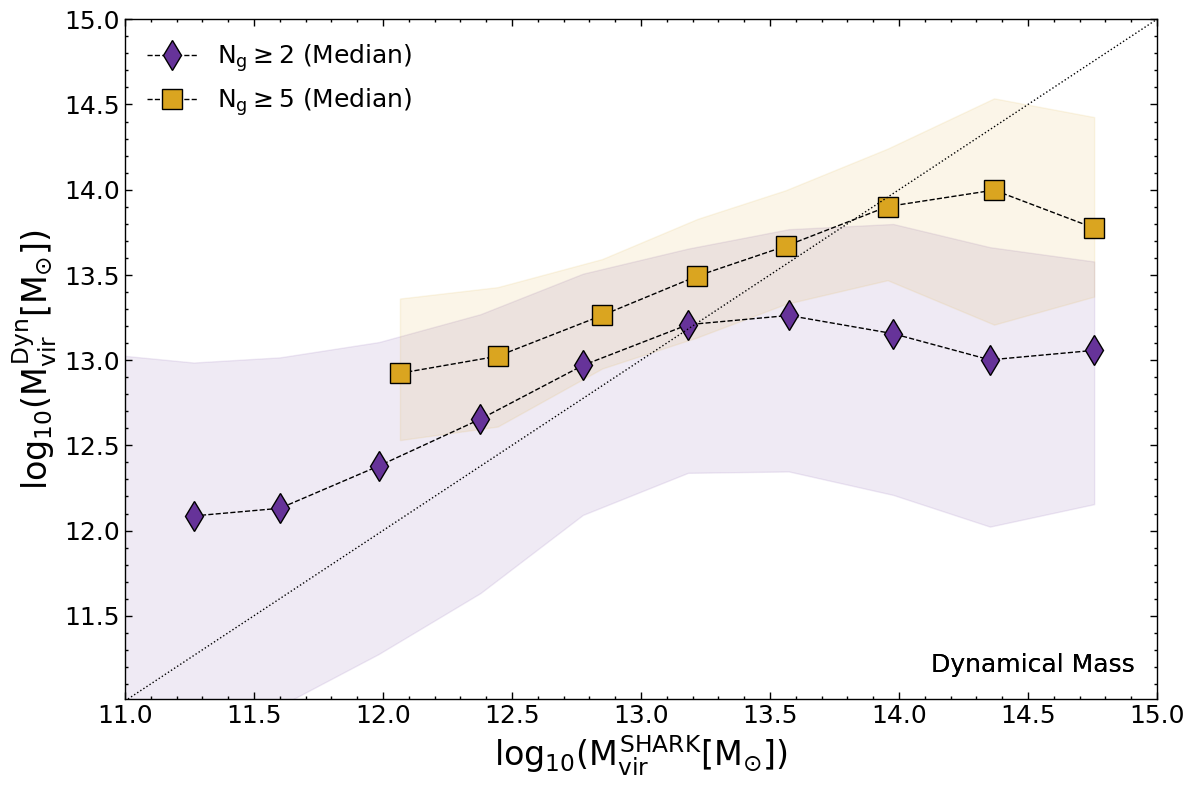}
  \end{minipage}

\caption{Comparison of halo mass estimate methods with the intrinsic halo masses in \shark. \textit{Upper panels:} Each point represents a galaxy group that has been allocated a halo mass by abundance matching (left) and the dynamical mass method (right), coloured according to the number of group members. \textit{Lower panels} shows the median for both \textit{all} (purple-diamonds) and \textit{high-membership} (yellow-square) groups, for both abundance matching (left) and the dynamical mass method (right), with the shaded region representing the \subsuperscript{16}{}{th}-\subsuperscript{84}{}{th} percentile range of the distribution. The dotted line in all the panels represents the 1:1 line. Abundance matching is seen to be closer to the 1:1 relation for \subsuperscript{M}{vir}{} $\lesssim$ \solarValue{13}, but deviates thereafter. Dynamical masses, on the other hand, are not as close to the 1:1 relation for lower halo masses, but tend to get closer as we move towards higher halo masses}
\label{fig:FOF-comparison-GAMA-SDSS-median}

\end{figure*}
\subsection{Using \textsc{shark} to assess how well halo masses are estimated}
\label{subsec:halo-mass-obs-comparison_chap5}

In order to reproduce the \hi-stacking results in our simulated lightcone, presented in Section \ref{sec:HI-stacking-obs}, we compute halo masses as done by the surveys (i.e. using abundance matching when comparing to SDSS-based stacking, and dynamical masses when comparing to GAMA-based stacking). To fully understand the systematic effects introduced by these different methods to measure the halo masses in the derived HIHM relation, we first investigate how well these estimates reproduce the intrinsic halo masses of the simulation.  

In this section, we investigate the differences between the halo masses inferred from abundance matching and dynamical mass estimates, to the intrinsic values in \shark. Here, we focus on galaxy groups with at least 2 or more members ($N_{\rm g} \geq 2$), as a dynamical mass estimate cannot be computed for an isolated galaxy (see Section \ref{subsubsec:dynamical-mass-explaination_chap5}).


\begin{figure*}
\begin{minipage}{0.49\textwidth}
    \includegraphics[width=\linewidth, trim=0.3cm 0.3cm 0.2cm 0.15cm, clip]{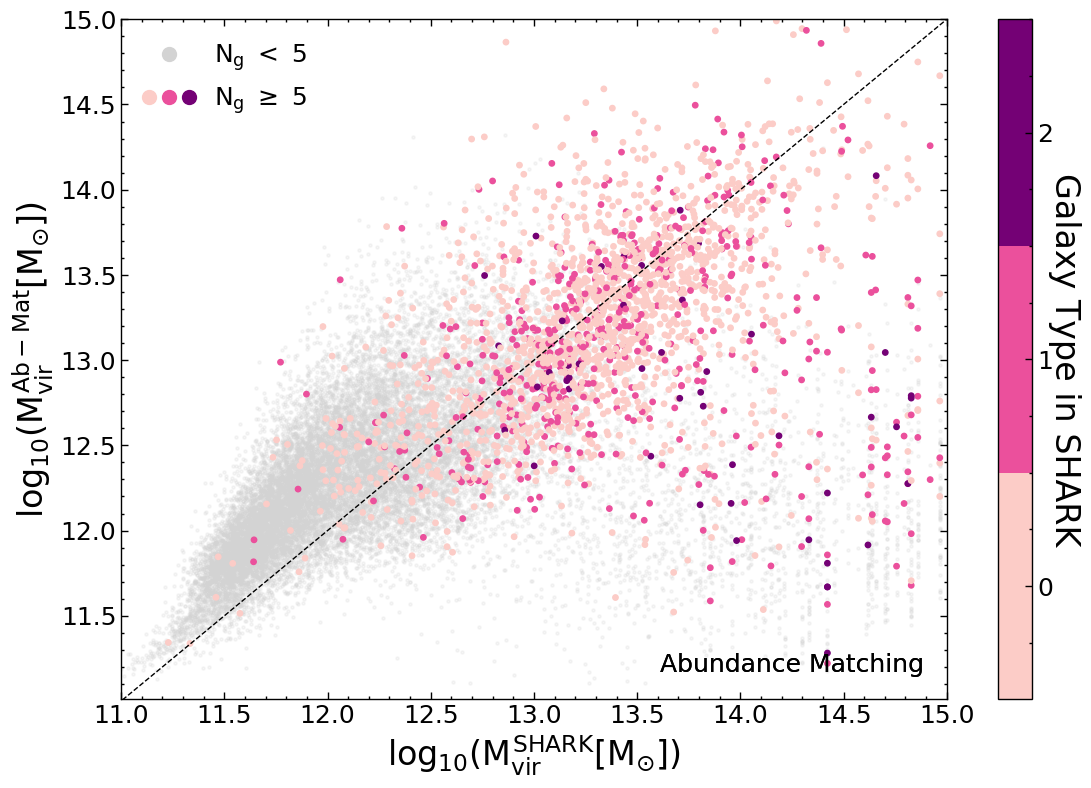}
  
  \end{minipage}
\hfill
  \begin{minipage}{0.49\textwidth}
    \includegraphics[width=\linewidth, trim=0.25cm 0.3cm 0.2cm 0.2cm, clip]{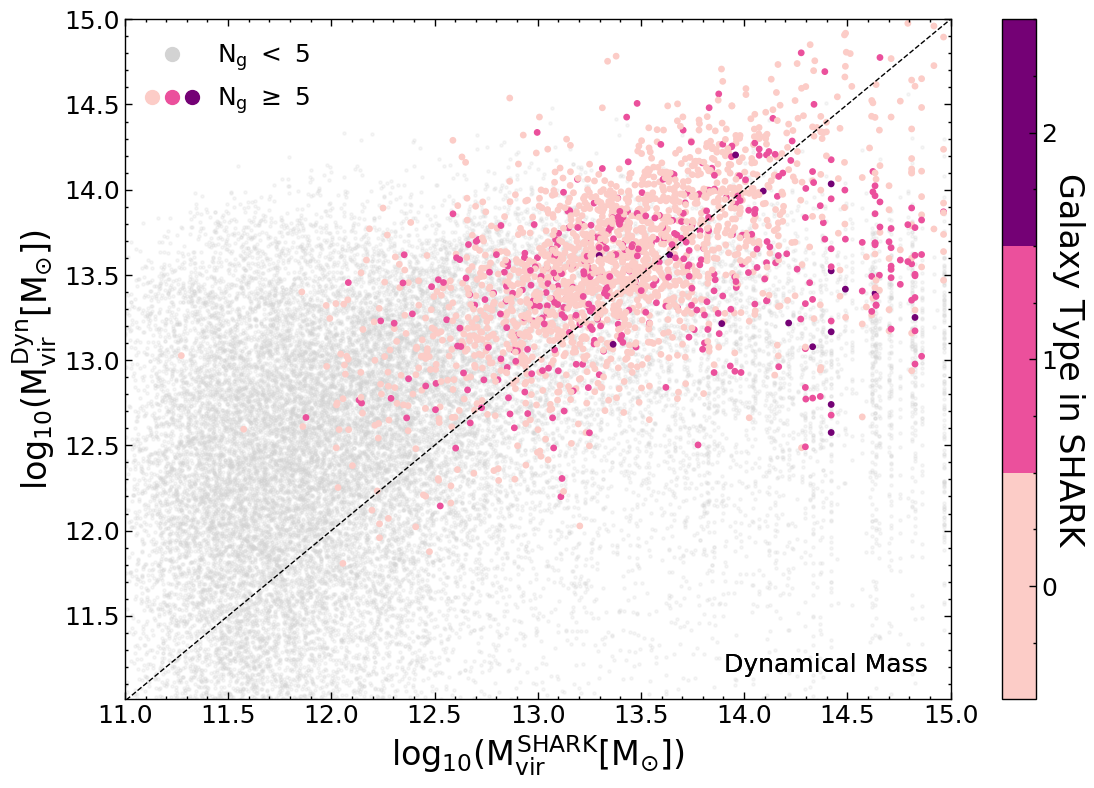}
    \end{minipage}

\caption{Similar to Figure \ref{fig:FOF-comparison-GAMA-SDSS-median}. Each point represents a galaxy group that has been allocated its halo mass by abundance matching (\textit{left panel}) and the dynamical mass method (\textit{right panel}), coloured according to the intrinsic galaxy type (as defined by \shark) of the central of the galaxy group. In \shark, centrals, satellite and orphan galaxies are termed \texttt{type = 0,1} and \texttt{2}, respectively. The dotted line in all the panels represents the 1:1 line.}
\label{fig:central-satellite-fof-comparison}
\end{figure*}




One key difference between the dynamical mass method and abundance matching is that the former requires at least several galaxies in a group to be reliable (preferably $N_{\rm g} \geq 5$, \citealp{Robotham2011GalaxyG3Cv1}). Abundance matching, on the other hand, relies on the mass-to-light ratio from the selected galaxies, which is less sensitive to the group membership. It is hence easier to apply, regardless of the occupation of groups. With this knowledge at hand, we divide the groups in two categories, \textit{all} and \textit{high membership}, with \textit{all} containing groups with two or more galaxies ($N_{\rm g} \geq 2$) and \textit{high membership} containing groups with at least five galaxies ($N_{\rm g} \geq 5$). To compare the halo mass estimations from abundance matching and the dynamical mass method with the intrinsic virial mass of haloes in \shark, we track the \textit{centrals} identified by the group finder back to the simulation box and use the host halo mass associated with that galaxy as our intrinsic virial mass for that galaxy group. Note that we do this regardless of whether that galaxy is classified as central or satellite in \shark.  

Figure \ref{fig: histograms-surveys-intrinsic-all} shows the distribution of halo masses assigned by the group finders (abundance matching and dynamical) against the true halo masses (intrinsic) of the haloes containing those groups. As can be seen from the \textit{left panels}, group membership plays an important role in the accuracy of the dynamical mass estimate. For $N_{\rm g} \geq 2$, the distribution obtained from the dynamical mass estimate is very different to the intrinsic one, with a peak at a higher mass compared to the intrinsic distribution and a tail towards low masses that is not seen in \shark. The abundance matching-derived distribution is more consistent with the intrinsic distribution at $N_{\rm g} \geq 2$, though with a significant underestimation in the number of groups for \subsuperscript{M}{vir}{} $\geq$ \solarValue{13.6}. We plot the same distribution for high-membership groups (in the \textit{right panels}) and see a remarkable improvement in the shape of the dynamical mass estimate distribution, as it now closely resembles the intrinsic distribution. The abundance-matching distribution follows the intrinsic distribution quite well (better than the \textit{all} distribution), which is to be expected as the \citet{Lim2017-SDSS-group} group masses were calibrated to reproduce a simulation (though, a different simulation and halo finder were used). Though the distributions are similar to each other, we do find discrepancies between the group finder allocations and the intrinsic distribution. The most prominent discrepancy is the underestimation of halo masses at the high-mass end, which is seen in both the all and high-membership distributions, for both the abundance matching and the dynamical mass estimates.

To investigate the effect of group membership further and to identify where it falters, Figure \ref{fig:FOF-comparison-GAMA-SDSS-median} compares the two halo mass estimate methods against the true halo masses in \shark, with the \textit{left panels} comparing abundance matching and \textit{right panels} looking at the dynamical mass. The \textit{lower panels} compare the running medians and percentiles of the distribution. We find that abundance matching, irrespective of member allocation, stays close to the 1:1 line, although there is significant scatter around the relation. The same effect is reflected in the median relations plotted below (\textit{lower-left panel}), where the purple and yellow points represent the median of all and high-membership groups, respectively, with the shaded region being the \subsuperscript{16}{}{th}--\subsuperscript{84}{}{th} percentile range of the distribution. As is evident from the left panel of Figure \ref{fig:FOF-comparison-GAMA-SDSS-median}, abundance matching underestimates the halo masses from \subsuperscript{M}{vir}{}$\gtrsim$\solarValue{13} and \subsuperscript{M}{vir}{}$\gtrsim$\solarValue{13.5}, and overestimates the halo masses for \subsuperscript{M}{vir}{}$\lesssim$\solarValue{12.5} and \subsuperscript{M}{vir}{}$\lesssim$\solarValue{13}, for all and high-membership groups, respectively. This underestimation explains the discrepancies seen in the upper panels of Figure \ref{fig: histograms-surveys-intrinsic-all}. Given that the abundance matching is based on a different simulation with a different halo mass function driven by the different assumed cosmology and halo mass definition, differences like these are to be expected.

For the dynamical mass estimates, we see a significantly larger scatter than the abundance matching comparison for all groups. This reverts for high-membership groups, where the scatter displayed by dynamical mass estimates is smaller than the one seen in abundance matching (\textit{right panel}). We show the medians for the same distribution (\textit{right bottom panel} in Figure \ref{fig:FOF-comparison-GAMA-SDSS-median}), and find that the dynamical masses overestimate the halo masses for \subsuperscript{M}{vir}{}$\lesssim$\solarValue{13.2} and \subsuperscript{M}{vir}{}$\lesssim$\solarValue{14}, for all and high-membership groups, respectively. Though, it can also be seen that for higher halo masses (\subsuperscript{M}{vir}{} $\sim$ \solarValue{14.5}), where abundance matching underestimates the halo masses by $\sim 2.4$ and $0.9$ dex for all and high-membership groups, respectively, dynamical masses only underestimate halo masses by $\sim 1.5$ and $0.6$ dex, respectively. As the dynamical mass estimate relies on the velocity dispersion calculated for the group, which becomes better constrained when there are more group members, we find dynamical mass estimates to be closer to the true halo masses of \shark\ for high-membership groups. Abundance matching, on the other hand, provides a better estimate for groups with low membership, as it primarily relies on the luminosity--halo relation, which is not so much affected by how the groups are partitioned.      

Both the methods to measure halo mass underestimate the halo masses at the high-mass end, which suggests that the systematic differences may come from the underlying group finder. The way galaxies are identified as being part of a group and flagged as centrals or satellites, might be the reason for the differences seen. The next step in our investigation is to quantify if the central recognised by the group finder is the true central for that halo or not.

In \shark\ we have three primary galaxy types: \textit{centrals}, \textit{satellites} and \textit{orphans}, which are assigned based on the merger trees and subhalo catalogues used by \shark\ as a skeleton. We define \textit{centrals} (\texttt{type = 0}) to be the central galaxy of the most massive subhalo in the group. \textit{Satellite} (\texttt{type = 1}) galaxies are the galaxies that are hosted by the other subhaloes of the group. \textit{Orphans} (\texttt{type = 2}) galaxies are the galaxies that cease to be tracked by \velociraptor\ (either because their number of particles drop below the threshold required to consider a detection, or because it becomes indistinguishable from the underlying density-velocity field), i.e. lack a subhalo entirely.  

Figure \ref{fig:central-satellite-fof-comparison} compares the halo mass estimates from the two methods against the true halo masses in \shark, in this case, coloured by the intrinsic galaxy type of the galaxies flagged as centrals by the group finder as they appear in \shark. During our analysis, we found that $\sim 22.7$ per cent of galaxies identified as centrals by the group finder are actually satellites or orphans (this is found to be in agreement with \citealp{Bravo2020}). When comparing this number with the scatter plot, we find that most of the groups with underestimated halo masses also correspond to the ones which have misidentified centrals (which in \shark\ are either satellites or orphans). Even though both abundance matching and dynamical mass estimates have similar percentages of mismatch (which is to be expected as the same group finder was run on them), it affects the abundance matching estimates more because of its reliance on the luminosity of centrals for the halo mass estimations. As satellites tend to be less massive than centrals in \shark\ and in general, they also tend to be less luminous. If a satellite is identified as a central, because of its low luminosity, it will be assigned a smaller halo mass, thus deviating from the 1:1 relation. This effect will be more prominent for halo mass estimations for higher halo masses.


\section{Towards a realistic \hi~stacking using mock catalogues}
\label{sec:HI-SIMULATION_chap5}

The strategy we follow below is to get closer to the way \hi~stacking is done in observations step by step, so that we can critically analyse the systematic and random effects introduced at each step of the process. We start by estimating the total \hi\ mass of the haloes defined by the group finder. We first obtain the HIHM relation of groups by using the true halo masses and then by using the estimated halo masses, using both abundance matching and dynamical masses.

\subsection{The dependence of the derived HIHM relation on the stacked volume and halo mass definition}
\label{subsec:shark-simulation-snapshots_chap5}

\begin{figure}
  \includegraphics[width=\linewidth, trim=0.3cm 0.3cm 0.3cm 0.2cm, clip]{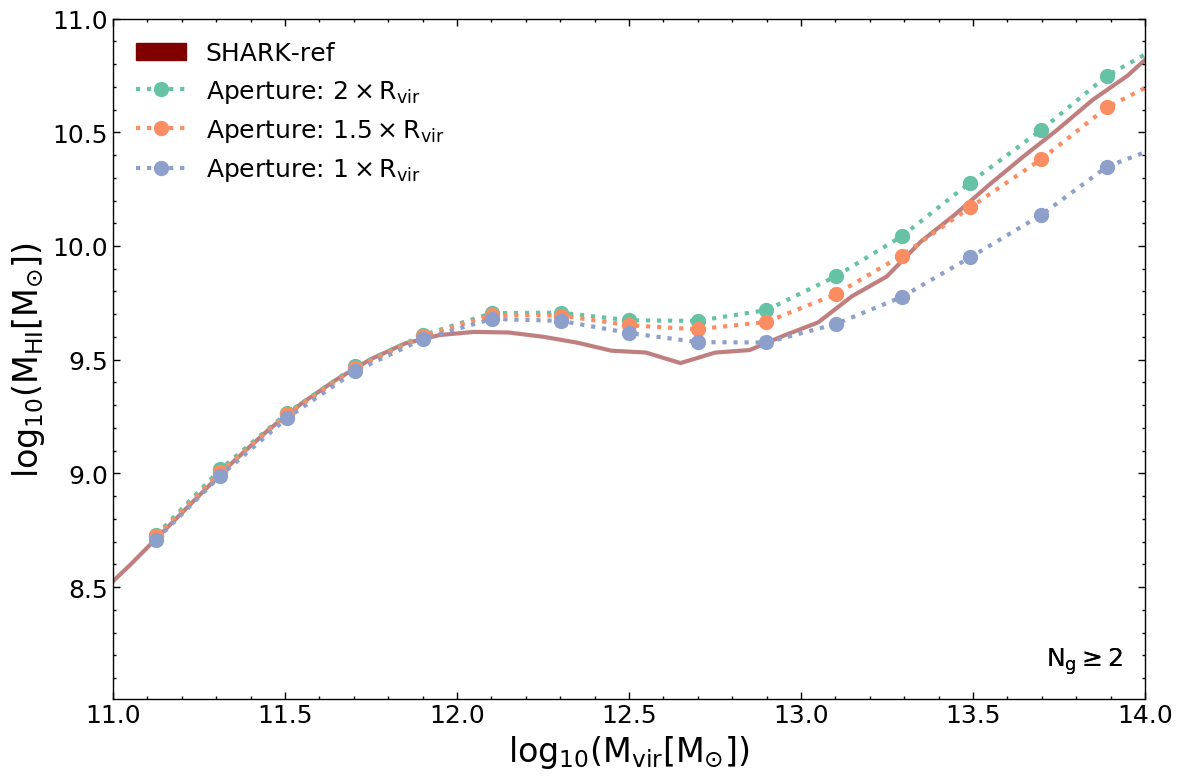}
\caption{Mean \hi\ content of groups identified by the group finder within a spherical aperture (as labelled) versus the intrinsic halo mass of the corresponding halo in \shark\ containing the galaxy flagged as a central by the group finder. 
These apertures are based on the intrinsic halo masses of the galaxy groups. It can be seen that as we move to larger aperture sizes, the \hi\ content of the haloes increases, though the characteristic shape of the intrinsic relation (maroon solid line) remains qualitatively the same.}

\label{fig:SphericalCoordinates_RVIR}
 \end{figure}

\citet{Chauhan2020} derived a mean HIHM relation directly from the \shark\ outputs and found it to have a distinct shape. Unlike the observed HIHM relation (see \citealp{Guo2020}), which was a monotonically increasing function, \citet{Chauhan2020} found a dip in their derived HIHM relation at \subsuperscript{M}{vir}{} $\sim$ \solarValue{12}, which plateaued up to \subsuperscript{M}{vir}{} $\sim$ \solarValue{12.8}, followed by an increase towards higher masses (see figure 1 in \citealp{Chauhan2020}). The characteristic shape of the intrinsic HIHM relation is a result of the \hi\ mass contribution from the central and satellite populations of the haloes. \citet[][]{Chauhan2020} found that central galaxy is the dominant \hi\ mass contributor for haloes with \subsuperscript{M}{vir}{} $\leq$ \solarValue{12.8}, with satellites taking over thereafter. The dip in the HIHM relation seen at \subsuperscript{M}{vir}{} $\sim$ \solarValue{12} is caused by the AGN feedback. One of the possible reasons given for the discrepancy between the observed and intrinsic \shark\ HIHM relation was the uncertainty in group definitions around that halo mass. 

In Figure \ref{fig:SphericalCoordinates_RVIR}, we compare the mean HIHM relation as obtained from \shark\ (\shark-ref) with the mean HIHM relation obtained when we use the groups defined by the group finder. \shark-ref refers to the intrinsic HIHM relation that was obtained for the haloes that have at least two subhaloes (dictated by the resolution of the DM-only simulation), irrespective of whether they are detectable by any survey or not, with the total \hi~mass being the sum of all the galaxies associated with the same host halo in the simulation box at $z=0$, regardless of their luminosity. 
We calculate the mean \hi~mass for the groups in our mock group catalogue by assuming the central, as defined by the group finder, to be the true central of the host halo, and then calculating the \hi~mass of all the galaxies that are within a spherical volume of $1-2$ times the virial radius (\subsuperscript{R}{vir}{}) of the host halo that galaxy belongs to. The virial mass assumed for this analysis is the true halo mass provided by \shark. From Figure \ref{fig:SphericalCoordinates_RVIR} we find that just by using the group finder defined groups, we overestimate the \hi~mass in the region \subsuperscript{M}{vir}{} $\sim$ \solarValue{12-13}. However, even though the uncertainty in group definitions cause a change in the \hi\ content of the group, it does not change the qualitative shape of the relation.

%

\begin{figure*}
  \centering
  \begin{minipage}{0.49\textwidth}
   \includegraphics[width=\linewidth, trim=0.3cm 0.3cm 0.3cm 0.2cm, clip]{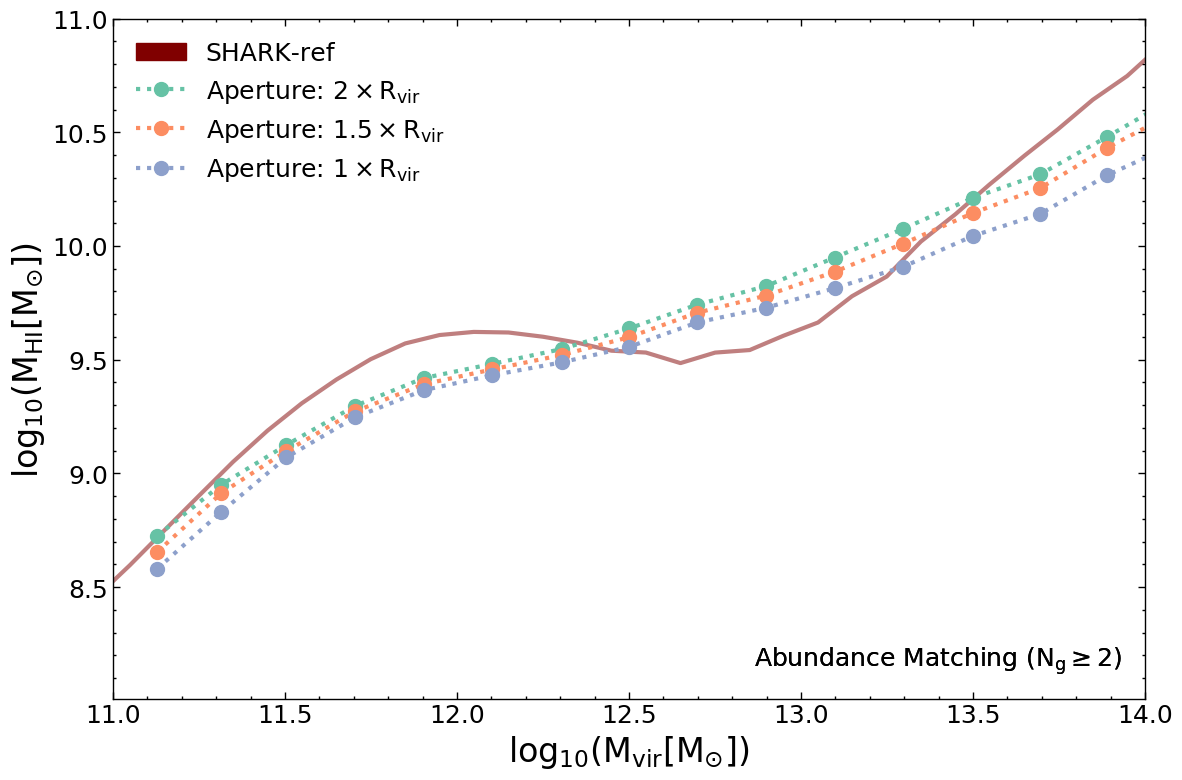}
\end{minipage}
  \hfill
  \begin{minipage}{0.49\textwidth}
    \includegraphics[width=\linewidth, trim=0.3cm 0.3cm 0.3cm 0.2cm, clip]{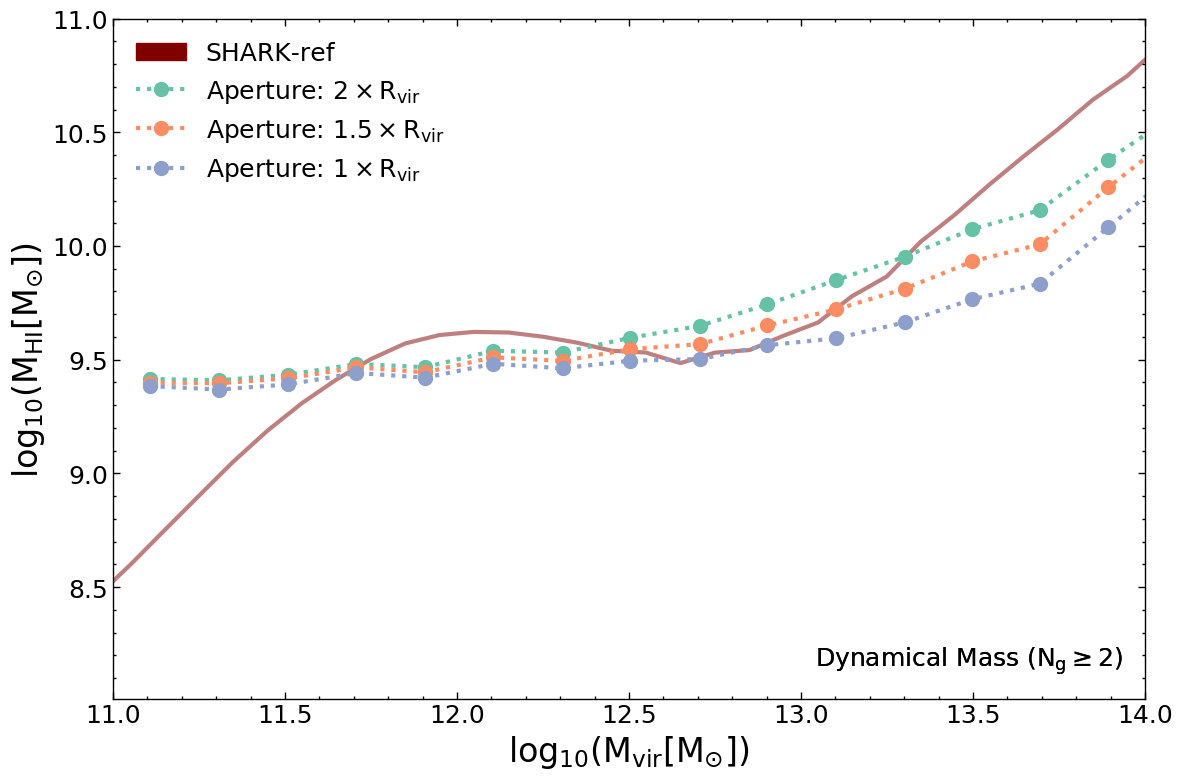}
  \end{minipage}
\caption{Mean \hi\ content of groups identified by the group finder within a spherical aperture (as labelled) versus the halo mass estimates from abundance matching (left panel) and dynamical mass (right panel) methods for the corresponding halo in \shark\ containing the galaxy flagged as a central by the group finder. The apertures are also based on the halo masses estimations made for the galaxy groups by abundance matching (left) and the dynamical mass method (right). The departure from the characteristic shape seen when using the halo mass estimations alludes to the fact that correct halo mass estimation plays a major role in determining the shape of the mean HIHM relation. }
\label{fig:SphericalCoordinates_halo-estimates-groupfinder}
\end{figure*}

As we move to \subsuperscript{M}{vir}{} $\geq$ \solarValue{13}, the total \hi~mass associated with the halo cannot be captured by an aperture of $1$ \subsuperscript{R}{vir}{} around the `central' galaxy (although, in this instance it can be captured with an aperture of $1.5$ \subsuperscript{R}{vir}{}). One of the probable reasons for this departure from the underlying intrinsic HIHM relation is the misidentification of satellite galaxies as centrals by the group finder. We find that almost $25$ ($9.5$)  per cent of satellites are misidentified as centrals for \subsuperscript{M}{vir}{} $\sim$ \solarValue{13}  (\solarValue{14}). If a satellite residing at the edge of the halo is misidentified as a central, we consider the position of that satellite as the halo centre, and calculate the \hi~mass of everything in the vicinity of that satellite, using the \subsuperscript{R}{vir}{} of the host halo the satellite resides in to define our sphere. This would lead us to miss many of the galaxies belonging to the original host halo. Another possible reason for the \hi\ mass not being captured by $1$ \subsuperscript{R}{vir}{} is that the haloes tend to be more elongated than spherical at high halo masses, making $1$ \subsuperscript{R}{vir}{} quite small for the major axis of a massive halo \citep[see figure 1 in ][]{Canas2020}. Both of these cases are remedied by using a larger aperture, as demonstrated by using $2$ \subsuperscript{R}{vir}{}, which captures a total \hi\ mass more in line with the \velociraptor\ haloes. Apart from these, there is also the possibility of the `true' central of that host halo being classified as a central in a different group. This will result in the same host halo being used twice (or maybe more) for calculating the \hi\ mass, which will end up being different each time.

Even though we find slight deviations from the intrinsic HIHM relation when using group-finder-defined groups, the characteristic shape of the relation remains qualitatively the same. We then investigate the systematic effect that halo mass definition has on the derived HIHM relation.  

In Figure \ref{fig:SphericalCoordinates_halo-estimates-groupfinder}, we repeat the same exercise as for Figure \ref{fig:SphericalCoordinates_RVIR}, but this time using the halo masses estimated by abundance matching (\textit{left panel}) and the dynamical mass method (\textit{right panel}), which appear in the x-axis and are also used to calculate  \subsuperscript{R}{vir}{} to estimate the total \hi~mass associated with that galaxy group. It becomes clear that \emph{the halo mass estimates causes a major deviation in the HIHM shape from the intrinsic one.} By using the abundance-matching halo mass estimates, we find that the characteristic shape of the relation is lost. It becomes a monotonically increasing relation, which underestimates the mean \hi~mass for \subsuperscript{M}{vir}{} $\leq$ \solarValue{12.4} and \subsuperscript{M}{vir}{} $\geq$ \solarValue{13.5}, and overestimates for the region in between. We do not recover the shape by using dynamical mass estimates either (see right panel) and find larger deviations from the intrinsic shape than seen when using abundance matching. The HIHM relation obtained by dynamical mass estimates remains almost constant for \subsuperscript{M}{vir}{} $\leq$ \solarValue{12.5}, above which it starts to increase. 

The median for all groups in the case of abundance matching (see lower-left panel in Figure \ref{fig:FOF-comparison-GAMA-SDSS-median}), stays close to the true halo masses for lower-mass haloes. As for the HIHM relation obtained using dynamical mass, the higher \hi\ mass estimate at the low-mass end could be a consequence of centrals being misidentified as satellites resulting in a higher total \hi\ mass for a group in that region; we found $\sim 35$ per cent of centrals in \subsuperscript{M}{vir}{} $\sim$ \solarValue{12} groups are actually satellites. Figure \ref{fig:SphericalCoordinates_halo-estimates-groupfinder} summarises the effect that uncertainties in the halo mass estimates have on the mean HIHM relation, which becomes paramount for our analysis of the parameters involved in \hi~stacking, covered in the following section.   



\subsection{Realistic geometrical stacking using mock catalogues}
\label{sec:HI-stacking-obs}

The general principle of \hi-stacking analysis is to identify the positions and redshifts of galaxies using an external source catalogue containing optical redshifts, extract \hi\ information at these coordinates and then co-add the data. In this section, we describe different \hi~stacking methods employed by surveys to recover the HIHM relation, along with analysing how tweaking their parameters changes the derived HIHM relation. For the latter we make use of our mock group catalogue and survey. Our main aim here is to highlight the differences between these stacking techniques and how well they recover the underlying intrinsic HIHM relation. 

\begin{figure}
\centering
  \includegraphics[width=\linewidth, frame]{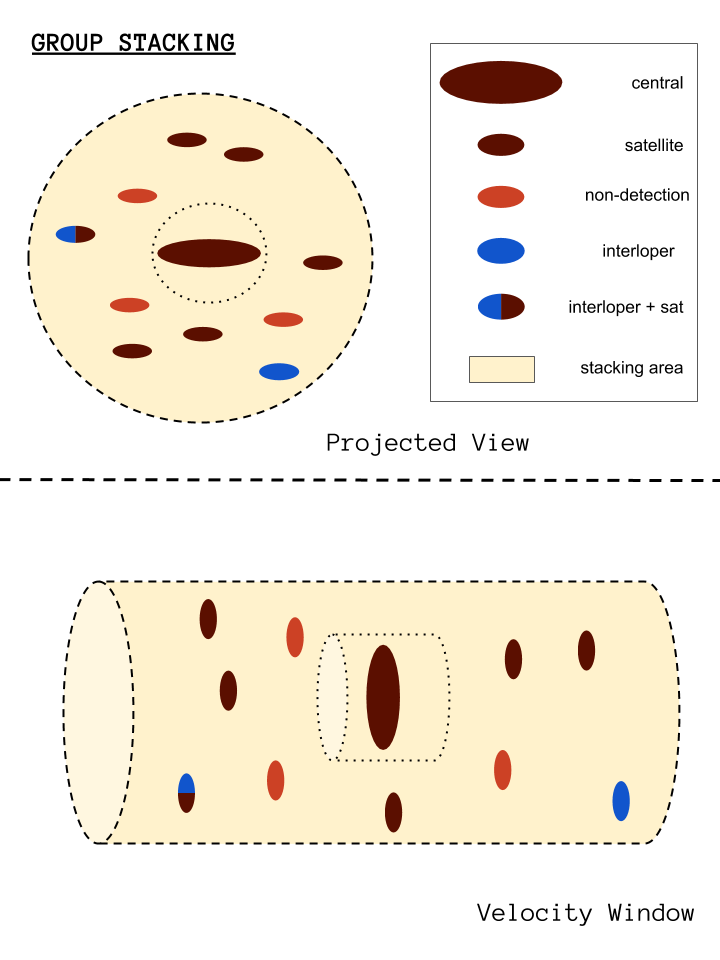}
\caption{Illustration of group \hi\ stacking, showing both the projected view (upper panel) and the orientation in redshift space (lower panel). The maroon ellipses symbolise the galaxies that are identified by the group finder as belonging to the group shown. The orange ellipses denote the galaxies that are missed by the group finder due to the magnitude limitations of the parent survey, which would include \hi\ gas in galaxies that are below the magnitude cut of the optical survey and the absence of \hi\ gas not in galaxies in the group environment. The blue ellipse represents a galaxy that does not belong to the group, but is still in the vicinity of the group, and would thus contribute towards (and contaminate) the stack. The half maroon-half blue ellipse represents a misidentified galaxy by the group finder, where the galaxy is not intrinsically a part of the common halo but is still assigned to the group by the group finder. The \hi\ content associated with this particular group will contain the \hi\ in all galaxies that are in the yellow shaded area.}
\label{fig:Illustration_GroupStacking}
 \end{figure}

\subsubsection{Group \hi~stacking}
\label{subsec:obs-stacking-ALFALFA}

Traditionally, \hi~stacking has been done on individual galaxies, using an optical catalogue to estimate the position and redshift of the target galaxy. One then extracts spectra associated with that area within a stacking velocity window and co-adds them, generally weighting each spectrum by its RMS noise \citep{Fabello2011-HI-stacking}.

\citet{Guo2020} extended this technique to a galaxy group scale, which we will refer to as `group \hi~stacking'. They use the SDSS group catalogue \citep{Lim2017-SDSS-group} as their optical spectroscopic catalogue and ALFALFA as their \hi\ survey. The ALFALFA (Arecibo Legacy Fast ALFA) survey is a `blind' \hi\ survey that has mapped nearly $6900$ \subsuperscript{\rm deg}{}{2} in the Northern Hemisphere, with $\sim 31,500$ direct detections \citep{Giovanelli2005, Haynes2018TheCatalog} out to redshift $z = 0.06$. The typical beam size is $\sim 3.3' \times 3.8'$ with a velocity resolution of $\sim 10$ \kms. \citet[][]{Guo2020} use $2R_{\rm 200}$ as the halo radius as the projected aperture over which \hi~stacking is done, with $\rm R_{200}$ being the radius containing a \textit{mean} mass density that is $200$ times the mean density of the universe at a given redshift (not to be confused with $R^{\rm crit}_{\rm 200}$, which corresponds to the radius containing a mean mass density that is $200$ times the critical density of the Universe). Using the software of \citet[][]{Fabello2011-HI-stacking}, they extracted a single spectrum for each group, which was then stacked with other group spectra in the same mass bin and, finally, fitted with a Gaussian profile.
The \hi\ mass was measured by integrating the signal within $\sim \pm 3\sigma$ width of the Gaussian profile fitted to the stacked spectra, centred on the central galaxy of the group. 
This method, in principle, should ensure that all the \hi\ associated with the target galaxy groups is captured. Figure \ref{fig:Illustration_GroupStacking} shows a schematic of how this stacking technique, which we refer to as ``group \hi~stacking", works. Notably, the \hi\ content of galaxies that were not detected in the parent optical catalogue would still contribute to the total \hi\ mass measured. But because of the large stacking velocity window and projected aperture, some contamination is expected, as the \hi\ content of galaxies that are not actually part of the group would still contribute to the stack. If the adopted apertures (both in projected and velocity space) are very large, then more contamination is expected. While very small apertures minimise such contamination, they come at the expense of excluding galaxies that belong to that group.

In this section, we test if we can reproduce the observationally derived HIHM relation using our mock survey, when we mimic the observational procedure.


Unlike observations, where they have to consider various observational limitations, like noise and flux sensitivity, our mock survey does not suffer from either of those. For keeping our analysis simple, we do not work with \hi\ emission spectra for mock-stacking our galaxies. Instead, we use the projected aperture (as used by observations considered here) and a stacking velocity window, to define the volume within which we sum the \hi\ mass of all the galaxies within. For approximating the  $\sim \pm 3\sigma$ width, we use $\Delta v = \pm 700$\kms\, as our stacking velocity window, which ends up being close to the $\sim \pm 3\sigma$ width for most of the haloes (barring the most massive ones). This choice was made after discussions with the authors of \citet[][]{Guo2020}.

\begin{figure*}
  \centering
  \begin{minipage}{0.49\textwidth}
    \includegraphics[width=\linewidth, trim=0.3cm 0.3cm 0.3cm 0.2cm, clip]{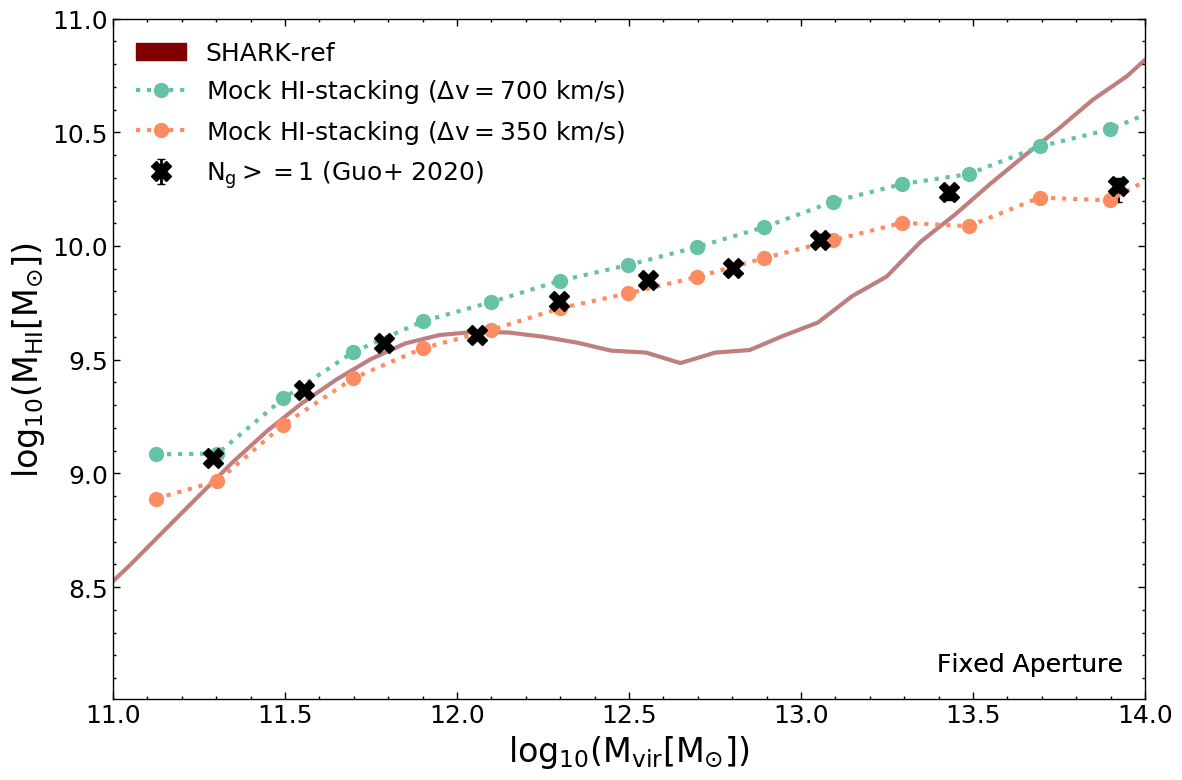}
  
  \end{minipage}
  \hfill
  \begin{minipage}{0.49\textwidth}
    \includegraphics[width=\linewidth, trim=0.3cm 0.3cm 0.3cm 0.2cm, clip]{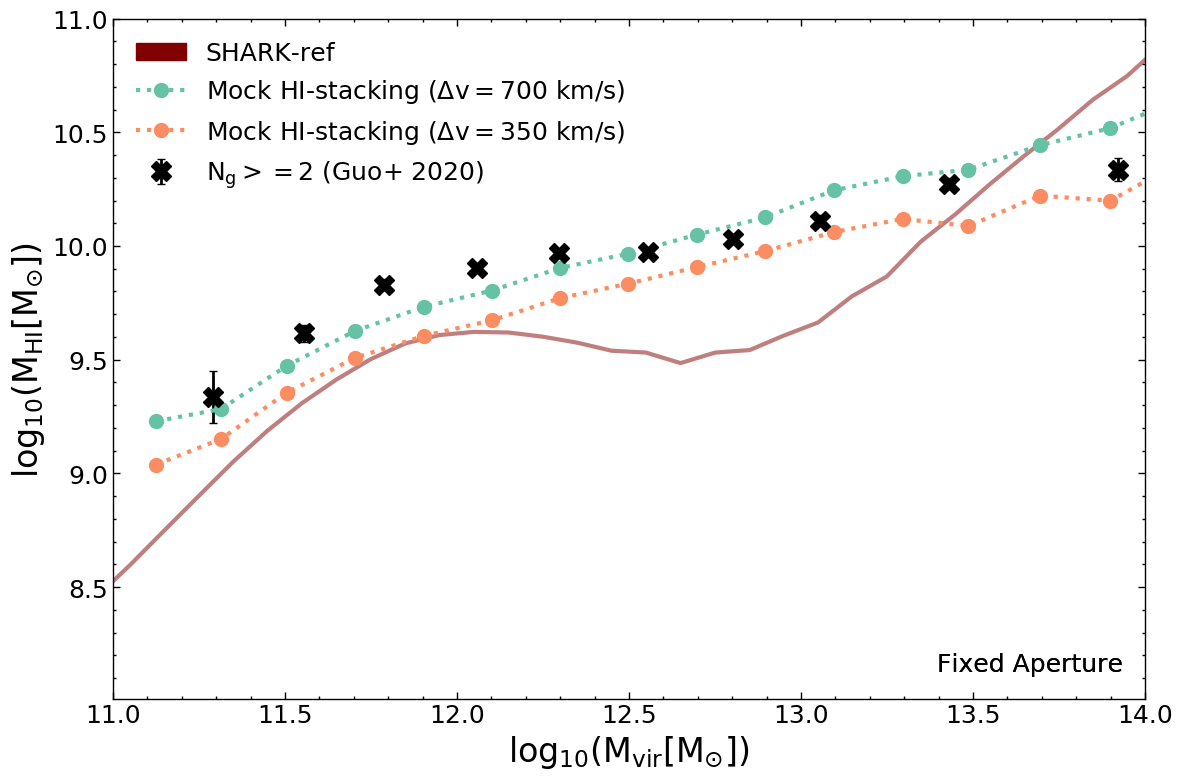}
  \end{minipage}
\caption{Mean \hi~content of haloes obtained by mock group stacking our mock group catalogue, over a fixed projected aperture, for groups with $N_{\rm g} \geq 1$ (left) and $N_{\rm g} \geq 2$ (right), compared to the intrinsic HIHM relation obtained from \shark. Crosses show is the observed HIHM relation from \citet{Guo2020}. Different coloured lines correspond to different `fixed velocity windows' adopted for the stacking, as labelled. Signs of confusion start to become evident at \subsuperscript{M}{vir}{} $\geq$ \solarValue{12} for $N_{\rm g} \geq 1$ groups, where the stacked \hi\ masses start deviating from the underlying intrinsic HIHM relation. Stacked results for $N_{\rm g} \geq 2$ are always higher than the intrinsic relation, showing that the contribution of confusion is prominent even at low halo masses. }
\label{fig:Zcut_window_groups}
\end{figure*}

\begin{figure*}
  \centering
  \begin{minipage}{0.49\textwidth}
    \includegraphics[width=\linewidth, trim=0.3cm 0.3cm 0.3cm 0.2cm, clip]{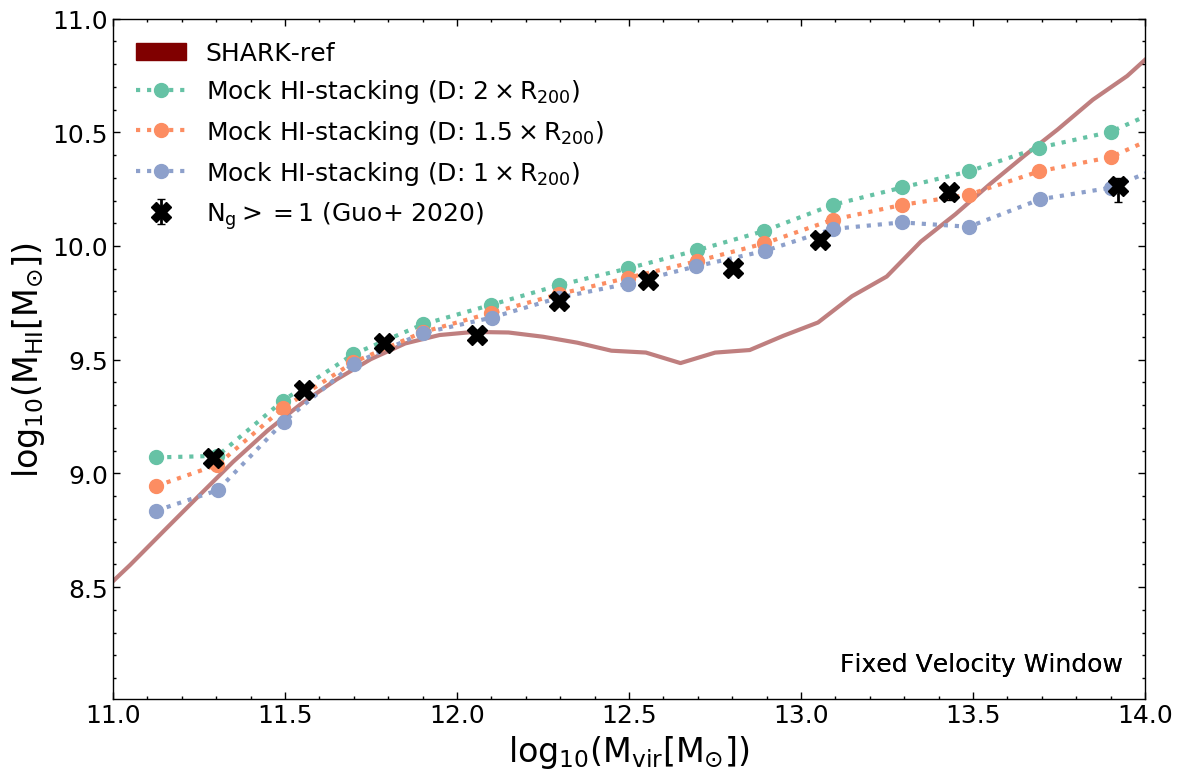}
  
  \end{minipage}
  \hfill
  \begin{minipage}{0.49\textwidth}
    \includegraphics[width=\linewidth, trim=0.3cm 0.3cm 0.3cm 0.2cm, clip]{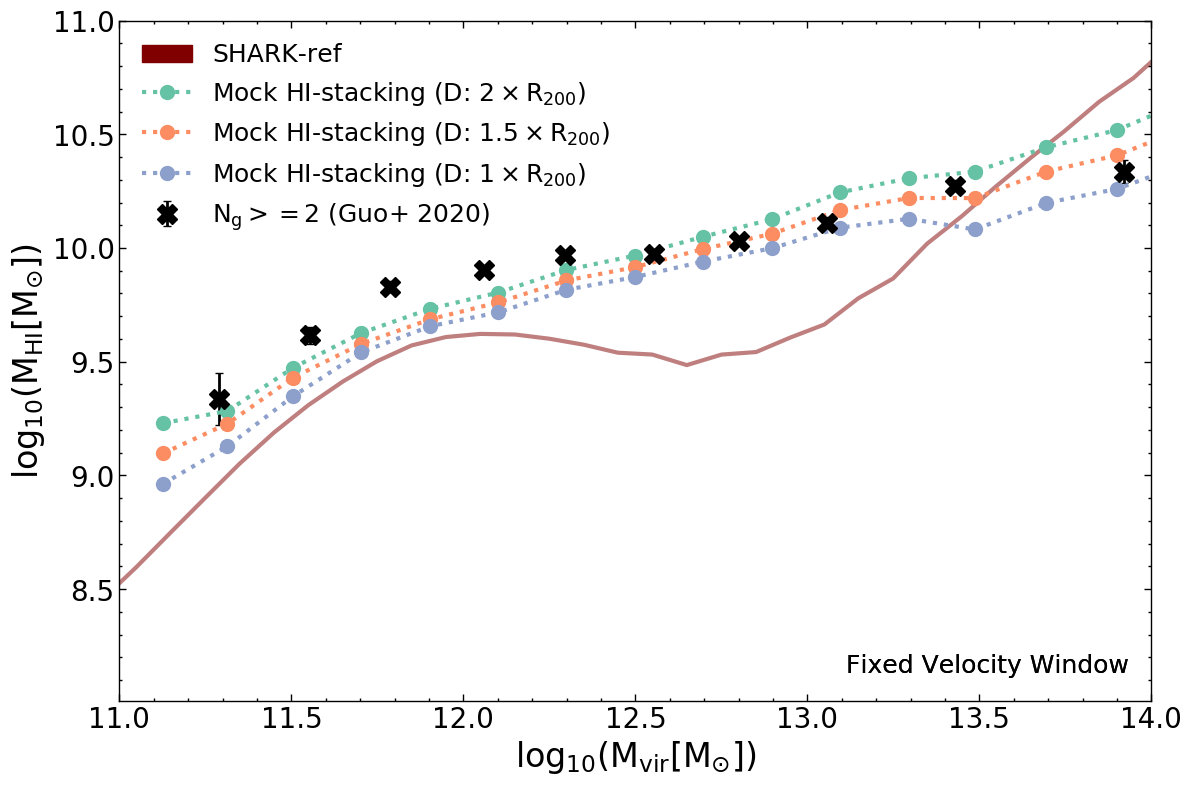}
  \end{minipage}
\caption{Similar to Figures \ref{fig:Zcut_window_groups}. Here we keep our stacking velocity window fixed at $\Delta v = \pm 700$ \kms, and change the projected aperture for stacking as labelled. The effect of changing the projected aperture is small at the low-mass end, but becomes prominent at higher halo masses.}
\label{fig:Zcut_radius_groups}
\end{figure*}

\begin{figure*}
  \centering
  \begin{minipage}{0.49\textwidth}
    \includegraphics[width=\linewidth, trim=0.3cm 0.3cm 0.3cm 0.2cm, clip]{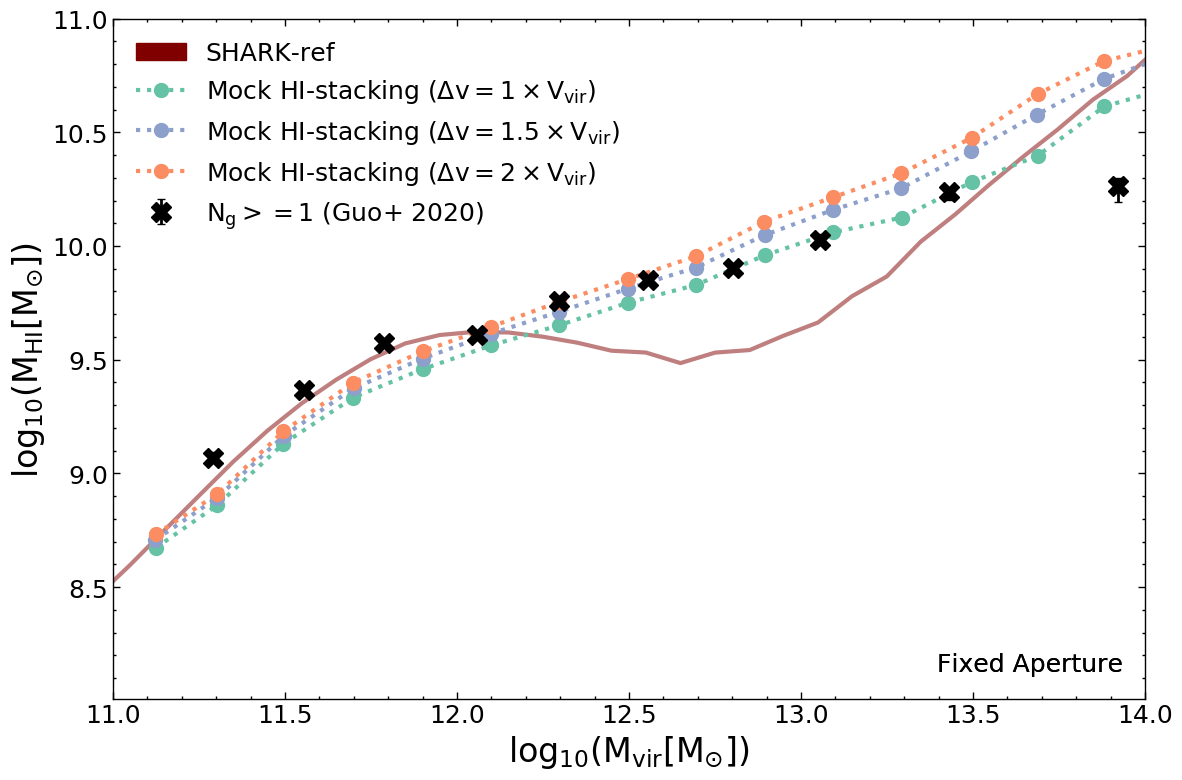}
  
  \end{minipage}
  \hfill
  \begin{minipage}{0.49\textwidth}
    \includegraphics[width=\linewidth, trim=0.3cm 0.3cm 0.3cm 0.2cm, clip]{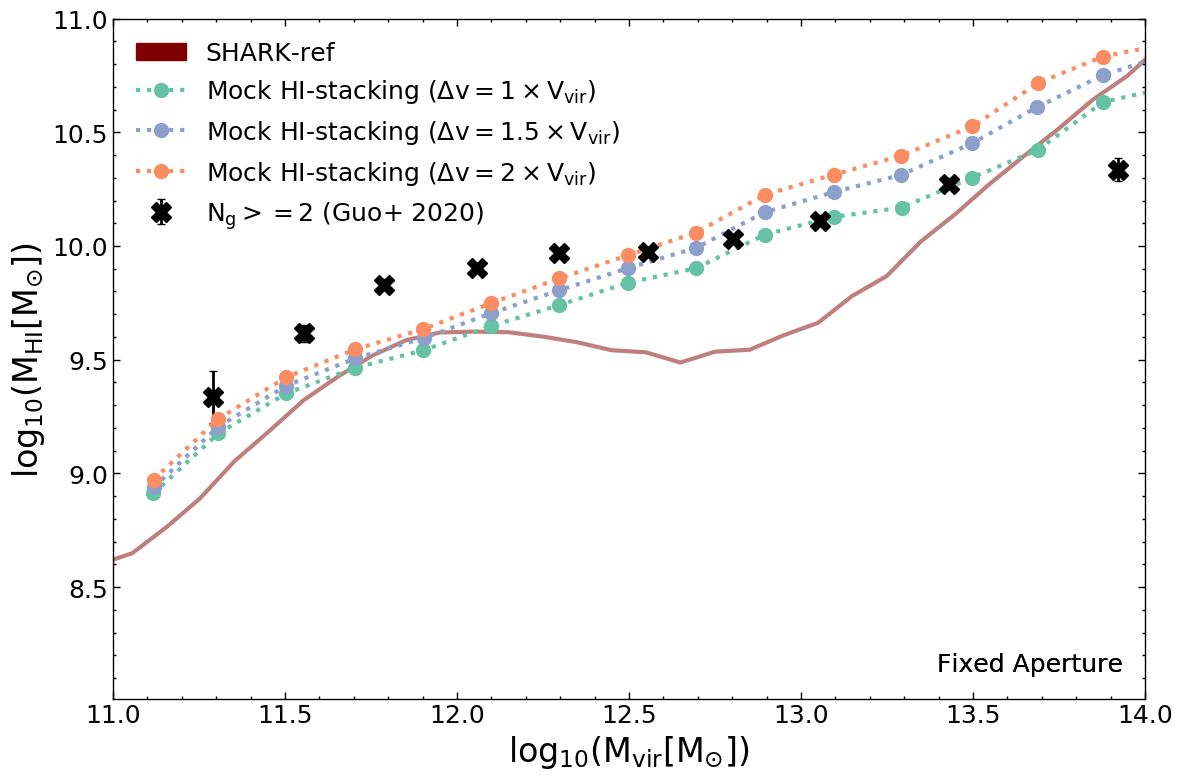}
  \end{minipage}
\caption{Mean \hi\ content of haloes obtained by group \hi~stacking of the groups identified by the group finder, over a fixed projected aperture, for groups with $N_{\rm g} \geq 1$ (left) and $N_{\rm g} \geq 2$ (right) in the mock catalogue, compared to the intrinsic HIHM relation obtained from \shark. Crosses show is the observed HIHM relation from \citet{Guo2020}. The stacking results shown here use an `adaptable velocity-window', which scales according to the virial velocity of the haloes. By using an adaptable velocity-window, the intrinsic HIHM relation for \subsuperscript{M}{vir}{} $\geq$ \solarValue{13} is recovered for both $N_{\rm g} \geq 1$ and $N_{\rm g} \geq 2$ groups.}
\label{fig:Vvircut_window_groups}
\end{figure*}

To understand the effect of the stacking velocity window, we repeat the same exercise, but this time adopting $\Delta v = \pm 350$\kms\ around the group central. For both of these \hi~stacking measurements, we use an aperture of $2$\subsuperscript{R}{200}{}, which is the aperture used by \citet{Guo2020}. The latter is based on the halo mass estimated via abundance matching (see Section \ref{subsubsec:abundance-matching-explaination_chap5}). In Figure \ref{fig:Zcut_window_groups}, we show the stacked results obtained for $\Delta v =\pm 700$\kms\ and $\pm 350$\kms, and compare them against the HIHM relation presented in \citet{Guo2020}. We notice that by using the $\Delta v = \pm 700$\kms\ for our stacking velocity window, we are able to recover the observed HIHM relation for groups that contain at least one galaxy ($N_{\rm g} \geq 1$; left panel) and for groups with two or more members ($N_{\rm g} \geq 2$; right panel) quite closely. A slight underestimation of \hi\ mass in the \subsuperscript{M}{vir}{} range \solarValue{11.5-12} is, however, seen. When we compare the derived HIHM relation obtained using $\Delta v = \pm 350$\kms\ as our stacking velocity window, we find an underestimation of the \hi~mass for haloes with \subsuperscript{M}{vir}{} $\geq$ \solarValue{13} and $N_{\rm g} \geq 1$, and for the entire halo mass range for $N_{\rm g} \geq 2$. We find that the HIHM relation we get from \hi~stacking starts to plateau after \subsuperscript{M}{vir}{} $\geq$ \solarValue{13}, which could be due to the stacking window not being large enough to encompass all the associated \hi~sources of the group. Despite these deviations, we can say that mimicking the stacking procedure of \citet[][]{Guo2020} yields very good agreement between simulations and their observations to within $0.15$ dex. This is not the conclusion we would have arrived at had we instead simply compared the intrinsic relation to the observational inferences of \citet[][]{Guo2020}, as shown in \citet[][]{Chauhan2020}. We notice that changing our stacking velocity window does make a difference in the amount of \hi\ that is measured for a certain group, but it is not a drastic change. An important conclusion here is that we are unable to recover the intrinsic HIHM shape by changing $\Delta v$, and we generally see that the derived HIHM relation starts deviating from the intrinsic one at \subsuperscript{M}{vir}{} $\geq$ \solarValue{12}. The latter is due to the systematic effects of halo mass definition as shown in Section \ref{subsec:shark-simulation-snapshots_chap5}.

In Figure \ref{fig:Zcut_radius_groups}, we keep our stacking velocity window fixed at $\Delta v = \pm 700$, but now change the projected aperture. Similar to Figure \ref{fig:Zcut_window_groups}, changing the aperture does make a difference in the amount of \hi\ measured for a group, but it does not affect the shape of the HIHM relation. By changing the aperture from $1$\subsuperscript{R}{vir}{} to $2$\subsuperscript{R}{vir}{}, we see a change of $\sim 0.5$ dex at the high-mass end (\subsuperscript{M}{vir}{} $\geq$ \solarValue{13}), for both $N_{\rm g} \geq 1$ and $2$. This is expected, as massive haloes contain many galaxies in the outskirts, and increasing/decreasing the aperture directly impacts the number of satellites that contribute to the \hi\ mass for that halo. However, at the low-mass end, haloes have very few or no satellites, so increasing the aperture has less of an effect.

In Figure \ref{fig:Vvircut_window_groups}, instead of using a `fixed velocity window' for stacking our groups, we use the virial velocity of the halo as our guide and produce `adaptable stacking velocity windows'. The adaptable stacking window would be closer to the $\pm 3 \sigma$ window adopted by \citet{Guo2020} for the higher halo masses, where $\Delta v = \pm 700$ \kms\ is too small compared to the velocity dispersion of the groups at that halo mass, which can be as high as, $\approx 1000$ \kms (for \subsuperscript{M}{vir}{} $\approx$ \solarValue{14}, which corresponds to the last halo mass bin). The velocity dispersion of a halo correlates strongly with the virial velocity. The halo mass that we consider here is obtained from abundance matching, which we use to estimate the virial velocity of that halo, thus, ending with a $\Delta v = \pm V_{\rm vir}$, where $V_{\rm vir}$ is the virial velocity of the halo. We aim to see if using a $\Delta v$ that adapts to the halo mass of the galaxy group will help recover the intrinsic shape of the HIHM relation. For our `adaptable stacking velocity window' test, we use the same aperture ($2\times R_{200}$) as we used for our fixed stacking velocity window test. We can see that by using an adaptable window, we are able to recover the HIHM relation in the low-mass  (\subsuperscript{M}{vir}{} $\leq$ \solarValue{12}) and high-mass (\subsuperscript{M}{vir}{} $\geq$ \solarValue{13.5}) end for $N_{\rm g} \geq 1$. We are also able to recover the HIHM relation for the high-mass end for $N_{\rm g} \geq 2$, even though we overestimate the \hi\ content at the low-mass end. We do find, however, that the derived HIHM using these stacking parameters in the mock catalogues is below the observationally derived one of \citet{Guo2020}. If we instead use an adaptable stacking velocity window of $1.5 V_{\rm vir}$, which is closer to the $\pm 3 \sigma$ window that was adopted for the observed HIHM relation, we find that there is still a difference of $\sim 0.3$ dex between the observed HIHM relation and the one we obtain in our mock catalogues at \subsuperscript{M}{vir}{} $\sim$ \solarValue{14}. This could be hinting at a limitation of our simulation, where we are producing satellites that are too \hi-rich in high-mass haloes. We discuss the limitation and effect this has in Section \ref{subsec:HI-physical-error}.  For the groups with \subsuperscript{M}{vir}{} $\leq$ \solarValue{12} (see \textit{right panels} in Figures 8, 9 and 10, which correspond the HIHM relation for groups with 2 or more galaxy members), which generally have few occupants ($N_{\rm g} \leq 5$), the mock-stacked mean \hi\ mass is higher than the intrinsic relation, which could be caused by group membership misallocation. When galaxies residing in two or more haloes are misidentified to be in a common group, group finders tend to misidentify centrals as satellites \citep{Campbell2015-groupfinder-comparison}. This, combined with the large projected aperture and stacking velocity window, results in a higher \hi\ mass estimation for the lower mass end, as seen.

An important takeaway conclusion from these tests is that estimating the halo mass correctly is very important, as the uncertain halo mass measures will wash out features in the shape of the HIHM relation. As the halo mass affects both the projected aperture and velocity considered for \hi~stacking, correctly identifying the halo mass becomes important for minimising contamination. We find that the deviation we see between the derived HIHM relation from the mock catalogues and the intrinsic one after \subsuperscript{M}{vir}{} $\geq$ \solarValue{12}, corresponds to the area where the scatter between the halo mass estimates from abundance matching against the intrinsic halo masses (see Figure \ref{fig:FOF-comparison-GAMA-SDSS-median}) is largest. This can cause many smaller groups to be assigned a larger halo mass than the one they reside in, leading to a bigger aperture size, which in turn can lead to an increase in the \hi~mass due to the inclusion of \hi~sources not associated with the group. Even in the absence of contamination, a systematically higher (lower) halo mass would lead to the \hi\ mass at fixed halo mass being lower (higher), leading to changes in the shape of the relation (as shown in Figure \ref{fig:SphericalCoordinates_halo-estimates-groupfinder}).

\subsubsection{Individual \hi~stacking}
\label{subsec:obs-stacking-DINGO}

\begin{figure}
\centering
  \includegraphics[width=\linewidth, frame]{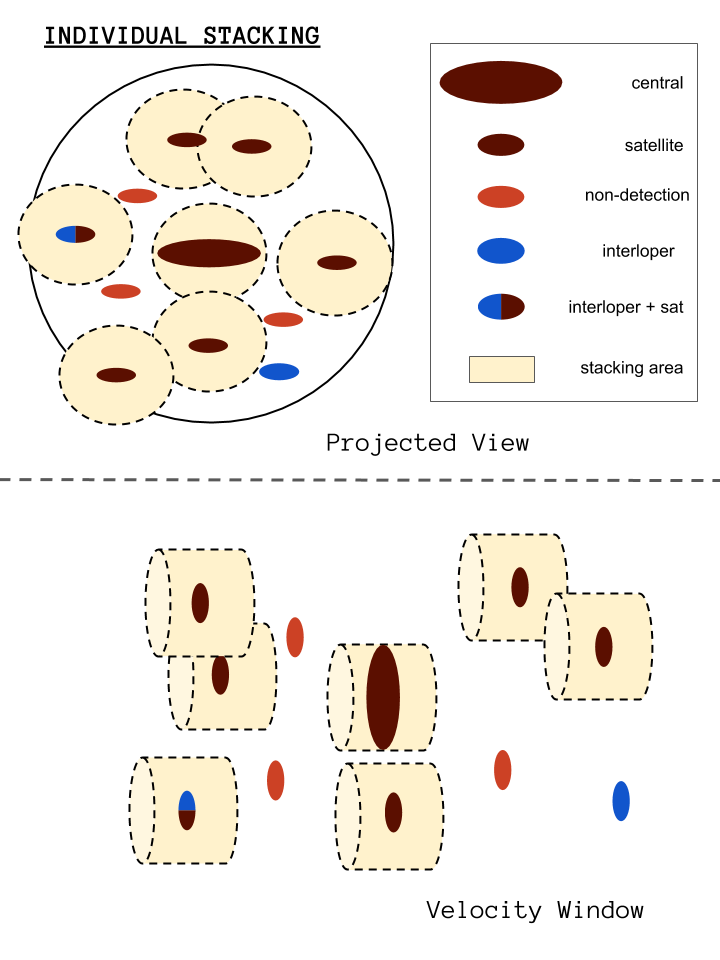}
\caption{Illustration of individual-galaxy \hi~stacking, showing both the projected view (upper panel) and the orientation in redshift space (lower panel). The galaxies in the illustration are the same as shown in Figure \ref{fig:Illustration_GroupStacking}. It can be seen that in the Individual Stacking technique, non-detections that are part of the group are missed, thus making individual stacking technique a way of providing a lower limit to the \hi~content of the group, when compared with group \hi\ stacking.}
\label{fig:Illustration_IndividualStacking}
 \end{figure}

As opposed to the group \hi~stacking, where whole group spectra are stacked, `individual \hi~stacking' involves stacking individual galaxies' \hi\ spectra that are part of the groups of interest. Figure \ref{fig:Illustration_IndividualStacking} shows a schematic of the individual \hi~stacking process. All the galaxies' spectra are stacked individually, and then combined in the halo mass bins of their respective groups, to determine the mean \hi\ mass of groups in that halo mass bin. Using this method ensures that the contamination from the \hi\ sources not part of the group (as recognised by the group finder) is kept at its bare minimum, as those sources would not be stacked. However, because to its reliance on the optical group catalogue, the \hi\ content of galaxies that are part of the group but below the detection limit of the optical survey will be missed.


\begin{figure}
\includegraphics[width=\linewidth, trim=0.3cm 0.3cm 0.3cm 0.2cm, clip]{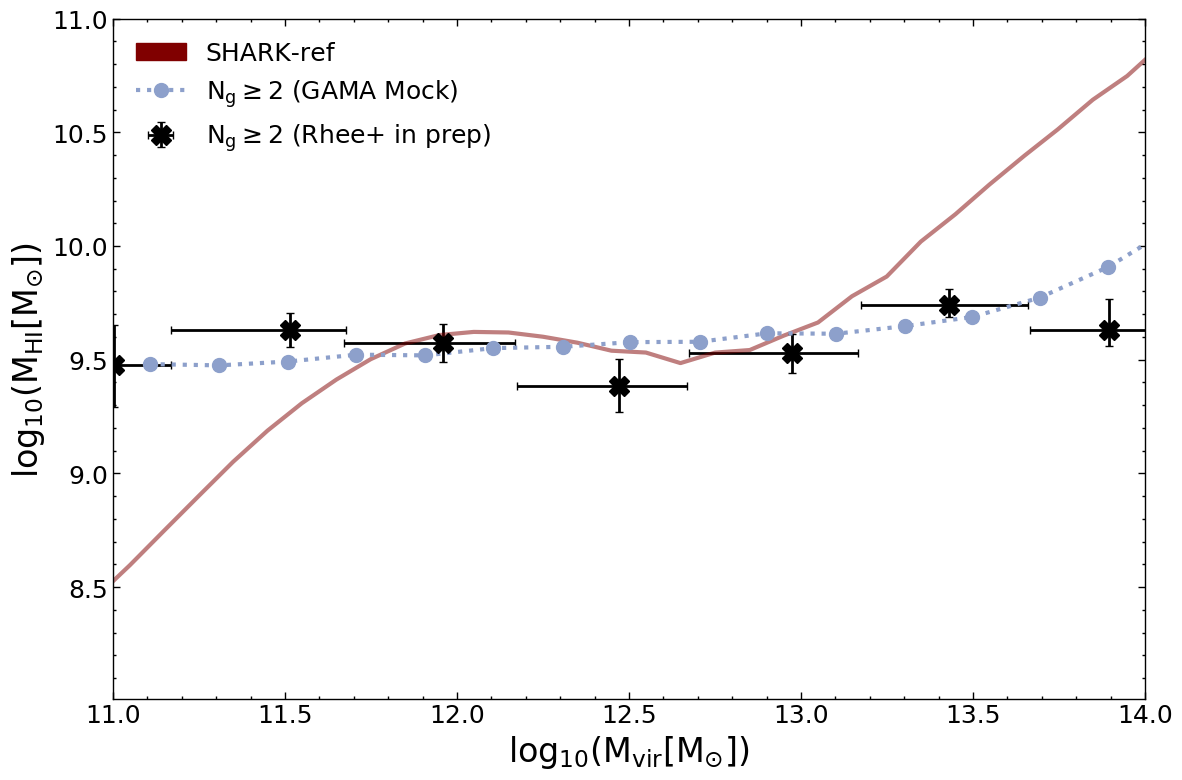}
\caption{Mean \hi~content of haloes obtained by individual \hi~stacking of the groups identified by the group finder, over a fixed aperture covering a square sky region of $49$ kpc, for groups with $N_{\rm g} \geq 2$, and comparing it against the intrinsic HIHM relation obtained from \shark. The scatter points show the observed HIHM relation as obtained from Rhee et. al. (in prep), by stacking the galaxy groups in G23 field. The observed, mock-stacked and intrinsic HIHM relations all agree for \solarValue{11.8} $\leq$ \subsuperscript{M}{vir}{} $\leq$ \solarValue{13}, at the level of observational uncertainties. }
\label{fig:DINGO_GROUPS}
\end{figure}

This method has been used by Rhee et al (in prep) to estimate the \hi~mass of galaxy groups in the G23 field of the GAMA survey, observed using the ASKAP \citep[Australian Square Kilometre Array Pathfinder;][]{Johnston-2008-ASKAP} radio telescope as a part of the early-science phase of DINGO\footnote{https://dingo-survey.org/} \citep[Deep Investigation of Neutral Gas Origin;][]{Meyer-2009-DINGO}. DINGO is a deep \hi\ survey that aims to observe the  \hi~gas content of galaxies out to $z \sim 0.43$ in the five GAMA regions. For their \hi~stacking, Rhee et.al. (in prep) used the early-science data for DINGO, which has a velocity resolution of $\sim 4$ \kms, with a synthesised beam size corresponding to $\sim 40'' \times 27''$. The external spectroscopic optical catalogue used was from \citet{Bellstedt2020-GAMA23}, which is an update from the initial \citet{Robotham2011GalaxyG3Cv1} catalogue, with the photometry performed using \textsc{ProFound} \citep{Robotham2018-Profound} for the GAMA G23 region. For the \hi~stacking, Rhee et.al. (in prep) extract the \hi~spectrum over the spatial pixels covering a square sky region of $\sim 49$ kpc centred on the target galaxy position, over a velocity window of $300$ \kms, calculating the flux density for each spectral channel using the method described in \citet{Shostak1980-HIprofiles}. The extracted spectra are then co-added, weighted by the RMS noise. In this analysis, we consider the \hi~stacking results of Rhee et al (in prep) for the redshift range of $0.039 < z < 0.088$ (Spectral Window ID 3 and 4 in Rhee et al, in prep). This range is slightly above our mock catalogue range (which is complete till $z=0.075$), but it would not have major impacts on our results (this has been discussed with the authors of Rhee et al., in prep). 

Here, we test if we can reproduce the individual stacking results as presented by Rhee et al. (in prep) using our mock survey and mimicking their observational procedure. Figure \ref{fig:DINGO_GROUPS} compares the average HIHM relation for galaxy groups ($N_{\rm g} \geq 2$) derived by Rhee et al. to our mock catalogue. For our measurements, we stack all the galaxies in our group catalogue, using a projected aperture covering a square sky region of $49$ kpc and adding all the \hi~sources around our target galaxy in a velocity window of $\pm 300$ \kms. We group these galaxies according to their corresponding group dynamical mass and calculate the average \hi\ mass in each bin. Our mock-stacked and the observed HIHM relations show similar features, in very good agreement, and remain fairly constant over the entire range. Our mock-stacking results are close to the \hi~mass predictions made by our intrinsic HIHM relation for \solarValue{12} $\leq$ \subsuperscript{M}{vir}{} $\leq$ \solarValue{13}. The observed HIHM points are close to the intrinsic HIHM relation in that region. Surprisingly, the HIHM relation obtained from group \hi~stacking (see Figure \ref{fig:Zcut_window_groups}) starts diverging from the intrinsic HIHM relation at \subsuperscript{M}{vir}{} $\geq$ \solarValue{12}. This hints at the fact that group \hi\ stacking at these halo masses starts to include too many \hi\ sources that are likely not part of the group. For the range, \solarValue{12} $\leq$ \subsuperscript{M}{vir}{} $\leq$ \solarValue{13}, the magnitude limit of GAMA is sufficient to recover all the relevant \hi~sources associated with the group, as can be seen from the agreement between the mock galaxy stacking and the intrinsic HIHM relation. However, as we move to \subsuperscript{M}{vir}{} $\geq$ \solarValue{13}, the GAMA magnitude limit ceases to be deep enough to recover all the \hi~sources. For the groups with \subsuperscript{M}{vir}{} $\leq$ \solarValue{12}, which are generally small groups ($N_{\rm g} \leq 5$), the \hi\ mass measured is higher than the intrinsic relation, which could be caused by group membership misallocation. This, combined with the uncertainty associated with the halo mass estimates from the dynamical mass method, sometimes leads to two centrals being grouped together and given a smaller halo mass (based on their uncertain velocity dispersion), and resulting in a higher \hi\ mass for that halo mass bin. A common conclusion between this test and the group stacking experiment is that by mimicking the observational procedure, \shark\ shows very good agreement with the observations. However, by reproducing the procedure in the observations, we are unable to recover the intrinsic HIHM relation, \textit{highlighting fundamental limitations of the current observational techniques.}

\begin{figure*}
  \centering
  \begin{minipage}{0.49\textwidth}
    \includegraphics[width=\linewidth, trim=0.3cm 0.3cm 0.3cm 0.2cm, clip]{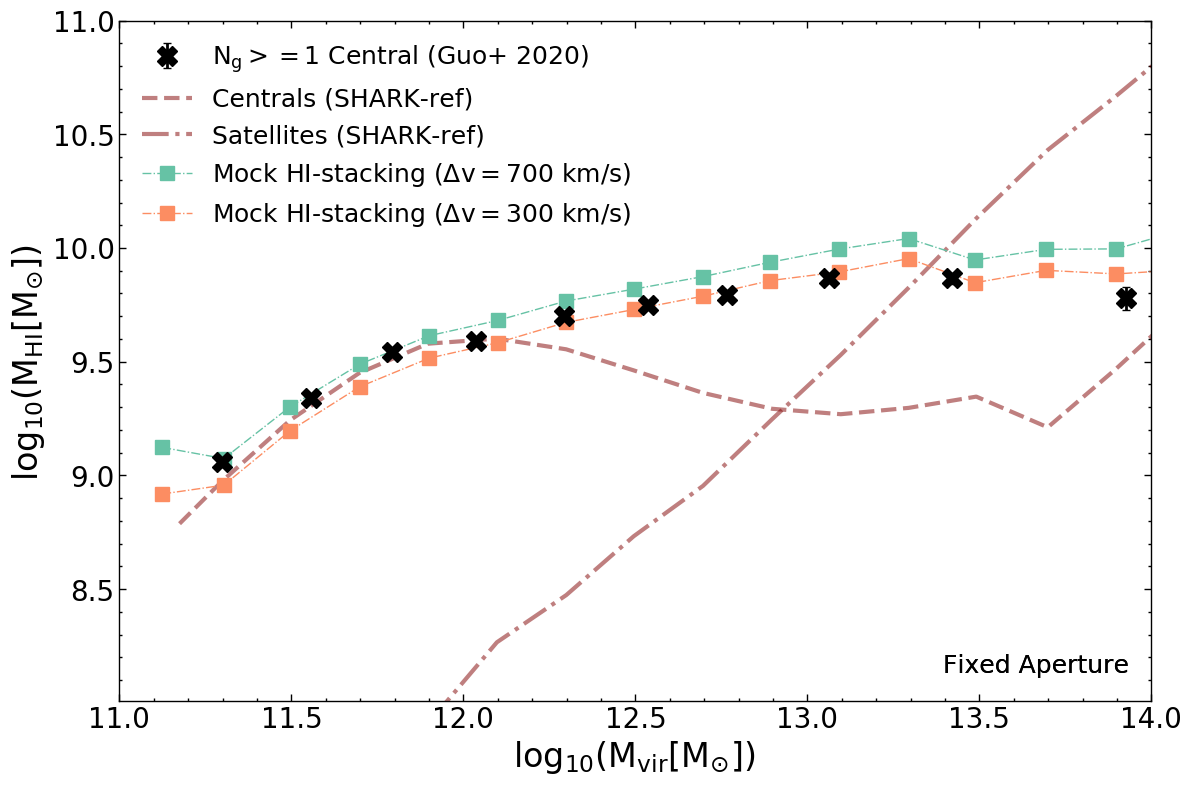}
    \end{minipage}
  \hfill
  \begin{minipage}{0.49\textwidth}
    \includegraphics[width=\linewidth, trim=0.3cm 0.3cm 0.3cm 0.2cm, clip]{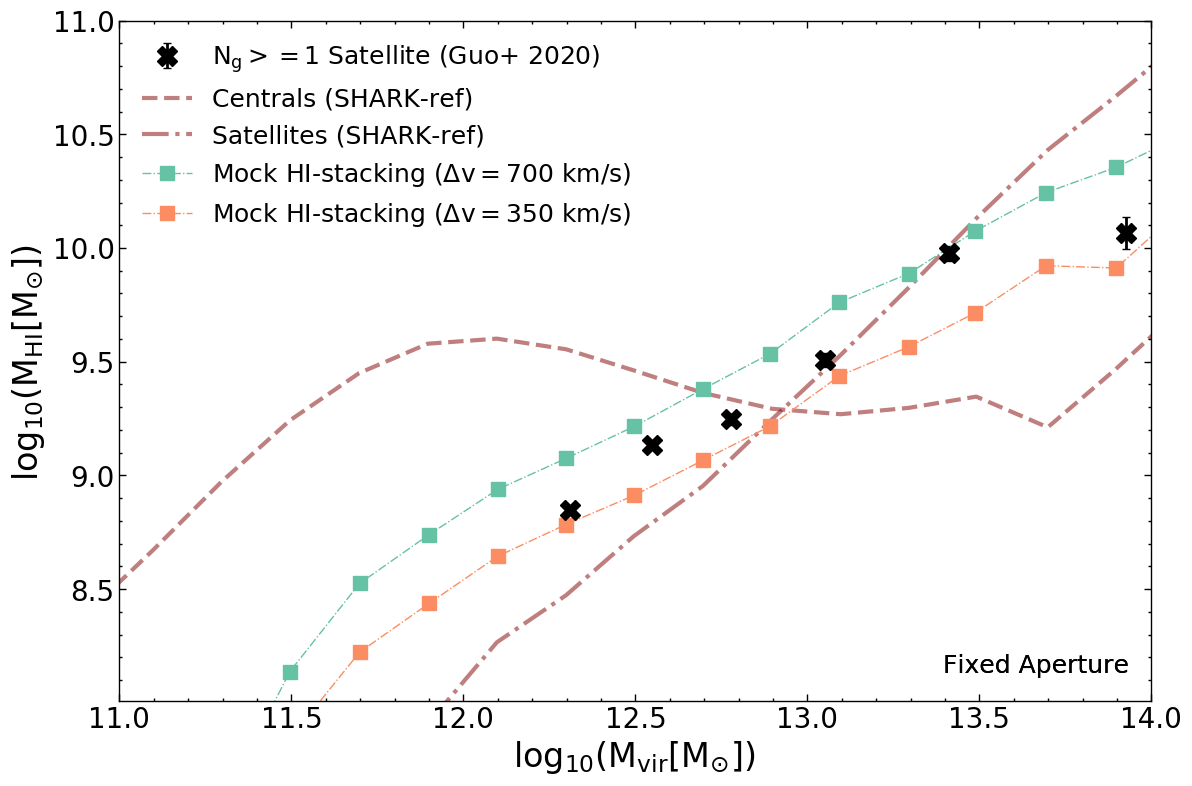}
  \end{minipage}

\caption{Central (left) and satellite (right) contributions to the total \hi~content of haloes obtained by stacking as shown in previous figures. The results are shown for two stacking windows $\Delta v = \pm 300$ \kms (initially used by \citealp{Guo2020}) and $\Delta v = \pm 700$ \kms, for comparison. The black crosses are the observed central and satellite contributions from \citet{Guo2020}. Our mock-stacking results are able to reproduce the observed group-stacked HIHM relation for centrals and satellites. Though, they only match with the intrinsic HIHM relation for \subsuperscript{M}{vir}{} $\leq$ \solarValue{12} and \subsuperscript{M}{vir}{} $\geq$ \solarValue{13} for central and satellite \hi\ contribution, respectively.  }
\label{fig:Vvir_Z_window_cen/sat}
\end{figure*}


\subsubsection{Central and satellite HI estimates}
\label{subsec:central-satellite-lightcone}

In the previous sections, we have discussed the effect that group stacking and individual stacking have on the total average \hi\ mass of a halo. Here, we analyse how the central and satellite contributions to the total \hi\ content of the halo are affected by the \hi~stacking technique used. 

In Figure \ref{fig:Vvir_Z_window_cen/sat}, we compare the \hi~stacking results for centrals (left panel) and satellites (right panel) for $N_{\rm g} \geq 1$, from \citet{Guo2020} with our mock group stacking results. We compare them against the contribution of \hi\ mass by centrals and satellites to the intrinsic HIHM relation. For stacking the centrals in our mock catalogue, we follow the approach adopted by \citet{Guo2020}. We take a projected aperture of $200$ kpc around the central and sum the \hi\ masses of all the galaxies that are within a velocity window of $\Delta v = \pm 300$ \kms. We find that we are able to reproduce the observed \hi\ mass contribution of centrals to the HIHM relation with our mock stacking quite well. In the region of \subsuperscript{M}{vir}{} $\leq$ \solarValue{12}, we find that our mock-stacked and observed \hi\ values of the centrals, are comparable with the intrinsic \hi\ mass contribution of centrals. This is not surprising, as the major contributor to the total \hi\ mass of haloes in this region are isolated centrals. Any contribution from confused satellites in this region will be small. As soon as we enter into the regime where centrals are typically no longer isolated (i.e. have satellites), \subsuperscript{M}{vir}{} $\geq$ \solarValue{12}, we start seeing the effect of confusion. That is, the large aperture and wide velocity window allows some satellites to contribute towards the \hi\ mass inferred for the centrals. The \hi~mass of centrals steadily increases as we go to higher halo masses (\subsuperscript{M}{vir}{} $\geq$ \solarValue{13}), and the distance between the intrinsic central \hi\ mass and the mock-stacked \hi\ mass of centrals keeps increasing (maximum being $0.7$ dex at \subsuperscript{M}{vir}{} $\approx$ \solarValue{13.2}), before coming closer again at \subsuperscript{M}{vir}{} $\approx$ \solarValue{14}. The latter happens because the satellites close to centrals at galaxy cluster scales are extremely gas-poor or even devoid of gas. To provide a comparison of how much difference a bigger stacking velocity window has on the total \hi~mass of the central, we repeat the same exercise, but this time with $\Delta v = \pm 700$ \kms, and find that the \hi\ mass measured for centrals is slightly higher than with the $\Delta v = \pm 300$ \kms\ counterpart. 

For estimating the satellite \hi~contribution (right panel) to our mock \hi-stacked groups, we subtract the \hi~mass measured for the central (with a stacking velocity window of $\Delta v = \pm 300$ \kms) from the total \hi\ content of the halo, which we had measured for a group stacking velocity window of $\Delta v = \pm 700\ \text{and}\ 350 $ \kms, mimicking the approach of \citet{Guo2020}. We are able to reproduce the observed \hi\ mass contribution of satellites by subtracting the central \hi\ contribution from the total \hi\ mass measured with a stacking window of $\Delta v = \pm 700$ \kms, which is closer to the values \citet{Guo2020} used. Our \hi\ estimates for satellites derived by mock stacking show a similar shape to the intrinsic \hi\ contribution of satellites, but display a different slope. Our mock-stacking results overestimate the underlying satellite intrinsic \hi\ contribution for \subsuperscript{M}{vir}{} $\leq$ \solarValue{13}, and underestimate thereafter. The difference between the \hi\ mass measured by observations and our mock-stacked results in the halo mass bin of \subsuperscript{M}{vir}{} $\approx$ \solarValue{14} is a physical effect. This is due to the lack of modelling of ISM stripping for \shark\ satellites. It is likely we would find less \hi\ in \shark\ satellites if the SAM included ISM stripping from environmental effects \citep[][]{Stevens2017PhysicalSage}. The lack of ISM-stripping modelling is most significant for haloes with \subsuperscript{M}{vir}{} $\geq$ \solarValue{14}. This is because generally the intra-group medium is not dense enough as to ram pressure strip significant amounts of the ISM content of satellites \citep[see ][for an analysis of this in hydrodynamical simulations]{Marasco2016-ISMstripping_EAGLE}.


\begin{figure*}

\begin{minipage}{0.49\textwidth}
    \includegraphics[width=\linewidth, trim=0.3cm 0.3cm 0.3cm 0.2cm, clip]{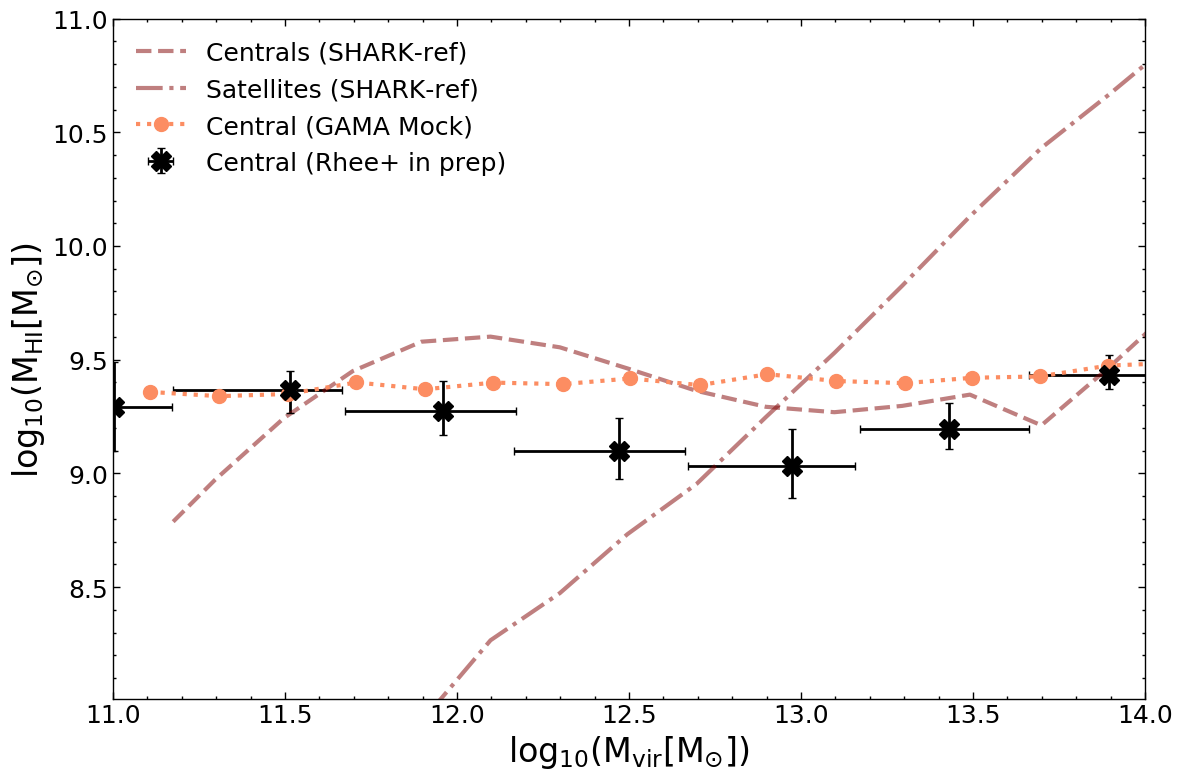}
  
  \end{minipage}
  \hfill
  \begin{minipage}{0.49\textwidth}
    \includegraphics[width=\linewidth, trim=0.3cm 0.3cm 0.3cm 0.2cm, clip]{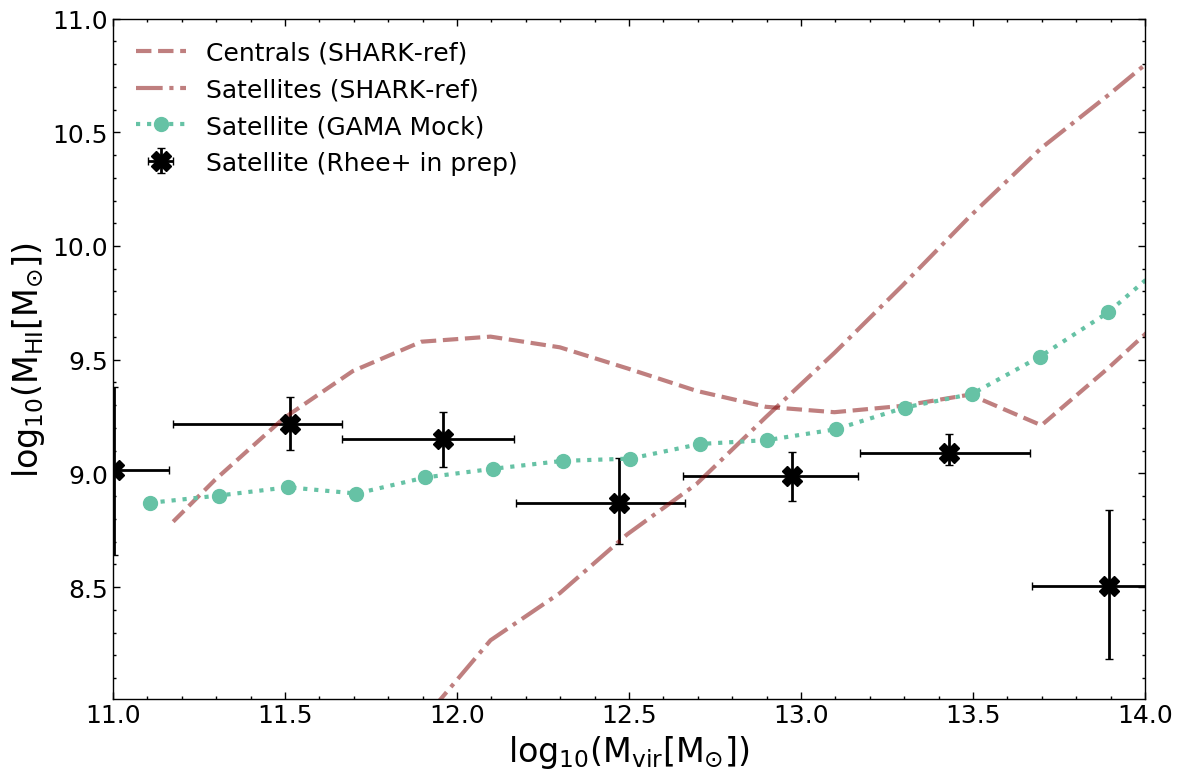}
  \end{minipage}
\caption{Similar to Figure \ref{fig:Vvir_Z_window_cen/sat}, the central (left) and satellite (right) contributions to the total \hi~content of haloes as measured by individual \hi~stacking. The black points with error-bars are the individual stacking results from Rhee et al. (in prep). Our mock-stacking results estimate higher \hi~masses for group centrals in the halo mass region of \solarValue{12} $\leq$ \subsuperscript{M}{vir}{} $\leq$ \solarValue{13}.}

\label{fig:DINGO_plots}
\end{figure*}
Unlike the central \hi\ measurement for group stacking, we find that with individual stacking the observed measurements remain lower than the intrinsic central \hi\ contribution. Figure \ref{fig:DINGO_plots} compares the \hi\ contributions of centrals (left panel) and satellites (right panel) to the total \hi\ mass of the group for \begin{enumerate*}[label=(\roman*)]
    \item observations as will be presented in Rhee et al. (in prep),
    \item our mock individual stacking results and
    \item the intrinsic contributions as calculated from \shark-ref.
\end{enumerate*} Despite being able to reproduce the group \hi\ mass measurements for the individual stacking observations via our mock-stacking technique, we predict higher \hi\ masses for centrals in the \solarValue{12} $\leq$ \subsuperscript{M}{vir}{} $\leq$ \solarValue{13.5} region. This difference is caused by the higher percentage of red centrals that are observed by Rhee et al. (in prep) in that mass bin. Rhee et al. (in prep) stack the \hi\ spectra from red and blue galaxies, separately, measuring different mean \hi\ masses for them and then using these to estimate the mean \hi\ mass of the centrals in a halo mass bin. Red galaxies tend to be gas-poor. Even when stacked, red centrals contribute relatively little  \hi\ compared to their blue counterparts. The fraction of red centrals in the region of disagreement between mock-stacked results and observations is higher than the blue centrals. This brings down the observed average \hi\ mass of centrals in that halo mass bin. As for our mock-stacked centrals, we find that the region of \solarValue{12} $\leq$ \subsuperscript{M}{vir}{} $\leq$ \solarValue{13.5} is dominated by lowly star-forming galaxies (which might or might not appear red when observed). The average \hi\ mass associated with them is \mhi $\approx 1.5 \times$\solarValue{9}. Compared to our values, Rhee et al. (in prep) find the average \hi\ mass of their \textit{red centrals} to be \mhi $\sim 0.707 \times$\solarValue{9}, and for their \textit{blue centrals} to be \mhi $\sim 1.21 \times$ \solarValue{9}.

Individual \hi~stacking uses both a small aperture and velocity window, thereby minimising the effect of confused sources. In our analysis for mock individual \hi~stacking, detections of a second source of \hi\ mass around the targeted galaxy (which could contribute to confusion) are negligible, with $\leq 1$ per cent of galaxies being confused and having their \hi\ mass added to the targeted central. In order to keep our mock individual \hi~stacking simplified, we accommodate the entire \hi\ mass of the galaxy that we target, which might end up with us measuring a higher \hi\ mass for centrals than what will be captured with a projected aperture covering a square sky region of $49$ kpc, if the \hi\ content of the central extends beyond that aperture. According to the observed \hi\ size--mass relation, for \mhi $\geq$ \solarValue{9.7}, the radius at which the density drops below $1$\M/pc$^2$ is larger than $49$ kpc \citep[see][]{Wang2016NewGalaxies, Stevens2019-HIsize-mass_relation}, thus a fixed projected aperture for galaxies at those \hi\ masses might not be enough to encompass all the \hi\ mass associated with them. Irrespective of the difference between observed and mock-stacked results, the characteristic shape of the central HIHM relation is lost. Our mock-stacked centrals lose the bump seen in the intrinsic relation at \subsuperscript{M}{vir}{} $\approx$ \solarValue{12}, though they start agreeing with the intrinsic central \hi\ mass from \subsuperscript{M}{vir}{} $\geq$ \solarValue{12.5}.  

In the right panel of Figure \ref{fig:DINGO_plots}, we find that the observed \hi~stacking measurements for satellites are very different from the intrinsic satellite \hi\ mass contribution. Our mock \hi~stacking results underestimate the \hi\ contribution of satellites for \subsuperscript{M}{vir}{} $\leq$ \solarValue{12} and overestimate thereafter, compared with Rhee et al. (in prep). Unlike the observed stacking points, which show a weak decline from low masses up to \subsuperscript{M}{vir}{} $\leq$ \solarValue{12.3} and rise thereafter (barring the last halo mass bin), our mock-stacking results show a steady increase in the satellite \hi\ contribution as we move towards higher halo masses. The satellite \hi\ mass contribution rises sharply after \subsuperscript{M}{vir}{} $\geq$ \solarValue{13.5} in our mock-stacking results, which is in tension with the observed stacking results. At these high halo masses, the difference between the observations and mock-stacking results is $\sim 1$ dex. The low \hi\ mass estimated for the observed satellites in the last bin is due to the almost equal fraction of blue and red satellites in that mass bin, which causes the average \hi\ mass in that region to be estimated lower than what actually would be. Though closer to the observed \hi\ mass of satellites, our mock stacking results are very different from the intrinsic \hi\ contribution of satellites. Relative to the intrinsic \shark\ relation, our mock stacked results consistently overestimate the satellite \hi\ contribution in \subsuperscript{M}{vir}{} $\leq$ \solarValue{12.7} and underestimate thereafter. We suspect the higher satellite contribution in our mock \hi~stacking result at lower halo masses (\subsuperscript{M}{vir}{} $\leq$ \solarValue{12.5}) could be a direct consequence of centrals being grouped together causing them to occasionally be misidentified as satellites.
This will result in a higher \hi\ mass contribution from the misidentified galaxy, resulting in increasing the mean \hi\ contribution from satellites in that halo mass bin.

\section{Discussion}
\label{sec:Why-parameters-cause-change}

As described in the previous sections, despite using the same underlying mock catalogue for our \hi~stacking experiments, changing the stacking technique leads to a remarkable change in the \hi~measured for galaxy groups. Our main aim in this section is to discuss what causes this change in the shape, and also assess the limitations of the HIHM relation derived using different stacking techniques.

\subsection{The effect of group finder fallibility on the HIHM relation}
\label{subsec:halo-mass-estimate-error-discussion}

The main difference between the intrinsic HIHM relation and the one measured via stacking is the group definitions. Intrinsic HIHM relation uses the halo merger tree catalogues to identify all the subhaloes associated with the host halo, and then calculates the \hi\ masses of the galaxies residing in those subhaloes. As for our \hi-stacking experiment, we rely on the groups as defined by our FoF based group finder, which links galaxies together based on some adopted linkage criteria. Where an ideal group finder would result in a perfect mapping to a simulation halo catalogue, group finders have to work within the limitations of observations in redshift space. According to \citet{Campbell2015-groupfinder-comparison}, there are three major challenges faced by a group finder: \begin{enumerate*}[label=(\roman*)]
    \item halo mass estimation,
    \item central/satellite designation error, and
    \item group-member allocation.
\end{enumerate*}

Already discussed in detail in Section \ref{subsec:halo-mass-obs-comparison_chap5}, comparing the halo mass assigned by a group finder with the intrinsic halo mass of the simulation results in a large scatter. Dynamical mass estimates show a particularly large scatter for \subsuperscript{M}{vir}{} $\leq$ \solarValue{13}. This region is mainly dominated by groups with a low occupancy, mostly $N_{\rm g} \leq 5$, for which the dynamical mass estimates are not reliable and are prone to errors. We demonstrate that for high-membership groups,  dynamical mass estimates show a significantly smaller scatter than the abundance matching estimates, and performs better at recovering the intrinsic halo mass. Although, abundance matching is more reliable in the low-halo-mass region, there is still a scatter of $\sim 0.5$ dex. The first effect of using the group mass estimates is the loss of the characteristic shape of the intrinsic HIHM relation, particularly at the critical \solarValue{12} $\leq$ \subsuperscript{M}{vir}{} $\leq$ \solarValue{12.5} mass region, which is washed out due to the uncertainties inherent to the method. This is a large limitation of the stacking techniques discussed here, as it is exactly around this mass region that AGN feedback imprints its effect, and is also the region that shows the largest differences among simulations (see figure 2 in \citealp{Chauhan2020}).

\begin{figure*}
  \centering
  \begin{minipage}{0.49\textwidth}
    \includegraphics[width=\linewidth, trim=0.3cm 0.3cm 0.3cm 0.2cm, clip]{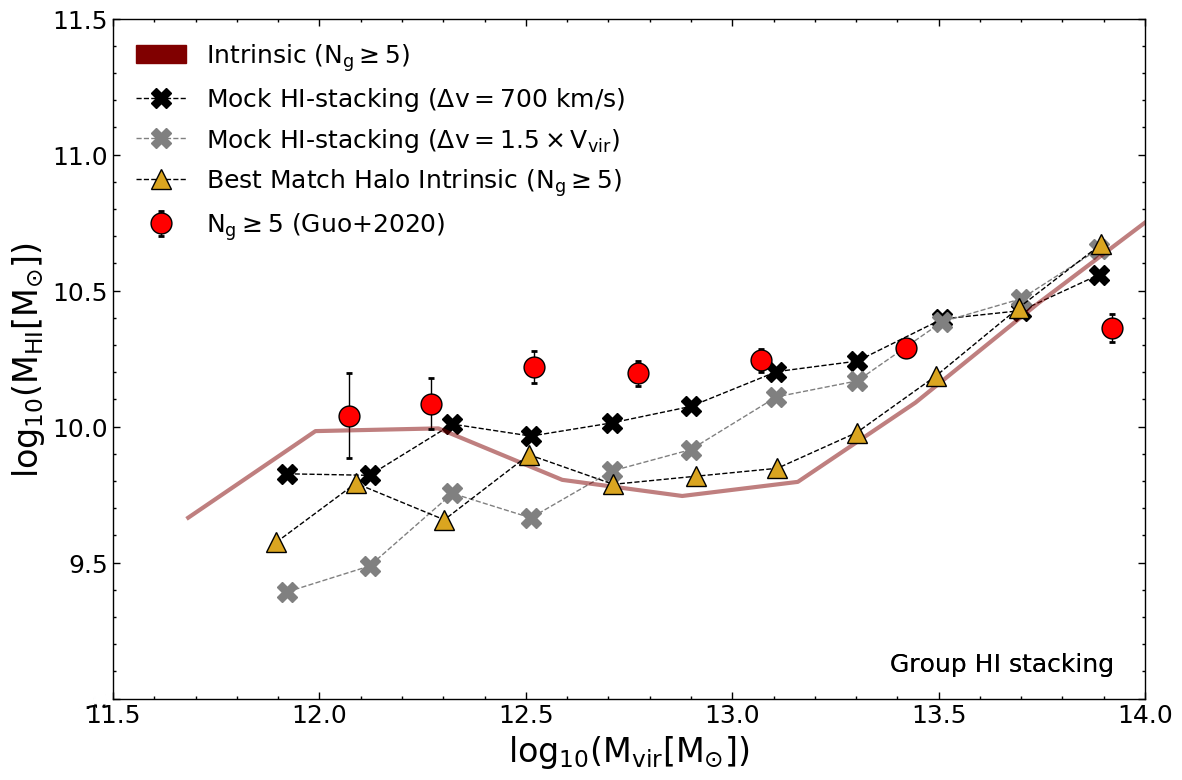}
    \end{minipage}
  \hfill
  \begin{minipage}{0.49\textwidth}
    \includegraphics[width=\linewidth, trim=0.3cm 0.3cm 0.3cm 0.2cm, clip]{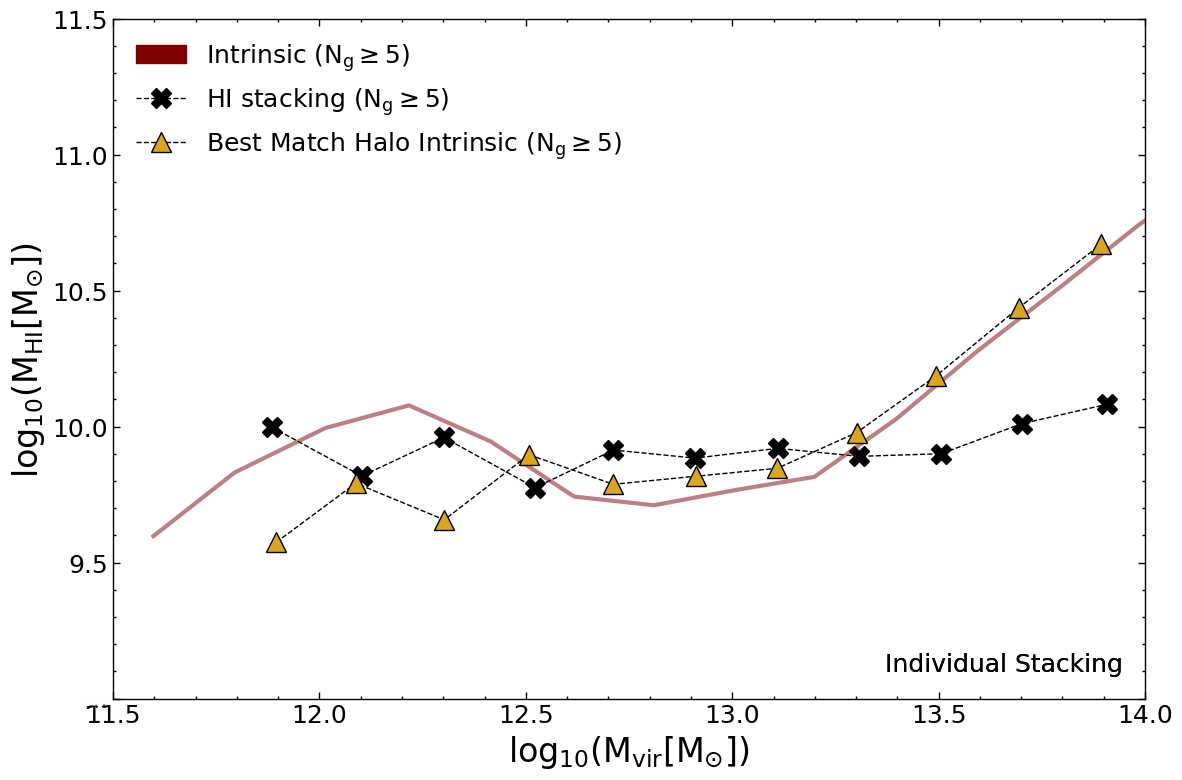}
    \end{minipage}
\caption{Comparing the contamination from \hi\ sources not part of the group for group (left panel) and individual (right panel) \hi~stacking measurements for high-membership groups. The red points are the observed HIHM relation from \citet{Guo2020} for $N_{\rm g} \geq 5$ groups. The yellow triangles and the maroon line represent the HIHM relation derived from the intrinsic galaxy groups and the best match haloes, respectively. The lines with crosses represent the mock \hi~stacking results as labelled. It becomes evident that group \hi~stacking measurements start being affected by contamination from \subsuperscript{M}{vir}{} $\geq$ \solarValue{12}, whereas individual \hi~stacking shows minimal contamination for \solarValue{12.3} $\leq$ \subsuperscript{M}{vir}{} $\leq$ \solarValue{13.2}. }
\label{fig:contamination}
\end{figure*}

The second challenge of the group finder is to identify the central galaxy of the galaxy group, which has paramount importance when estimating the halo mass for abundance matching. The brightest member of a galaxy group is normally recognised as a central galaxy of the group, based on the idea that central galaxies grow in mass by cannibalising their satellites \citep{Dubinski1998-centrals-cluster, Cooray2005-central-assembly}. This way of identifying centrals works in a statistical sense, but it has been shown by \citet{vandenBosch2005-satellite-distribution, Skibba2011-bright-satellites} that in a significant fraction of dark matter haloes the brightest group member is a satellite rather than a true central. This is usually the case in dynamically young groups, in which the most massive galaxies have not yet had time to merge. \citet{Campbell2015-groupfinder-comparison} quantify this fraction and show that the fraction of brightest galaxies not being centrals ranges from $\sim 10$ per cent to $\sim 30$ per cent for haloes of masses \solarValue{13} and \solarValue{14}, respectively. Misidentification of the central galaxy would affect the luminosity and thus, the halo mass estimates made by the abundance matching technique. This is highlighted by the fact that the scatter in the virial mass comparison made for the abundance-matching allocation increased with increasing halo mass (see Figure \ref{fig:central-satellite-fof-comparison}) -- which also coincides with the increase in the fraction of bright satellites. As dynamical mass is not dependent on the luminosities for their halo mass estimates, they do not suffer from it, but still tags a fraction of satellites as centrals.

The final challenge of the group finder is to correctly partition galaxies into their groups. A group finder can fail at this by either assigning galaxies from different haloes to the same group or by assigning galaxies from the same halo to different groups. The former case will lead to two \textit{true} centrals being assigned to a single group, which will lead to one of them being recognised as a satellite and thus provide spurious results. For the latter case, a true satellite could be identified as a central, leading again, to errors. The effect of both these phenomena results in deviations in halo mass estimates, which in turn would lead to uncertainty in the \hi\ mass measured by \hi\ stacking of those groups.

\subsection{The effect of contamination from \hi\ sources not part of the group}
\label{subsec:contamination-groups-error-discussion}

In Section \ref{sec:HI-stacking-obs}, we mentioned that group \hi~stacking is expected to have some contamination from the \hi\ content of galaxies not part of the group, due to those galaxies falling within the large projected aperture and velocity window employed for the stacking. Individual \hi~stacking, by virtue of using a smaller aperture and stacking velocity window, should not suffer from as much contamination. 

To quantify the contamination, we compare the \hi\ mass measured via \hi~stacking with the total \hi\ mass contained in the ``best match halo" counterpart of the group being stacked in our simulation box. We define a halo to be the best match counterpart of our galaxy group by comparing the number of galaxies residing in that group with the number of GAMA/SDSS detectable galaxies residing in a common halo. We use a `purity fraction' \citep[][]{Robotham2011GalaxyG3Cv1} for our analysis, which is defined as \begin{equation}
    f_{\rm pf} = \frac{ (N^{\rm gal}_{\rm shared})^2}{ N^{\rm gal}_{\rm group}\times N^{\rm gal}_{\rm halo}}, 
\end{equation}
\noindent where \subsuperscript{N}{shared}{gal}, \subsuperscript{N}{group}{gal} and \subsuperscript{N}{halo}{gal} is the number of GAMA or SDSS detectable galaxies that are shared between the intrinsic halo in \shark\ and the FoF group in the mock group catalogue, the total number of galaxies in the FoF mock group, and the total number of GAMA or SDSS detectable galaxies in the intrinsic \shark\ halo, respectively. We consider the halo with the highest purity fraction to be the best match counterpart of our galaxy group.   
For example, if a galaxy group defined by the group finder, consists of five member galaxies, where three of these galaxies share a halo that contains nine detectable members, while the other two share a halo with three detectable member galaxies, the purity fraction for these two haloes will be $3/5 \times 3/9 = 0.2$ and $2/5 \times 2/3 \approx 0.27$, respectively. Therefore the latter is considered the best match. 
Once we have identified a best-match halo, we compute the total \hi\ mass content of that halo by simply adding the \hi\ content of all galaxies belonging to it and use that as our `true' \hi\ mass associated with the group. Any deviation from this result (by stacking) is considered to be caused by contamination from \hi\ sources not part of the halo. We also produce an intrinsic HIHM line for comparison, where we consider the total \hi\ mass of the haloes in our lightcone that have at least five or more GAMA or SDSS detectable galaxies. For this comparison, we mock stack only the high-membership groups which have had their best match halo identified ($94$ per cent of the total high-membership groups).  

In Figure \ref{fig:contamination}, we compare the contamination in the measured \hi\ mass of high-membership groups by both group (left panel) and individual (right panel) \hi~stacking. We can see that the \citet[][]{Guo2020} observed HIHM relation for high-membership groups is higher from \subsuperscript{M}{vir}{} $\geq$ \solarValue{12.2} onwards, relative to both the intrinsic and the best-match relations. The mock-stacked (group) measurements (see left panel) underestimate the \hi\ mass for \subsuperscript{M}{vir}{} $\leq$ \solarValue{13}, agree for \subsuperscript{M}{vir}{} $\sim$ \solarValue{13-13.7} and overestimate for \subsuperscript{M}{vir}{} $\geq$ \solarValue{13.7}, compared to the observed HIHM relation. Irrespective of the agreement with the observed HIHM relation, the HIHM relation derived from the mock-stacked (group) \hi\ measurements starts suffering the effects of contamination from \subsuperscript{M}{vir}{} $\geq$ \solarValue{12.2} and \subsuperscript{M}{vir}{} $\geq$ \solarValue{12.6} for the fixed and adaptable velocity stacking window, respectively. The level of contamination -- that is, the difference between the \hi\ mass of the best-matched haloes (yellow triangles) and the derived \hi\ mass via group stacking -- is largest at \solarValue{12.5} $\leq$ \subsuperscript{M}{vir}{} $\leq$ \solarValue{13.2}, according to our mock catalogue. In our analysis we find that, on average, the fraction of contaminants that are central galaxies from other smaller haloes is $\sim 81$ per cent. These centrals have \solarValue{6} $\leq$ \subsuperscript{M}{\star}{} $\leq$ \solarValue{11.8} and are hosted in haloes of \solarValue{10.5} $\leq$ \subsuperscript{M}{vir}{} $\leq$ \solarValue{13.4}. The \textit{average} stellar mass and halo mass weighted by the \hi\ mass of the contaminant galaxies are \solarValue{10.9} and \solarValue{12.8}, respectively. We find that these galaxies have a median \hi-to-stellar mass ratio $\simeq 0.34$.


\begin{figure}
  \centering
    \includegraphics[width=\linewidth, trim=0.3cm 0.3cm 0.3cm 0.2cm, clip]{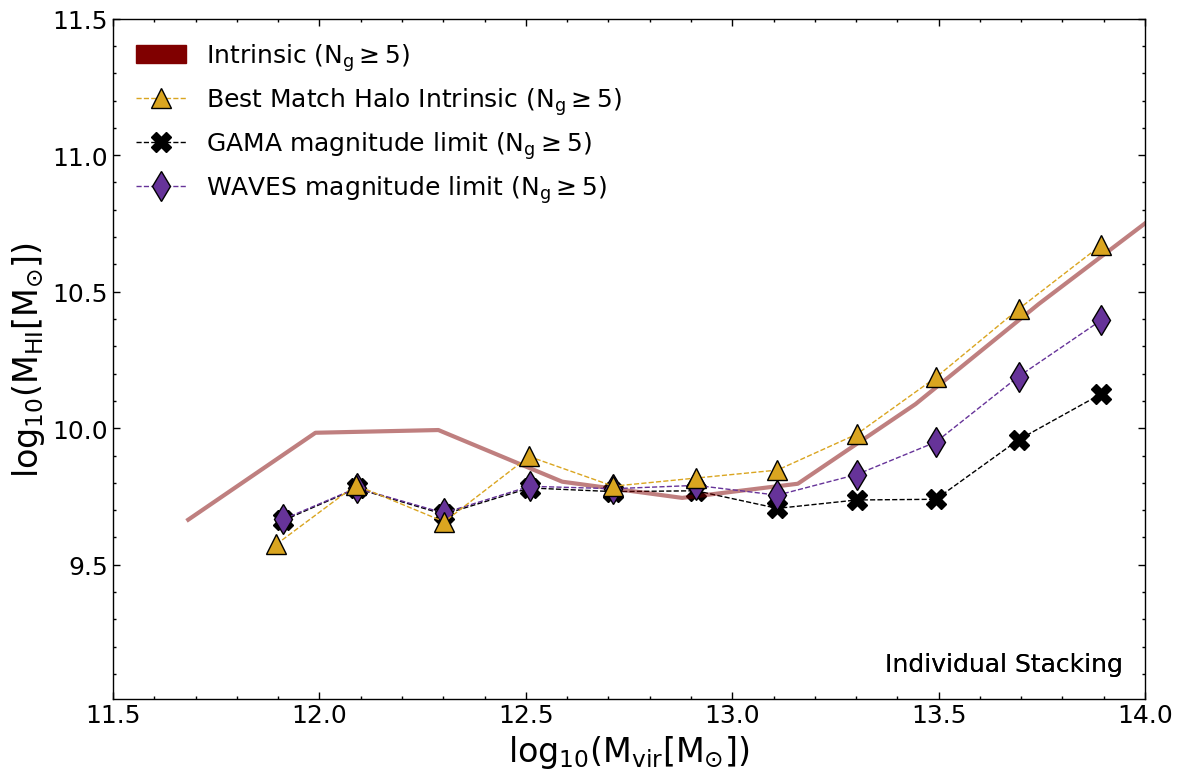}
  \caption[Comparing the \hi\ contribution from the galaxies in the best-match haloes above the GAMA detection limit, against the expected \hi\ mass measurements from WAVES for the best-match haloes in our simulation for high-membership groups.]{Comparing the \hi\ mass contribution from the galaxies in the best-match haloes above the GAMA detection limit (black crosses), against the expected \hi\ mass measurements from WAVES for the best-match haloes in our simulation for high-membership groups. It becomes evident that both the GAMA and WAVES detection limits are able to recover the major \hi\ sources for \solarValue{12.5} $\lesssim$ \subsuperscript{M}{vir}{} $\lesssim$ \solarValue{13.1}, and remain equivalent to the true mean \hi\ mass of the best-match haloes. For \subsuperscript{M}{vir}{} $>$ \solarValue{13.1}, the expected \hi\ mass from WAVES starts diverging from both the GAMA estimates and the true \hi\ mass of the best-match haloes, where it recovers more \hi\ than GAMA but not all. An increase of $0.3$ dex in the \hi\ mass estimates by WAVES is seen over GAMA for \subsuperscript{M}{vir}{} $\geq$ \solarValue{13.6}. We use the virial mass from \shark\ as our halo mass in this plot. }
\label{fig:waves_HI_mass}
\end{figure}

Individual \hi-stacked measurements, on the other hand, agree with the best-match halo \hi\ mass for \solarValue{12}$\leq$ \subsuperscript{M}{vir}{} $\leq$ \solarValue{13.3}, signalling that the contamination is at its bare minimum when using this \hi\ stacking technique. This also signifies that the magnitude limit of GAMA (apparent $r$-band magnitude $\geq 19.5$) is able to recover all the prominent \hi\ sources in that halo mass region that contribute meaningfully to the total \hi\ content of groups. As we move towards \subsuperscript{M}{vir}{} $\geq$ \solarValue{13.3}, our mock-stacked (individual)  \hi\ measurements are lower than the total \hi\ mass of the corresponding best-match halo. This is due to GAMA magnitude limit not being sufficient to detect all the major \hi\ contributing galaxies, as the majority of them (from this analysis) appear to be under the detection limit of GAMA. 



\begin{table*}
\centering

\begin{tabular}{|p{1.5cm}|p{1cm}|p{1cm}|p{1cm}|p{1cm}|p{1cm}|p{1cm}|p{1.25cm}|p{1.25cm}|p{1.25cm}|p{1.25cm}|}
\toprule
\multicolumn{1}{|c|}{Halo Mass} &
  \multicolumn{3}{|c|}{$r$-band [mag]}  &
  \multicolumn{3}{|c|}{$z$-band [mag]} &
  \multicolumn{2}{|c|}{GAMA galaxies} &
  \multicolumn{2}{|c|}{WAVES galaxies}\\
\cmidrule{2-4} \cmidrule{5-7}  \cmidrule{8-9} \cmidrule{10-11}
  {[\M]} & {$5^{th}$} & {$50^{th}$} & {$95^{th}$} & {$5^{th}$} & {$50^{th}$} & {$95^{th}$} & {Satellites} & {\hi\ mass} & {Satellites} & {\hi\ mass}\\
\midrule
\subsuperscript{10}{}{12.5}  & $17.6$ & $20.7$ & $22.8$ & $17.2$ & $20.2$ & $22.5$ & $45.4$\% & $84.8$\% & $60$\% & $88.5$\% \\
\subsuperscript{10}{}{13}  & $19.0$ & $22.0$ & $23.5$ & $18.6$ & $21.8$ & $23.5$ & $27.6$\% & $68.8$\% & $46$\% & $78.8$\% \\
\subsuperscript{10}{}{13.5}  & $21.3$ & $22.7$ & $23.3$ & $21.0$ & $22.5$ & $23.0$ & $10.7$\% & $34.4$\% & $28$\% & $54.2$\% \\
\subsuperscript{10}{}{14}  & $21.4$ & $22.8$ & $23.3$ & $21.4$ & $20.5$ & $23.0$ & $7.6$\% & $21.5$\% & $25$\% & $51.8$\% \\
\bottomrule
\end{tabular}
\caption{Table shows the \subsuperscript{5}{}{th}--\subsuperscript{95}{}{th} percentile distribution of the galaxies of our mock-survey, along with the median value (\subsuperscript{50}{}{th} percentile), in $r$- and $z$-bands residing in different halo mass bins. The column `Satellites' in the GAMA- and WAVES-galaxies section refer to the percentage of galaxies that are detected by the respective surveys to the total number of galaxies residing in that halo. The `\hi\ mass' column refers to the percentage \hi\ mass contribution of these detectable galaxies to the total \hi\ mass of the halo they reside in. We can see that with a deeper survey, such as WAVES, there is a significant increase in the detection of galaxies which will provide better constrains for the HIHM relation. }
\label{tab:magnitudes}
\end{table*}

During our analysis, we found that the median $r$-band and $z$-band magnitude of satellites residing in haloes of masses \subsuperscript{M}{vir}{} $\sim$ (\solarValue{12.5}, \solarValue{13}, \solarValue{13.5}, \solarValue{14}) is (20.65, 22.04, 22.72, 22.83) mag and (20.18, 21.76, 22.46, 22.49), respectively. In Table \ref{tab:magnitudes} we list the \subsuperscript{5}{}{th}--\subsuperscript{95}{}{th} percentile distribution of the galaxies for $r$- and $z$-band magnitudes for different halo mass bins. We find that from the total number of satellites belonging to the haloes of masses \subsuperscript{M}{vir}{} $\sim$ (\solarValue{12.5}, \solarValue{13}, \solarValue{13.5}, \solarValue{14}), GAMA is able to detect $\sim$ (45.4, 27.6, 10.7, 7.6) per cent of satellites, which make up for (84.8, 68.8, 34.4, 21.5) of the total \hi\ mass of those haloes (see Table \ref{tab:magnitudes}). Even though the GAMA magnitude limit is sufficient for detecting all the major \hi\ sources in the \solarValue{12}$\leq$ \subsuperscript{M}{vir}{} $\leq$ \solarValue{13.3} range, it is too bright to recover the major \hi\ contributing galaxies for \subsuperscript{M}{vir}{} $\geq$ \solarValue{13.3}. The satellite population also spreads about $4$ mag in both $r$-band and $z$-band magnitudes, though the majority of them might not be the major \hi\ contributors. Nevertheless, the need for a deeper spectroscopic survey arises to capture more of the missed faint satellites.

The Wide-Area VISTA Extragalactic Survey \citep[WAVES;][]{Driver2019-Waves} is an upcoming spectroscopic survey that aims to map $1200$ \subsuperscript{\rm deg}{}{2} of sky area up to redshift $z=0.2$, complete to an apparent $z$-band magnitude of $21.1$ mag (WAVES-Wide). In Figure \ref{fig:waves_HI_mass}, the purple diamonds represent the total \hi\ mass contribution from satellites that would be detectable by WAVES, for the best match haloes. For comparison, the black crosses here represent the \hi\ mass contribution from the group members that are above the GAMA magnitude limit, with the yellow triangles comprising the \hi\ mass from all galaxy members of the best-match halo, irrespective of whether they are above the GAMA or WAVES magnitude limit. We use the intrinsic \shark\ virial masses as our halo masses for this analysis. An increase of about $\sim 0.3$ dex in \hi\ mass measurements is expected for a WAVES-like selection compared to a GAMA-like one at \subsuperscript{M}{vir}{} $\geq$ \solarValue{13.6}. We find that with the deeper spectroscopic range of WAVES, we will be able to detect $\sim$ (60, 46, 28, 25) per cent of all satellites belonging to the haloes at \subsuperscript{M}{vir}{} $\sim $ (\solarValue{12.5}, \solarValue{13}, \solarValue{13.5}, \solarValue{14}), and recovering $\sim$ (88.5, 78.8, 54.2, 51.8) per cent of the total \hi\ mass at those halo masses, according to our mock catalogue (see Table \ref{tab:magnitudes}). The difference between the intrinsic \hi\ mass predicted by \shark\ for \subsuperscript{M}{vir}{} $\sim$ \solarValue{13.8}, and the \hi\ mass measured by GAMA and a survey of the depth of WAVES is $0.6$ and $0.3$ dex, respectively, which in itself is a significant improvement. In addition to recovering a larger fraction of the satellite in massive groups, WAVES will also allow for the measurement of dynamical masses down to lower halo masses compared to GAMA. This improvement is key to recover the true shape of the HIHM relation around the fundamental range of \solarValue{12} $\leq$ \subsuperscript{M}{vir}{} $\leq$ \solarValue{12.5}.

\subsection{The effect of lack of satellite ISM stripping modelling}
\label{subsec:HI-physical-error}

The final piece to solving the puzzle of the discrepancy between the intrinsic and observed HIHM relation lies in the modelling of gas stripping in satellites in \shark. In \shark, the model of ``instantaneous ram-pressure stripping" \citep[as described in][]{Lagos2014-stripping} is used, which assumes that as soon as galaxies become satellites, their halo gas is instantaneously stripped and transferred to the hot gas of the central galaxy. Thus, gas is only allowed to accrete onto the central galaxy in the halo and not onto the satellites. In addtion, satellite galaxies are cut off cosmological accretion as soon as they become satellites, which prevents their halo gas from being replinished.

Though the halo gas of the satellites is stripped in \shark, the cold gas in the discs of the galaxies is not stripped. This lack of modelling of ``ISM stripping" does not affect the total \hi\ in haloes in the regime where the major \hi\ contributor is the central galaxy. As we move into the regime of halo masses where satellites are the major \hi\ reservoir, we see the effects of ram pressure stripping. \citet{Stevens2017PhysicalSage, Stevens2019AtomicSurveys} in their analysis, find that the ram pressure stripping will start affecting the total \hi\ mass of the haloes as soon as we enter the regime where satellites dominate the \hi\ content of haloes (over centrals). The effect of ram pressure stripping continuously increases with halo mass, with satellites residing in more massive haloes displaying lower average gas content relative to those in lower-mass haloes. 
Other simulations, however, show that ram pressure stripping becomes effective only at \subsuperscript{M}{vir}{} $\geq$ \solarValue{14} \citep[][]{Marasco2016-ISMstripping_EAGLE}. We find that the hot gas stripping model employed by \shark\ is sufficient to match observations well for satellites residing in haloes at \subsuperscript{M}{vir}{} $\leq$ \solarValue{14}, as displayed by our intrinsic satellite \hi\ mass contribution matching with the observed satellite \hi\ mass in the group \hi\ stacking results (see Figure \ref{fig:Vvir_Z_window_cen/sat}).  Though, at higher halo masses, a higher \hi\ mass is seen in both mock-stacked as well the intrinsic HIHM relation for the last halo mass bin (corresponding to \subsuperscript{M}{vir}{} $\approx$ \solarValue{14}). This is when the tension between the mock stacked and observed \hi\ mass is seen. This shows that this is an important area that requires revision in the \shark\ model, which we leave for future work.


\section{Conclusions}
\label{sec:conclusion_chap5}
In this work, we have used the \shark\ semi-analytic galaxy formation model to create a mock survey with the area $6900$ \subsuperscript{\rm deg}{}{2} and redshift $z \leq 0.1$. We have also produced a corresponding mock group catalogue for galaxy groups till redshift $z=0.075$, complete for the GAMA magnitude limit. We use this mock survey to analyse the effect that group finding and different \hi\ stacking techniques have on the inferred HIHM relation, and to determine whether the tension between the intrinsic and observed HIHM relation, as reported in \citet[][]{Chauhan2020}, is due to the systematics involved in making the measurements or is physical in nature. We have presented how we mock stack our survey, mimicking the approach adopted by \citet[][]{Guo2020} and Rhee et al. (in prep), to compare different \hi~stacking techniques. 

Our key results can be summarised as follows:

\begin{itemize}
    \item The correct estimation of halo masses for galaxy groups plays a major role in defining the shape of the derived HIHM relation. Irrespective of a group being well recovered or not, as soon as we start using the halo mass estimates from abundance matching or the dynamical mass method, we lose the characteristic intrinsic HIHM shape. Abundance-matching halo mass estimates are more reliable for isolated centrals and small group ($N_g \leq 5$) compared to those derived by dynamical estimates, as they follow a 1:1 relation with the virial masses of haloes in \shark, albeit with significant scatter ($\sim 0.5$ dex). For \textit{higher-membership} groups, the dynamical mass method provides a more reliable halo mass estimate and shows less scatter than the abundance matching estimates. The difference of the median halo mass estimates for high-membership groups between the \shark\ intrinsic values and, those derived from dynamical and abundance-matching estimates is $0.6$ and $0.9$ dex, respectively.
    
    \item We find that by mimicking the \hi~stacking procedure used by different surveys, we are able to reproduce all the observed HIHM relation. We also find that despite making changes to the \hi~stacking projected apertures and stacking windows, used by the surveys, we are unable to recover the intrinsic HIHM relation, which again points to the importance of correct halo mass estimation.
    
    \item We find that the group \hi~stacking suffers from contamination from \subsuperscript{M}{vir}{} $\geq$ \solarValue{12}, due to its reliance on large projected aperture and stacking velocity window. This contamination amounts up to $0.6$ dex in \hi\ mass for groups with \subsuperscript{N}{g}{} $\geq 5$ (Figure \ref{fig:contamination}). Group \hi~stacking, though, successfully recovers the intrinsic HIHM shape for \subsuperscript{M}{vir}{} $\leq$ \solarValue{12}, and is able to recover the total \hi\ mass associated with groups in the halo mass range \subsuperscript{M}{vir}{} $\gtrsim$ \solarValue{14}. Individual \hi~stacking shows minimal contamination and recovers the intrinsic total \hi\ mass for groups residing in \subsuperscript{M}{vir}{} $\approx$ \solarValue{12-13}. Due to the detection limit of GAMA, the \hi\ masses of groups thereafter are underestimated, as the major \hi\ contributing galaxies lie below the detection limit.
    
    \item We estimate that a deeper spectroscopic survey, such as WAVES, will be able to recover $\sim 51-88$ per cent of the total \hi\ mass of the haloes. This will lead to an improvement of $0.3-0.4$ dex in the \hi\ mass measurement for galaxy groups at halo masses \subsuperscript{M}{vir}{} $\sim$ \solarValue{13.5-14} and reliable dynamical mass estimates (with $N_{\rm g} \geq 5$) for haloes down to masses of \solarValue{12}. This improvement is likely sufficient to unveil the true shape of the HIHM relation at the critical halo mass range of $10^{12}$--\solarValue{12.5}, where differences are largest among simulations and the effect of AGN feedback becomes most prevalent.
    
    \item We note that the \hi\ mass estimates from the intrinsic HIHM relation and the results of the mock \hi~stacked HIHM relation for \subsuperscript{M}{vir}{} $\approx$ \solarValue{14} are higher than their observed counterparts. This is likely a result of lack of modelling of ISM stripping in \shark\ satellites, which results in satellites being more \hi-rich in the \subsuperscript{M}{vir}{} $\geq$ \solarValue{14} halo mass range than they need to be to recover observations.
\end{itemize}

The current paucity of observational constraints on the shape, scatter and evolution of the HIHM relation is likely to change in the coming decade, ultimately with the Square Kilometre Array \citep[SKA;][]{Abdalla2005-SKA-paper}, but also with its pathfinders. With the coming of Wide-Area VISTA Extragalactic Survey \citep[WAVES;][]{Driver2019-Waves}, a new era of optical surveys will start, with surveys going as deep as an apparent $z$-band magnitude of $21.1$ mag. This will open a new range of galaxies and \hi\ sources, thus providing data for measuring constraints on the HIHM relation which will likely yield inferences very close to the total \hi\ content of haloes when applying individual stacking. Having complete, deep spectroscopic surveys, such as WAVES, will also allow reliable measurements of halo masses down to \subsuperscript{M}{vir}{} $\sim$ \solarValue{11}, which we find in this work to be required to recover the true shape of the underlying HIHM relation. The depth of these surveys will certainly lead to improvements over the previous \hi\ and optical surveys; however, careful consideration of systematic effects such as those described here will be necessary to make measurements that can be robustly compared with the simulation predictions. 

\section*{Acknowledgements}

We would like to thank Hong Guo and Michael Jones for their constructive comments, guidance and useful discussions. We also thank Aaron Robotham, Rodrigo Tobar and Pascal Elahi for their contribution towards \surfs\ and \shark, and Mark Boulton for his IT help.
GC is funded by the MERAC Foundation, through the Postdoctoral Research Award of CL, and the University of Western Australia. Parts of this research were carried out by the ARC Centre of Excellence for All Sky Astrophysics in 3 Dimensions (ASTRO 3D), through project number CE170100013. CL is funded by ASTRO 3D. ARHS acknowledges receipt of the Jim Buckee Fellowship at ICRAR-UWA. DO is a recipient of an Australian Research Council Future Fellowships (FT190100083) funded by the Australian Government. MB acknowledges the support of the University of Western Australia through a Scholarship for International Research and Ad Hoc Postgraduate Scholarship. This work was supported by resources provided by the Pawsey Supercomputing Centre with funding from the Australian Government and the Government of Western Australia.

\section*{Data Availability}
The data that support the findings of this study are available upon request from the corresponding author, GC. The \surfs\ simulations used in this work can be freely accessed from
\url{https://tinyurl.com/y4pvra87} (micro-\surfs) and \url{https://tinyurl.com/y6ql46d4} (medi-\surfs).



\bibliographystyle{mnras}
\bibliography{Bibliography} 

\begin{thebibliography}{}
\makeatletter
\relax
\def\mn@urlcharsother{\let\do\@makeother \do\$\do\&\do\#\do\^\do\_\do\%\do\~}
\def\mn@doi{\begingroup\mn@urlcharsother \@ifnextchar [ {\mn@doi@}
  {\mn@doi@[]}}
\def\mn@doi@[#1]#2{\def\@tempa{#1}\ifx\@tempa\@empty \href
  {http://dx.doi.org/#2} {doi:#2}\else \href {http://dx.doi.org/#2} {#1}\fi
  \endgroup}
\def\mn@eprint#1#2{\mn@eprint@#1:#2::\@nil}
\def\mn@eprint@arXiv#1{\href {http://arxiv.org/abs/#1} {{\tt arXiv:#1}}}
\def\mn@eprint@dblp#1{\href {http://dblp.uni-trier.de/rec/bibtex/#1.xml}
  {dblp:#1}}
\def\mn@eprint@#1:#2:#3:#4\@nil{\def\@tempa {#1}\def\@tempb {#2}\def\@tempc
  {#3}\ifx \@tempc \@empty \let \@tempc \@tempb \let \@tempb \@tempa \fi \ifx
  \@tempb \@empty \def\@tempb {arXiv}\fi \@ifundefined
  {mn@eprint@\@tempb}{\@tempb:\@tempc}{\expandafter \expandafter \csname
  mn@eprint@\@tempb\endcsname \expandafter{\@tempc}}}

\bibitem[\protect\citeauthoryear{{Abdalla} \& {Rawlings}}{{Abdalla} \&
  {Rawlings}}{2005}]{Abdalla2005-SKA-paper}
{Abdalla} F.~B.,  {Rawlings} S.,  2005, \mn@doi [\mnras]
  {10.1111/j.1365-2966.2005.08650.x}, \href
  {https://ui.adsabs.harvard.edu/abs/2005MNRAS.360...27A} {360, 27}

\bibitem[\protect\citeauthoryear{{Albareti} et~al.,}{{Albareti}
  et~al.}{2017}]{Albareti2017-SDSS_DR7}
{Albareti} F.~D.,  et~al., 2017, \mn@doi [\apjs] {10.3847/1538-4365/aa8992},
  \href {https://ui.adsabs.harvard.edu/abs/2017ApJS..233...25A} {233, 25}

\bibitem[\protect\citeauthoryear{{Amarantidis} et~al.,}{{Amarantidis}
  et~al.}{2019}]{Amarantidis2019}
{Amarantidis} S.,  et~al., 2019, \mn@doi [\mnras] {10.1093/mnras/stz551}, \href
  {https://ui.adsabs.harvard.edu/abs/2019MNRAS.485.2694A} {485, 2694}

\bibitem[\protect\citeauthoryear{Baugh}{Baugh}{2006}]{Baugh2006}
Baugh C.~M.,  2006, \mn@doi [Reports on Progress in Physics]
  {10.1088/0034-4885/69/12/R02}, 69

\bibitem[\protect\citeauthoryear{Baugh et~al.,}{Baugh
  et~al.}{2019}]{Baugh2018-PMillennium}
Baugh C.~M.,  et~al., 2019, \mn@doi [\mnras] {10.1093/mnras/sty3427}, 483, 4922

\bibitem[\protect\citeauthoryear{Behroozi, Conroy  \& Wechsler}{Behroozi
  et~al.}{2010}]{Behroozi2010-Stellar-halo-mass}
Behroozi P.~S.,  Conroy C.,   Wechsler R.~H.,  2010, \mn@doi [\apj]
  {10.1088/0004-637x/717/1/379}, 717, 379

\bibitem[\protect\citeauthoryear{{Bellstedt} et~al.,}{{Bellstedt}
  et~al.}{2020}]{Bellstedt2020-GAMA23}
{Bellstedt} S.,  et~al., 2020, \mn@doi [\mnras] {10.1093/mnras/staa1466}, \href
  {https://ui.adsabs.harvard.edu/abs/2020MNRAS.496.3235B} {496, 3235}

\bibitem[\protect\citeauthoryear{Benson}{Benson}{2010}]{Benson2010}
Benson A.~J.,  2010, \mn@doi [Physics Reports] {10.1016/j.physrep.2010.06.001}

\bibitem[\protect\citeauthoryear{Blitz \& Rosolowsky}{Blitz \&
  Rosolowsky}{2006}]{Blitz2006}
Blitz L.,  Rosolowsky E.,  2006, \mn@doi [\apj] {10.1086/505417}

\bibitem[\protect\citeauthoryear{{Bravo}, {Lagos}, {Robotham}, {Bellstedt}  \&
  {Obreschkow}}{{Bravo} et~al.}{2020}]{Bravo2020}
{Bravo} M.,  {Lagos} C. d.~P.,  {Robotham} A. S.~G.,  {Bellstedt} S.,
  {Obreschkow} D.,  2020, arXiv e-prints, \href
  {https://ui.adsabs.harvard.edu/abs/2020arXiv200311258B} {p. arXiv:2003.11258}

\bibitem[\protect\citeauthoryear{Brown, Catinella, Cortese, Kilborn, Haynes  \&
  Giovanelli}{Brown et~al.}{2015}]{Brown2015TheGalaxies}
Brown T.,  Catinella B.,  Cortese L.,  Kilborn V.,  Haynes M.~P.,   Giovanelli
  R.,  2015, \mn@doi [\mnras] {10.1093/mnras/stv1311}, 452, 2479

\bibitem[\protect\citeauthoryear{Brown et~al.,}{Brown
  et~al.}{2017}]{Brown2017ColdClusters}
Brown T.,  et~al., 2017, \mn@doi [\mnras] {10.1093/mnras/stw2991}, 466, 1275

\bibitem[\protect\citeauthoryear{{Bruzual} \& {Charlot}}{{Bruzual} \&
  {Charlot}}{2003}]{Bruzual2003}
{Bruzual} G.,  {Charlot} S.,  2003, \mn@doi [\mnras]
  {10.1046/j.1365-8711.2003.06897.x}, \href
  {http://adsabs.harvard.edu/abs/2003MNRAS.344.1000B} {344, 1000}

\bibitem[\protect\citeauthoryear{{Ca{\~n}as}, {Elahi}, {Welker}, {del P Lagos},
  {Power}, {Dubois}  \& {Pichon}}{{Ca{\~n}as}
  et~al.}{2019}]{Canas2019-VELOCIRAPTOR}
{Ca{\~n}as} R.,  {Elahi} P.~J.,  {Welker} C.,  {del P Lagos} C.,  {Power} C.,
  {Dubois} Y.,   {Pichon} C.,  2019, \mn@doi [\mnras] {10.1093/mnras/sty2725},
  \href {https://ui.adsabs.harvard.edu/abs/2019MNRAS.482.2039C} {482, 2039}

\bibitem[\protect\citeauthoryear{{Ca{\~n}as}, {Lagos}, {Elahi}, {Power},
  {Welker}, {Dubois}  \& {Pichon}}{{Ca{\~n}as} et~al.}{2020}]{Canas2020}
{Ca{\~n}as} R.,  {Lagos} C. d.~P.,  {Elahi} P.~J.,  {Power} C.,  {Welker} C.,
  {Dubois} Y.,   {Pichon} C.,  2020, \mn@doi [\mnras] {10.1093/mnras/staa1027},
  \href {https://ui.adsabs.harvard.edu/abs/2020MNRAS.494.4314C} {494, 4314}

\bibitem[\protect\citeauthoryear{Campbell, {Van Den Bosch}, Hearin,
  Padmanabhan, Berlind, Mo, Tinker  \& Yang}{Campbell
  et~al.}{2015}]{Campbell2015-groupfinder-comparison}
Campbell D.,  {Van Den Bosch} F.~C.,  Hearin A.,  Padmanabhan N.,  Berlind A.,
  Mo H.~J.,  Tinker J.,   Yang X.,  2015, \mn@doi [\mnras]
  {10.1093/mnras/stv1091}, 452, 444

\bibitem[\protect\citeauthoryear{{Catinella} et~al.,}{{Catinella}
  et~al.}{2010}]{Catinella2010}
{Catinella} B.,  et~al., 2010, \mn@doi [\mnras]
  {10.1111/j.1365-2966.2009.16180.x}, \href
  {http://adsabs.harvard.edu/abs/2010MNRAS.403..683C} {403, 683}

\bibitem[\protect\citeauthoryear{{Chabrier}}{{Chabrier}}{2003}]{Chabrier2003_IMF}
{Chabrier} G.,  2003, \mn@doi [\pasp] {10.1086/376392}, \href
  {https://ui.adsabs.harvard.edu/abs/2003PASP..115..763C} {115, 763}

\bibitem[\protect\citeauthoryear{{Charlot} \& {Fall}}{{Charlot} \&
  {Fall}}{2000}]{Charlot2000}
{Charlot} S.,  {Fall} S.~M.,  2000, \mn@doi [\apj] {10.1086/309250}, \href
  {http://adsabs.harvard.edu/abs/2000ApJ...539..718C} {539, 718}

\bibitem[\protect\citeauthoryear{Chauhan, Lagos, Obreschkow, Power, Oman  \&
  Elahi}{Chauhan et~al.}{2019}]{Chauhan2019}
Chauhan G.,  Lagos C. d.~P.,  Obreschkow D.,  Power C.,  Oman K.,   Elahi
  P.~J.,  2019, \mn@doi [\mnras] {10.1093/mnras/stz2069}, 488, 5898

\bibitem[\protect\citeauthoryear{{Chauhan}, {Lagos}, {Stevens}, {Obreschkow},
  {Power}  \& {Meyer}}{{Chauhan} et~al.}{2020}]{Chauhan2020}
{Chauhan} G.,  {Lagos} C. d.~P.,  {Stevens} A. R.~H.,  {Obreschkow} D.,
  {Power} C.,   {Meyer} M.,  2020, \mn@doi [\mnras] {10.1093/mnras/staa2251},
  \href {https://ui.adsabs.harvard.edu/abs/2020MNRAS.tmp.2341C} {}

\bibitem[\protect\citeauthoryear{{Chung}, {van Gorkom}, {Kenney}, {Crowl}  \&
  {Vollmer}}{{Chung} et~al.}{2009}]{Chung:2009}
{Chung} A.,  {van Gorkom} J.~H.,  {Kenney} J.~D.~P.,  {Crowl} H.,   {Vollmer}
  B.,  2009, \mn@doi [\aj] {10.1088/0004-6256/138/6/1741}, \href
  {http://adsabs.harvard.edu/abs/2009AJ....138.1741C} {138, 1741}

\bibitem[\protect\citeauthoryear{{Colless} et~al.,}{{Colless}
  et~al.}{2001}]{Colless2001-2dfgrs}
{Colless} M.,  et~al., 2001, \mn@doi [\mnras]
  {10.1046/j.1365-8711.2001.04902.x}, \href
  {https://ui.adsabs.harvard.edu/abs/2001MNRAS.328.1039C} {328, 1039}

\bibitem[\protect\citeauthoryear{{Conroy}}{{Conroy}}{2013}]{Conroy2013-review_SED}
{Conroy} C.,  2013, \mn@doi [\araa] {10.1146/annurev-astro-082812-141017},
  \href {https://ui.adsabs.harvard.edu/abs/2013ARA&A..51..393C} {51, 393}

\bibitem[\protect\citeauthoryear{{Cooray} \& {Milosavljevi{\'c}}}{{Cooray} \&
  {Milosavljevi{\'c}}}{2005}]{Cooray2005-central-assembly}
{Cooray} A.,  {Milosavljevi{\'c}} M.,  2005, \mn@doi [\apjl] {10.1086/432257},
  \href {https://ui.adsabs.harvard.edu/abs/2005ApJ...627L..85C} {627, L85}

\bibitem[\protect\citeauthoryear{{Crain} et~al.,}{{Crain}
  et~al.}{2015}]{Crain2015-EAGLE-reference}
{Crain} R.~A.,  et~al., 2015, \mn@doi [\mnras] {10.1093/mnras/stv725}, \href
  {https://ui.adsabs.harvard.edu/abs/2015MNRAS.450.1937C} {450, 1937}

\bibitem[\protect\citeauthoryear{{Dale}, {Helou}, {Magdis}, {Armus},
  {D{\'{\i}}az-Santos}  \& {Shi}}{{Dale} et~al.}{2014}]{Dale2014}
{Dale} D.~A.,  {Helou} G.,  {Magdis} G.~E.,  {Armus} L.,  {D{\'{\i}}az-Santos}
  T.,   {Shi} Y.,  2014, \mn@doi [\apj] {10.1088/0004-637X/784/1/83}, \href
  {http://adsabs.harvard.edu/abs/2014ApJ...784...83D} {784, 83}

\bibitem[\protect\citeauthoryear{{Davies} et~al.,}{{Davies}
  et~al.}{2019}]{Davies2019}
{Davies} L.~J.~M.,  et~al., 2019, \mn@doi [\mnras] {10.1093/mnras/sty2957},
  \href {https://ui.adsabs.harvard.edu/abs/2019MNRAS.483.1881D} {483, 1881}

\bibitem[\protect\citeauthoryear{{D{\'e}nes}, {Kilborn}, {Koribalski}  \&
  {Wong}}{{D{\'e}nes} et~al.}{2016}]{Denes:2016}
{D{\'e}nes} H.,  {Kilborn} V.~A.,  {Koribalski} B.~S.,   {Wong} O.~I.,  2016,
  \mn@doi [\mnras] {10.1093/mnras/stv2391}, \href
  {http://adsabs.harvard.edu/abs/2016MNRAS.455.1294D} {455, 1294}

\bibitem[\protect\citeauthoryear{{Driver} et~al.,}{{Driver}
  et~al.}{2011}]{Driver2011-GAMA}
{Driver} S.~P.,  et~al., 2011, \mn@doi [\mnras]
  {10.1111/j.1365-2966.2010.18188.x}, \href
  {https://ui.adsabs.harvard.edu/abs/2011MNRAS.413..971D} {413, 971}

\bibitem[\protect\citeauthoryear{{Driver} et~al.,}{{Driver}
  et~al.}{2019}]{Driver2019-Waves}
{Driver} S.~P.,  et~al., 2019, \mn@doi [The Messenger]
  {10.18727/0722-6691/5126}, \href
  {https://ui.adsabs.harvard.edu/abs/2019Msngr.175...46D} {175, 46}

\bibitem[\protect\citeauthoryear{{Dubinski}}{{Dubinski}}{1998}]{Dubinski1998-centrals-cluster}
{Dubinski} J.,  1998, \mn@doi [\apj] {10.1086/305901}, \href
  {https://ui.adsabs.harvard.edu/abs/1998ApJ...502..141D} {502, 141}

\bibitem[\protect\citeauthoryear{{Elahi}, {Welker}, {Power}, {Lagos},
  {Robotham}, {Ca{\~n}as}  \& {Poulton}}{{Elahi} et~al.}{2018}]{Elahi_SURFS}
{Elahi} P.~J.,  {Welker} C.,  {Power} C.,  {Lagos} C.~d.~P.,  {Robotham}
  A.~S.~G.,  {Ca{\~n}as} R.,   {Poulton} R.,  2018, \mn@doi [\mnras]
  {10.1093/mnras/sty061}, \href
  {http://adsabs.harvard.edu/abs/2018MNRAS.475.5338E} {475, 5338}

\bibitem[\protect\citeauthoryear{{Elahi}, {Ca{\~n}as}, {Poulton}, {Tobar},
  {Willis}, {Lagos}, {Power}  \& {Robotham}}{{Elahi}
  et~al.}{2019a}]{Elahi2019-Velociraptor}
{Elahi} P.~J.,  {Ca{\~n}as} R.,  {Poulton} R. J.~J.,  {Tobar} R.~J.,  {Willis}
  J.~S.,  {Lagos} C. d.~P.,  {Power} C.,   {Robotham} A. S.~G.,  2019a, \mn@doi
  [\pasa] {10.1017/pasa.2019.12}, \href
  {https://ui.adsabs.harvard.edu/abs/2019PASA...36...21E} {36, e021}

\bibitem[\protect\citeauthoryear{{Elahi}, {Poulton}, {Tobar}, {Ca{\~n}as},
  {Lagos}, {Power}  \& {Robotham}}{{Elahi} et~al.}{2019b}]{Elahi2019-TreeFrog}
{Elahi} P.~J.,  {Poulton} R. J.~J.,  {Tobar} R.~J.,  {Ca{\~n}as} R.,  {Lagos}
  C. d.~P.,  {Power} C.,   {Robotham} A. S.~G.,  2019b, \mn@doi [\pasa]
  {10.1017/pasa.2019.18}, \href
  {https://ui.adsabs.harvard.edu/abs/2019PASA...36...28E} {36, e028}

\bibitem[\protect\citeauthoryear{{Fabello}, {Catinella}, {Giovanelli},
  {Kauffmann}, {Haynes}, {Heckman}  \& {Schiminovich}}{{Fabello}
  et~al.}{2011}]{Fabello2011-HI-stacking}
{Fabello} S.,  {Catinella} B.,  {Giovanelli} R.,  {Kauffmann} G.,  {Haynes}
  M.~P.,  {Heckman} T.~M.,   {Schiminovich} D.,  2011, \mn@doi [\mnras]
  {10.1111/j.1365-2966.2010.17742.x}, \href
  {https://ui.adsabs.harvard.edu/abs/2011MNRAS.411..993F} {411, 993}

\bibitem[\protect\citeauthoryear{{Giovanelli} et~al.,}{{Giovanelli}
  et~al.}{2005}]{Giovanelli2005}
{Giovanelli} R.,  et~al., 2005, \mn@doi [\aj] {10.1086/497431}, \href
  {https://ui.adsabs.harvard.edu/abs/2005AJ....130.2598G} {130, 2598}

\bibitem[\protect\citeauthoryear{{Gunn} \& {Gott}}{{Gunn} \&
  {Gott}}{1972}]{Gunn:1972}
{Gunn} J.~E.,  {Gott} III J.~R.,  1972, \mn@doi [\apj] {10.1086/151605}, \href
  {http://adsabs.harvard.edu/abs/1972ApJ...176....1G} {176, 1}

\bibitem[\protect\citeauthoryear{Guo, Li, Zheng, Mo, Jing, Zu, Lim  \& Xu}{Guo
  et~al.}{2017}]{Guo2017ConstrainingClustering}
Guo H.,  Li C.,  Zheng Z.,  Mo H.~J.,  Jing Y.~P.,  Zu Y.,  Lim S.~H.,   Xu H.,
   2017, \mn@doi [\apj] {10.3847/1538-4357/aa85e7}, 846, 61

\bibitem[\protect\citeauthoryear{{Guo}, {Jones}, {Haynes}  \& {Fu}}{{Guo}
  et~al.}{2020}]{Guo2020}
{Guo} H.,  {Jones} M.~G.,  {Haynes} M.~P.,   {Fu} J.,  2020, arXiv e-prints,
  \href {https://ui.adsabs.harvard.edu/abs/2020arXiv200404762G} {p.
  arXiv:2004.04762}

\bibitem[\protect\citeauthoryear{{Haynes} et~al.,}{{Haynes}
  et~al.}{2018}]{Haynes2018TheCatalog}
{Haynes} M.~P.,  et~al., 2018, \mn@doi [\apj] {10.3847/1538-4357/aac956}, \href
  {http://adsabs.harvard.edu/abs/2018ApJ...861...49H} {861, 49}

\bibitem[\protect\citeauthoryear{{Johnston} et~al.,}{{Johnston}
  et~al.}{2008}]{Johnston-2008-ASKAP}
{Johnston} S.,  et~al., 2008, \mn@doi [Experimental Astronomy]
  {10.1007/s10686-008-9124-7}, \href
  {http://adsabs.harvard.edu/abs/2008ExA....22..151J} {22, 151}

\bibitem[\protect\citeauthoryear{{Lagos}, {Davis}, {Lacey}, {Zwaan}, {Baugh},
  {Gonzalez-Perez}  \& {Padilla}}{{Lagos} et~al.}{2014}]{Lagos2014-stripping}
{Lagos} C. d.~P.,  {Davis} T.~A.,  {Lacey} C.~G.,  {Zwaan} M.~A.,  {Baugh}
  C.~M.,  {Gonzalez-Perez} V.,   {Padilla} N.~D.,  2014, \mn@doi [\mnras]
  {10.1093/mnras/stu1209}, \href
  {https://ui.adsabs.harvard.edu/abs/2014MNRAS.443.1002L} {443, 1002}

\bibitem[\protect\citeauthoryear{{Lagos}, {Tobar}, {Robotham}, {Obreschkow},
  {Mitchell}, {Power}  \& {Elahi}}{{Lagos} et~al.}{2018}]{Lagos2018-Shark}
{Lagos} C.~d.~P.,  {Tobar} R.~J.,  {Robotham} A.~S.~G.,  {Obreschkow} D.,
  {Mitchell} P.~D.,  {Power} C.,   {Elahi} P.~J.,  2018, \mn@doi [\mnras]
  {10.1093/mnras/sty2440}, \href
  {http://adsabs.harvard.edu/abs/2018MNRAS.481.3573L} {481, 3573}

\bibitem[\protect\citeauthoryear{Lagos et~al.,}{Lagos
  et~al.}{2019}]{Lagos2019-SED}
Lagos C. d.~P.,  et~al., 2019, \mn@doi [\mnras] {10.1093/mnras/stz2427}, 489,
  4196

\bibitem[\protect\citeauthoryear{{Lagos}, {da Cunha}, {Robotham}, {Obreschkow},
  {Valentino}, {Fujimoto}, {Magdis}  \& {Tobar}}{{Lagos}
  et~al.}{2020}]{Lagos2020}
{Lagos} C. d.~P.,  {da Cunha} E.,  {Robotham} A. S.~G.,  {Obreschkow} D.,
  {Valentino} F.,  {Fujimoto} S.,  {Magdis} G.~E.,   {Tobar} R.,  2020, arXiv
  e-prints, \href {https://ui.adsabs.harvard.edu/abs/2020arXiv200709853L} {p.
  arXiv:2007.09853}

\bibitem[\protect\citeauthoryear{{Lim}, {Mo}, {Lu}, {Wang}  \& {Yang}}{{Lim}
  et~al.}{2017}]{Lim2017-SDSS-group}
{Lim} S.~H.,  {Mo} H.~J.,  {Lu} Y.,  {Wang} H.,   {Yang} X.,  2017, \mn@doi
  [\mnras] {10.1093/mnras/stx1462}, \href
  {https://ui.adsabs.harvard.edu/abs/2017MNRAS.470.2982L} {470, 2982}

\bibitem[\protect\citeauthoryear{{Liske} et~al.,}{{Liske}
  et~al.}{2015}]{Liske2015-GAMA}
{Liske} J.,  et~al., 2015, \mn@doi [\mnras] {10.1093/mnras/stv1436}, \href
  {https://ui.adsabs.harvard.edu/abs/2015MNRAS.452.2087L} {452, 2087}

\bibitem[\protect\citeauthoryear{{Lu} et~al.,}{{Lu}
  et~al.}{2016}]{Lu2016-GAP_correction}
{Lu} Y.,  et~al., 2016, \mn@doi [\apj] {10.3847/0004-637X/832/1/39}, \href
  {https://ui.adsabs.harvard.edu/abs/2016ApJ...832...39L} {832, 39}

\bibitem[\protect\citeauthoryear{{Marasco}, {Crain}, {Schaye}, {Bah{\'e}}, {van
  der Hulst}, {Theuns}  \& {Bower}}{{Marasco}
  et~al.}{2016}]{Marasco2016-ISMstripping_EAGLE}
{Marasco} A.,  {Crain} R.~A.,  {Schaye} J.,  {Bah{\'e}} Y.~M.,  {van der Hulst}
  T.,  {Theuns} T.,   {Bower} R.~G.,  2016, \mn@doi [\mnras]
  {10.1093/mnras/stw1498}, \href
  {https://ui.adsabs.harvard.edu/abs/2016MNRAS.461.2630M} {461, 2630}

\bibitem[\protect\citeauthoryear{{Meyer}}{{Meyer}}{2009}]{Meyer-2009-DINGO}
{Meyer} M.,  2009, in Panoramic Radio Astronomy: Wide-field 1-2 GHz Research on
  Galaxy Evolution. p.~15 (\mn@eprint {arXiv} {0912.2167})

\bibitem[\protect\citeauthoryear{Moster, Somerville, Maulbetsch, van~den Bosch,
  Macci{\`{o}}, Naab  \& Oser}{Moster
  et~al.}{2010}]{Moster2010-StellarHalo-mass}
Moster B.~P.,  Somerville R.~S.,  Maulbetsch C.,  van~den Bosch F.~C.,
  Macci{\`{o}} A.~V.,  Naab T.,   Oser L.,  2010, \mn@doi [\apj]
  {10.1088/0004-637x/710/2/903}, 710, 903

\bibitem[\protect\citeauthoryear{{Obreschkow}, {Kl{\"o}ckner}, {Heywood},
  {Levrier}  \& {Rawlings}}{{Obreschkow}
  et~al.}{2009}]{Obreschkow_2009_ligthcone}
{Obreschkow} D.,  {Kl{\"o}ckner} H.~R.,  {Heywood} I.,  {Levrier} F.,
  {Rawlings} S.,  2009, \mn@doi [\apj] {10.1088/0004-637X/703/2/1890}, \href
  {https://ui.adsabs.harvard.edu/abs/2009ApJ...703.1890O} {703, 1890}

\bibitem[\protect\citeauthoryear{Obuljen, Alonso, Villaescusa-Navarro, Yoon  \&
  Jones}{Obuljen et~al.}{2019}]{Obuljen2019TheFromALFALFA}
Obuljen A.,  Alonso D.,  Villaescusa-Navarro F.,  Yoon I.,   Jones M.,  2019,
  \mn@doi [\mnras] {10.1093/mnras/stz1118}, 486, 5124

\bibitem[\protect\citeauthoryear{Padmanabhan \& Refregier}{Padmanabhan \&
  Refregier}{2017}]{Padmanabhan2017ConstrainingHydrogen}
Padmanabhan H.,  Refregier A.,  2017, \mn@doi [\mnras] {10.1093/mnras/stw2706},
  464, 4008

\bibitem[\protect\citeauthoryear{{Planck Collaboration} et~al.,}{{Planck
  Collaboration} et~al.}{2016}]{PlanckXIII}
{Planck Collaboration} et~al., 2016, \mn@doi [A\&A]
  {10.1051/0004-6361/201525830}, 594, A13

\bibitem[\protect\citeauthoryear{{Poulton}, {Robotham}, {Power}  \&
  {Elahi}}{{Poulton} et~al.}{2018}]{Poulton_Treefrog2018}
{Poulton} R.~J.~J.,  {Robotham} A.~S.~G.,  {Power} C.,   {Elahi} P.~J.,  2018,
  \mn@doi [\pasa] {10.1017/pasa.2018.34}, \href
  {http://adsabs.harvard.edu/abs/2018PASA...35...42P} {35}

\bibitem[\protect\citeauthoryear{Robotham et~al.,}{Robotham
  et~al.}{2011}]{Robotham2011GalaxyG3Cv1}
Robotham A. S.~G.,  et~al., 2011, \mn@doi [\mnras]
  {10.1111/j.1365-2966.2011.19217.x}, 416, 2640

\bibitem[\protect\citeauthoryear{{Robotham}, {Davies}, {Driver}, {Koushan},
  {Taranu}, {Casura}  \& {Liske}}{{Robotham}
  et~al.}{2018}]{Robotham2018-Profound}
{Robotham} A.~S.~G.,  {Davies} L.~J.~M.,  {Driver} S.~P.,  {Koushan} S.,
  {Taranu} D.~S.,  {Casura} S.,   {Liske} J.,  2018, \mn@doi [\mnras]
  {10.1093/mnras/sty440}, \href
  {https://ui.adsabs.harvard.edu/abs/2018MNRAS.476.3137R} {476, 3137}

\bibitem[\protect\citeauthoryear{{Robotham}, {Bellstedt}, {Lagos}, {Thorne},
  {Davies}, {Driver}  \& {Bravo}}{{Robotham}
  et~al.}{2020}]{Robotham2020-prospect}
{Robotham} A.~S.~G.,  {Bellstedt} S.,  {Lagos} C. d.~P.,  {Thorne} J.~E.,
  {Davies} L.~J.,  {Driver} S.~P.,   {Bravo} M.,  2020, \mn@doi [\mnras]
  {10.1093/mnras/staa1116}, \href
  {https://ui.adsabs.harvard.edu/abs/2020MNRAS.495..905R} {495, 905}

\bibitem[\protect\citeauthoryear{{Sargent} et~al.,}{{Sargent}
  et~al.}{2014}]{Sargent2014-MolecularGasRedshiftIndependence}
{Sargent} M.~T.,  et~al., 2014, \mn@doi [\apj] {10.1088/0004-637X/793/1/19},
  \href {https://ui.adsabs.harvard.edu/abs/2014ApJ...793...19S} {793, 19}

\bibitem[\protect\citeauthoryear{{Schaye} et~al.,}{{Schaye}
  et~al.}{2015}]{Schaye2015-EAGLE-reference}
{Schaye} J.,  et~al., 2015, \mn@doi [\mnras] {10.1093/mnras/stu2058}, \href
  {https://ui.adsabs.harvard.edu/abs/2015MNRAS.446..521S} {446, 521}

\bibitem[\protect\citeauthoryear{{Shostak} \& {Allen}}{{Shostak} \&
  {Allen}}{1980}]{Shostak1980-HIprofiles}
{Shostak} G.~S.,  {Allen} R.~J.,  1980, \aap, \href
  {https://ui.adsabs.harvard.edu/abs/1980A&A....81..167S} {81, 167}

\bibitem[\protect\citeauthoryear{{Skibba}, {van den Bosch}, {Yang}, {More},
  {Mo}  \& {Fontanot}}{{Skibba} et~al.}{2011}]{Skibba2011-bright-satellites}
{Skibba} R.~A.,  {van den Bosch} F.~C.,  {Yang} X.,  {More} S.,  {Mo} H.,
  {Fontanot} F.,  2011, \mn@doi [\mnras] {10.1111/j.1365-2966.2010.17452.x},
  \href {https://ui.adsabs.harvard.edu/abs/2011MNRAS.410..417S} {410, 417}

\bibitem[\protect\citeauthoryear{Spinelli, Zoldan, Lucia, Xie  \&
  Viel}{Spinelli et~al.}{2019}]{Spinelli2019_Marta}
Spinelli M.,  Zoldan A.,  Lucia G.~D.,  Xie L.,   Viel M.,  2019, The atomic
  Hydrogen content of the post-reionization Universe (\mn@eprint {arXiv}
  {1909.02242})

\bibitem[\protect\citeauthoryear{Stevens \& Brown}{Stevens \&
  Brown}{2017}]{Stevens2017PhysicalSage}
Stevens A. R.~H.,  Brown T.,  2017, \mn@doi [\mnras] {10.1093/mnras/stx1596},
  471, 447

\bibitem[\protect\citeauthoryear{Stevens et~al.,}{Stevens
  et~al.}{2019a}]{Stevens2019AtomicSurveys}
Stevens A. R.~H.,  et~al., 2019a, \mn@doi [\mnras] {10.1093/mnras/sty3451},
  483, 5334

\bibitem[\protect\citeauthoryear{{Stevens}, {Diemer}, {Lagos}, {Nelson},
  {Obreschkow}, {Wang}  \& {Marinacci}}{{Stevens}
  et~al.}{2019b}]{Stevens2019-HIsize-mass_relation}
{Stevens} A. R.~H.,  {Diemer} B.,  {Lagos} C. d.~P.,  {Nelson} D.,
  {Obreschkow} D.,  {Wang} J.,   {Marinacci} F.,  2019b, \mn@doi [\mnras]
  {10.1093/mnras/stz2513}, \href
  {https://ui.adsabs.harvard.edu/abs/2019MNRAS.490...96S} {490, 96}

\bibitem[\protect\citeauthoryear{{Trayford}, {Lagos}, {Robotham}  \&
  {Obreschkow}}{{Trayford} et~al.}{2020}]{Trayford2020-EAGLE}
{Trayford} J.~W.,  {Lagos} C. d.~P.,  {Robotham} A. S.~G.,   {Obreschkow} D.,
  2020, \mn@doi [\mnras] {10.1093/mnras/stz3234}, \href
  {https://ui.adsabs.harvard.edu/abs/2020MNRAS.491.3937T} {491, 3937}

\bibitem[\protect\citeauthoryear{{Vazdekis}, {Koleva}, {Ricciardelli},
  {R{\"o}ck}  \& {Falc{\'o}n-Barroso}}{{Vazdekis} et~al.}{2016}]{Vazdekis2016}
{Vazdekis} A.,  {Koleva} M.,  {Ricciardelli} E.,  {R{\"o}ck} B.,
  {Falc{\'o}n-Barroso} J.,  2016, \mn@doi [\mnras] {10.1093/mnras/stw2231},
  \href {http://adsabs.harvard.edu/abs/2016MNRAS.463.3409V} {463, 3409}

\bibitem[\protect\citeauthoryear{{Villaescusa-Navarro}
  et~al.,}{{Villaescusa-Navarro} et~al.}{2018}]{Villaescusa-Navarro2018}
{Villaescusa-Navarro} F.,  et~al., 2018, \mn@doi [\apj]
  {10.3847/1538-4357/aadba0}, \href
  {https://ui.adsabs.harvard.edu/abs/2018ApJ...866..135V} {866, 135}

\bibitem[\protect\citeauthoryear{Wang, Koribalski, Serra, van~der Hulst,
  Roychowdhury, Kamphuis  \& N.~Chengalur}{Wang
  et~al.}{2016}]{Wang2016NewGalaxies}
Wang J.,  Koribalski B.~S.,  Serra P.,  van~der Hulst T.,  Roychowdhury S.,
  Kamphuis P.,   N.~Chengalur J.,  2016, \mn@doi [\mnras]
  {10.1093/mnras/stw1099}, 460, 2143

\bibitem[\protect\citeauthoryear{{Wechsler} \& {Tinker}}{{Wechsler} \&
  {Tinker}}{2018}]{Wechsler2018-galaxyDMhaloes}
{Wechsler} R.~H.,  {Tinker} J.~L.,  2018, \mn@doi [\araa]
  {10.1146/annurev-astro-081817-051756}, \href
  {https://ui.adsabs.harvard.edu/abs/2018ARA&A..56..435W} {56, 435}

\bibitem[\protect\citeauthoryear{{Yang}, {Mo}, {van den Bosch}  \&
  {Jing}}{{Yang} et~al.}{2005}]{Yang2005-SDSS-groupfinder}
{Yang} X.,  {Mo} H.~J.,  {van den Bosch} F.~C.,   {Jing} Y.~P.,  2005, \mn@doi
  [\mnras] {10.1111/j.1365-2966.2005.08560.x}, \href
  {https://ui.adsabs.harvard.edu/abs/2005MNRAS.356.1293Y} {356, 1293}

\bibitem[\protect\citeauthoryear{{York} et~al.,}{{York}
  et~al.}{2000}]{York2000-SDSS}
{York} D.~G.,  et~al., 2000, \mn@doi [\aj] {10.1086/301513}, \href
  {https://ui.adsabs.harvard.edu/abs/2000AJ....120.1579Y} {120, 1579}

\bibitem[\protect\citeauthoryear{{van den Bosch}, {Yang}, {Mo}  \&
  {Norberg}}{{van den Bosch}
  et~al.}{2005}]{vandenBosch2005-satellite-distribution}
{van den Bosch} F.~C.,  {Yang} X.,  {Mo} H.~J.,   {Norberg} P.,  2005, \mn@doi
  [\mnras] {10.1111/j.1365-2966.2004.08407.x}, \href
  {https://ui.adsabs.harvard.edu/abs/2005MNRAS.356.1233V} {356, 1233}

\makeatother
\end{thebibliography}



\appendix


\bsp	
\label{lastpage}
\end{document}